\documentclass[letterpaper,11pt,fleqn]{article}
\pdfoutput=1
\usepackage{jheppub}

\usepackage{epstopdf}
\usepackage[utf8]{inputenc}
\usepackage{extarrows}
\usepackage{bm,amsmath,amssymb}
\usepackage[mathscr]{eucal}
\usepackage{graphicx}
\usepackage{cancel}
\usepackage{subcaption}

\long\def\comment#1{ }

\newcommand{\eqn}[1]{Eq.~\eqref{#1}}
\newcommand{\beq}{\begin{equation}}
\newcommand{\eeq}{\end{equation}}
\newcommand{\bal}{\begin{align}}
\newcommand{\eal}{\end{align}}
\newcommand{\tr}{{\rm tr}}

\newcommand{\order}[1]{\mathcal{O}{(#1)}}
\newcommand{\abar}{\bar{\alpha}_s}
\newcommand{\nn}{\nonumber\\}
\newcommand{\rmd}{{\rm d}}
\newcommand{\dif}{{\rm d}}

\newcommand{\bk}{\bm{k}}

\newcommand{\bx}{\bm{x}}
\newcommand{\by}{\bm{y}}

\newcommand{\bz}{\bm{z}}

\newcommand{\br}{\bm{r}}
\newcommand{\bb}{\bm{b}}
\newcommand{\xbj}{x_{_{\rm Bj}}}
\newcommand{\Ybj}{Y_{_{\rm Bj}}}
\newcommand{\mcal}{\mathcal}
\newcommand{\bK}{\bm{K}}
\newcommand{\bP}{\bm{P}}

\title{Gluon  dipole factorisation for diffractive dijets}

\author[a]{E.~Iancu,}

\author[b]{A.H.~Mueller,}

\author[c]{D.N.~Triantafyllopoulos\,}

\author[d]{and S.Y.~Wei\,}

\affiliation[a]{Universit\'{e} Paris-Saclay, CNRS, CEA, Institut de physique th\'{e}orique, F-91191, Gif-sur-Yvette, France}

\affiliation[b]{Department of Physics, Columbia University, New York, NY 10027, USA}

\affiliation[c]{European Centre for Theoretical Studies in Nuclear Physics and Related Areas (ECT*)\\and Fondazione Bruno Kessler, Strada delle Tabarelle 286, I-38123 Villazzano (TN), Italy}

\affiliation[d]{Key Laboratory of Particle Physics and Particle Irradiation (MOE), Institute of frontier and interdisciplinary science, Shandong University, Qingdao, Shandong 266237, China}

\emailAdd{edmond.iancu@ipht.fr}
\emailAdd{ahm4@columbia.edu}
\emailAdd{trianta@ectstar.eu}
\emailAdd{shuyi@sdu.edu.cn}

\abstract{Within the colour dipole picture for deep inelastic scattering at small Bjorken $x$, we study
the production of a pair of relatively hard jets via coherent diffraction. By ``relatively hard'' we mean that
the transverse momenta of the two jets --- the quark ($q$) and the antiquark ($\bar q$) generated by the 
decay of the virtual photon --- are much larger than
the target saturation momentum $Q_s(Y_{\mathbb{P}})$ evaluated at the rapidity gap $Y_{\mathbb{P}}$.
We argue that the typical final-state configurations  are such that the hard $q\bar q$ dijets are accompanied 
by a semi-hard gluon jet, with a transverse momentum of the order of $Q_s(Y_{\mathbb{P}})$.
The presence of this third jet ensures that the scattering is strong and thus avoids the strong suppression
of exclusive (hard) dijet production due to colour transparency. For such  ``2+1'' jet configurations, we demonstrate that both the emission of the semi-hard gluon and its
scattering with the hadronic target can be factorised in terms of an effective gluon-gluon dipole.
This effective description, originally proposed in
 \cite{Wusthoff:1997fz,GolecBiernat:1999qd, Hebecker:1997gp,Buchmuller:1998jv}, 
 builds a bridge between the colour dipole picture and collinear 
 factorisation: the cross-section  for diffractive 2+1 jets can be written
as the product between a hard factor describing the $q\bar q$ dijets and a semi-hard factor expressing the 
unintegrated gluon distribution of the Pomeron. The latter is controlled by 
gluon dipole scattering in the black disk limit and hence is strongly sensitive to gluon saturation.
By integrating out the kinematics of the 3 jets, we obtain the $q\bar q g$ contribution to the diffractive structure
function in collinearly-factorised form.

}

\keywords{Perturbative QCD, Deep Inelastic Scattering, Diffraction, Jets, Gluon Saturation}

\begin{document}
\maketitle

\section{Introduction}
\label{sec:intro}

A general question motivating our present study is whether gluon saturation 
\cite{Iancu:2002xk,Iancu:2003xm,Gelis:2010nm,Kovchegov:2012mbw}  can be probed via
{\it relatively hard} photon-nucleus interactions, as in deep inelastic scattering (DIS), or ultraperipheral nucleus-nucleus 
collisions (UPCs). By ``relatively hard'' we mean high-energy processes where at least one of the intervening 
transverse-momentum scales,
like the virtuality of the exchanged photon in DIS, or the transverse momenta of some of the final
particles (hadrons or jets), is much larger than the nuclear saturation momentum $Q_s$ 
 (itself assumed to be semi-hard, i.e. of the order of few GeV). 
Hard processes are appealing from a theoretical standpoint in that they
can be studied via controlled calculations in perturbative QCD. Also, they are often experimentally easier
to measure. Yet, as a general rule, one knows that saturation effects are especially important for {\it semi-hard}
processes, where the exchanged transverse momenta are of the order of $Q_s$. That said, there are a couple
of examples, to be briefly reviewed below, where gluon saturation has an impact on relatively hard processes.
Inspired by these examples, we shall then propose a new such a process, that we believe to be a good
laboratory for studies of gluon saturation in QCD at high energy. This process was briefly discussed
in \cite{Iancu:2021rup}, but our subsequent study will be more complete and also more detailed.

Our first example is diffraction in electron-hadron DIS at high energy, or small Bjorken-$x$ ($\xbj\ll 1$). It is by now
well appreciated that diffraction is controlled by strong scattering in the black disk limit --- hence, by gluon
saturation --- including for hard virtualities $Q^2\gg Q_s^2$ (see e.g. \cite{GolecBiernat:1999qd}). 
The diffractive $\gamma^*p$ (or  $\gamma^*A$) scattering
is mainly driven by very asymmetric, ``aligned-jet'', quark-antiquark  ($q\bar q$) fluctuations of the virtual photon,
where one of the two fermions carries a small  longitudinal
momentum fraction $\vartheta\sim Q_s^2/Q^2\ll 1$. 
 Such asymmetric $q\bar q$ colour dipoles have a relatively large transverse size
$r\sim 1/Q_s$, so they strongly scatter, with an amplitude $\mathcal{T}_{q\bar q}(r)$
of order one. This situation should be contrasted to the case of inclusive DIS at high $Q^2\gg Q_s^2$, 
where the dominant $q\bar q$ configurations are asymmetric as well ($\vartheta(1-\vartheta)\ll 1$), 
yet they are relatively small\footnote{The dominant contribution  to the
total DIS cross-section comes from intermediate dipole sizes within the range $1/Q_s\ll r\ll 1/Q$, 
since the integration over this range yields a large logarithm  $\ln(Q^2/Q_s^2)$ \cite{GolecBiernat:1999qd}.}
 ($r\ll 1/Q_s$) and scatter only weakly  ($\mathcal{T}_{q\bar q}(r)\ll 1$), due to the colour transparency of small dipoles.
This difference stems from the fact that diffraction proceeds via {\it elastic} scattering, hence the 
respective cross-section is proportional to the square $|\mathcal{T}_{q\bar q}(r)|^2$ of the dipole amplitude,
unlike the total  cross-section, which (via the optical theorem) is linear in $\mathcal{T}_{q\bar q}(r)$.

More recently, it has been realised that gluon saturation can also be probed in hard {\it inclusive} processes, provided
one measures particle production in the final state --- more precisely,  {\it multi-particle correlations} \cite{Marquet:2007vb,Albacete:2010pg,Dominguez:2011wm,Metz:2011wb,Dominguez:2011br,Stasto:2011ru,Lappi:2012nh,Iancu:2013dta,Zheng:2014vka,Dumitru:2015gaa,Kotko:2015ura,Marquet:2016cgx,vanHameren:2016ftb,Marquet:2017xwy,Albacete:2018ruq,Dumitru:2018kuw,Boussarie:2021ybe,Kotko:2017oxg,Klein:2019qfb,Hatta:2021jcd,Iancu:2020mos,Caucal:2021ent,Taels:2022tza,Bergabo:2022tcu}. 
A standard example is the production of a pair of hard jets in ``dilute--dense'' collisions (DIS, proton-nucleus collisions,
UPCs) in the ``correlation limit'' \cite{Dominguez:2011wm,Metz:2011wb,Dominguez:2011br,Dumitru:2015gaa,Kotko:2015ura,Marquet:2016cgx,vanHameren:2016ftb,Albacete:2018ruq,Dumitru:2018kuw,Taels:2022tza}. This is the limit 
where the transverse momenta  $k_{1\perp}$ and $k_{2\perp}$ of the two jets are much larger than both the target saturation
momentum $Q_s$ and the dijet imbalance $K_\perp\equiv |\bk_1+\bk_2|$, so the two jets propagate nearly 
back-to-back in the transverse plane.  
Hard dijets are naturally produced in DIS at high virtuality, $Q^2\gg Q_s^2$, but they can also appear in  the final state of low virtuality processes, like $pA$ collisions
(where a quark collinear with the proton can split into a hard quark-gluon pair),
or UPCs (where the exchanged photon is quasi-real).  In all these cases, the
dijet relative momentum $P_\perp\equiv |\bk_1-\bk_2|/2$ provides a natural hard scale.

To leading order in perturbation theory, the dijet imbalance $K_\perp$ 
is equal to the transverse momentum transferred by the target, which can be determined by the physics of saturation
(via multiple soft scattering and the associated transverse momentum broadening). This opens the possibility to 
measure $Q_s$ from the dijet distribution in azimuthal angle (i.e. from the broadening of the back-to-back peak). 
In practice though, this is complicated by the fact that
the $K_\perp$--distribution exhibits a slowly decaying tail $\propto 1/K_\perp^2$ at large 
momenta $K_\perp\gg Q_s$,  as produced via hard scattering off the dilute part of the target gluon distribution.
Accordingly, the {\it typical} dijet events are characterised by a large momentum imbalance, which obscures the effects of saturation on the azimuthal distribution.
The problem is further amplified by the ``Sudakov effect''  \cite{Mueller:2013wwa,Taels:2022tza}, 
i.e. by the recoil due to the final-state radiation, which brings an additional contribution to the dijet imbalance.

To summarise the two previous examples, diffraction is sensitive to gluon saturation because it is controlled by strong scattering,
whereas for hard inclusive dijets, the overall scattering is weak, yet saturation could leave an imprint on the
dijet azimuthal distribution. At this point, it is natural to ask whether one can imagine a process which ``takes the best''
of these examples, that is, a process where a pair of {\it hard} jets is produced {\it diffractively} and 
in such a way that gluon saturation is probed via {\it both} strong scattering and final-state correlations. 
This process cannot be the {\it exclusive} dijet production, which involves the elastic scattering of a small $q\bar q$ dipole
with size $r\sim 1/P_\perp$: besides being insensitive to gluon 
saturation\footnote{Of course, exclusive dihadron production can be sensitive 
to gluon saturation provided the two produced hadrons are semi-hard,
 $P_\perp\sim Q_s$. This case has been thoroughly studied in the recent literature \cite{Altinoluk:2015dpi,Hatta:2016dxp,Hagiwara:2017fye,Mantysaari:2019csc,Salazar:2019ncp,Mantysaari:2019hkq,Hatta:2021jcd}, but it is not the most interesting regime for jet production, since semi-hard jets are difficult to measure.},
this exclusive process is strongly suppressed at large $P_\perp$,
 by colour transparency (see \cite{Salazar:2019ncp} and App.~\ref{sec:el} for an explicit calculation).  

As observed in \cite{Iancu:2021rup}, a more interesting process 
is the diffractive production of {\it three} jets in a special configuration,
which is reminiscent of the correlation limit: two of the jets --- the quark and the antiquark 
 produced by the decay of the virtual photon --- are
relatively hard and nearly back-to-back, $k_{1\perp}\simeq k_{2\perp}\gg Q_s$, whereas the third jet
---  a gluon emitted by either the quark, or the antiquark --- is semi-hard, with $k_{3\perp}\sim Q_s$.
This process was originally addressed within $k_T$--factorisation  \cite{Bartels:1999tn}, 
but that early paper overlooked the key role of  gluon saturation (see also \cite{Kovchegov:2001ni} for a related
study, where the $q\bar q$ dijets are not measured in the final state).

As implicit in the above discussion, we describe this DIS process within the colour dipole picture,
as appropriate at high energy and in a Lorentz frame where the virtual photon itself is very energetic: 
the three partons ($q\bar q g$) are all produced via successive emissions within the photon light-cone wavefunction
(LCWF). Furthermore, $Q_s\equiv Q_s(A, Y_{\mathbb{P}})$ is the target saturation momentum 
 evaluated at a rapidity scale  $Y_{\mathbb{P}}$ equal to the rapidity gap between the  three partons and the target
(see Sect.~\ref{sec:kin} for details on the kinematics).
This ``2+1 jet'' configuration satisfies our above requirements:
\texttt{(i)} it suffers strong elastic scattering, due to
the large transverse separation $R\sim 1/Q_s$ between the gluon and the $q\bar q$ pair;
\texttt{(ii)}  the transverse momentum imbalance $K_\perp=|\bk_1+\bk_2|$ between the two hard jets is of order $Q_s$.
Indeed, we consider the case of  {\it coherent} diffraction, where the net transverse momentum transfer from the target is 
negligible (of the order of the QCD confinement scale $\Lambda$), so $K_\perp\simeq k_{3\perp}$ by 
transverse momentum conservation.
This discussion shows that, even if the semi-hard jet is not measured (this might be difficult in practice), it should 
still have observable consequences on the final state.

It is important to stress that such 2+1 jet configurations are by no means rare events: in fact, they should
be the {\it typical} (diffractive) events with a pair of hard jets in the final state. Indeed, unlike the exclusive dijets,
whose cross-section is rapidly decreasing at $P_\perp\gg Q_s$, the diffractive 2+1 jets are of
leading-twist order, i.e. their cross-section has the same high--$P_\perp$ behaviour  as for 
inclusive dijets. Also, the cross-section is peaked at $K_\perp\sim Q_s$ due to the elastic nature of
the scattering.

The special colour and transverse structure of the $q\bar q g$ system --- after emitting the gluon,
the $q\bar q$ pair forms a colour octet with nearly zero transverse size ($r\sim 1/P_\perp\ll 1/Q_s$)
which is separated from the gluon by a large distance $R\sim 1/Q_s$ --- 
makes it natural to describe this 3-parton system as an {\it effective gluon-gluon dipole}.
This effective picture has been recognised long time ago and used to compute
$F_{2D}^{q\bar q g}$ (the $q\bar q g$ contribution to the diffractive structure function)
at large $Q^2$ and small $x$
\cite{Wusthoff:1997fz,GolecBiernat:1999qd,Hebecker:1997gp,Buchmuller:1998jv}.
Although insightful and inspiring, these early approaches are strictly speaking heuristic in that
they combined ingredients from two different approaches, which in general are not equivalent to each
other: they assumed collinear factorisation between a hard photon-parton cross-section and
diffractive parton distribution functions (DPDFs), but at the same time have treated the produced
partons as projectiles which scatter of the target gluon distribution (in the spirit of the colour dipole 
picture). Yet, as we shall see their final results for $F_{2D}^{q\bar q g}$ are correct and consistent
with a {\it bona fidae} calculation in perturbative QCD, as performed in the colour dipole picture.

Motivated by the problem of diffractive 2+1 jet production, we have recently revisited the gluon dipole
picture and shown that it can be unambiguously established within perturbative QCD \cite{Iancu:2021rup}. 
Namely, we have recovered this picture via a study
of the light-cone wavefunction for the  $q\bar q g$ Fock component of the virtual photon, 
as computed to leading-order in pQCD. On the same occasion, we demonstrated that this picture
comes together with an interesting factorisation:  the  $q\bar q g$  diffractive amplitude factorises
between a ``hard'' component describing the formation of the $q\bar q$ pair and its coupling to the gluon, 
and a ``semi-hard'' component describing the gluon emission from the $q\bar q$ pair, followed by the elastic scattering
of an effective $gg$ dipole. The ``semi-hard'' component turns out to be identical to the ``gluon dipole wavefunction'' postulated by 
Wüsthoff \cite{Wusthoff:1997fz}.

Our original argument in Ref.~\cite{Iancu:2021rup} focused on the case of a very soft gluon,
i.e. a gluon with very small longitudinal momentum fraction $\xi\ll Q_s^2/Q^2$, whose emission can be 
computed in the eikonal approximation. This case, to be discussed in Sect.~\ref{sec:eik} below,
is technically simpler, but already instructive: it exhibits the remarkable, traceless,
tensorial structure of the gluon dipole wavefunction \cite{Wusthoff:1997fz,Hebecker:1997gp}. 
One of our main objectives in this paper is to extend 
the argument in \cite{Iancu:2021rup} to the more interesting case where the gluon is  {\it less} soft, such that $\xi\sim Q_s^2/Q^2$.
This case is considerably more complicated --- it requires a careful treatment of the sub-eikonal corrections 
---, but is also more interesting, since it encompasses the regime of low 
diffractive mass (or moderately large $\beta$; cf. Sect.~\ref{sec:kin}), 
where the rapidity gap $Y_{\mathbb{P}}$ is as large as possible. This is particularly interesting
for a study of gluon saturation, since $Y_{\mathbb{P}}$ is also the longitudinal phase-space for the high-energy 
evolution of the ``Pomeron''  --- the colourless exchange between the $q\bar q g$ system and the target. 
So, by maximising $Y_{\mathbb{P}}$, one ensures that the saturation momentum $Q_s(Y_{\mathbb{P}})$
is as large as possible.

As we shall demonstrate in Sect.~\ref{sect:factx}, the gluon-dipole factorisation of the amplitude remains valid,
up to small changes, for generic values of $\xi$ (i.e.  for a generic diffractive mass).
In turn, this implies a factorised structure for the diffractive (2+1)-jet cross-section, 
which is quite similar to the transverse-momentum dependent (TMD) factorisation 
for inclusive dijets \cite{Dominguez:2011wm}. Specifically, this cross-section 
can be written as the product between a ``hard factor'', which is exactly the same as for inclusive dijets
(this encodes the dependence upon $P_\perp$), and a ``semi-hard factor'', that plays the role of a  {\it diffractive} 
TMD (a transverse-momentum dependent DPDF).
The  ``semi-hard factor'' encodes the dependence upon the momentum imbalance $K_\perp$ 
and upon the rapidity gap $Y_{\mathbb{P}}$.
It is built with the scattering amplitude  $\mathcal{T}_{gg}(R)$ of the $gg$ dipole and is naturally interpreted as
the unintegrated gluon distribution (UGD) of the Pomeron. 

This interpretation is supported by the recent analysis in \cite{Hatta:2022lzj}, which for the first time provided
operatorial definitions for the (quark and gluon) diffractive TMDs, via appropriate generalisations of
the standard definitions for  dependent diffractive parton distributions  (DPDFs) \cite{Trentadue:1993ka,Berera:1995fj,Collins:1997sr}. By evaluating these definitions within  small--$x$ approximations, 
  the authors of \cite{Hatta:2022lzj} obtained a result for the gluon diffractive TMD  (our Pomeron UGD)
  which is indeed consistent\footnote{So far, this agreement has been explicitly verified only via numerical
  comparisons. It would be interesting to establish this equivalence also at analytic level.}
   with our original results in \cite{Iancu:2021rup} and their refined version
 to be presented in this paper.

This ``Pomeron UGD'' already appeared in the pioneering papers  \cite{Wusthoff:1997fz,GolecBiernat:1999qd, Hebecker:1997gp,Buchmuller:1998jv}, but only as an ingredient of a particular contribution to the diffractive
structure function $F_{2D}^{q\bar q g}$ --- that enhanced by the large logarithm $\ln(Q^2/K_\perp^2)$. 
And indeed, by integrating our (2+1)-jet cross-section over the kinematics of the final jets, we shall
recover these previous results for $F_{2D}^{q\bar q g}$, see e.g. Eq.~(39) in 
Ref.~\cite{GolecBiernat:1999qd} and Eq.~(16) in Ref.~\cite{Buchmuller:1998jv} (cf. Sect.~\ref{sec:qqgFD} below).
Very recently, by the time where this paper was essentially completed, an independent derivation of this result
has been presented by Beuf {\it et al.} \cite{Beuf:2022kyp}. This derivation is
similar in spirit to ours, but it directly focuses on $F_{2D}^{q\bar q g}$, so it misses a couple of points which are important for jet production:
the fact that the gluon dipole
factorisation already applies at the amplitude level and that it also holds for relatively symmetric $q\bar q$ pairs\footnote{In
 \cite{Beuf:2022kyp}, the gluon dipole factorisation is demonstrated only at the level of the cross-section, i.e. for the
amplitude squared, and only in the aligned jet limit, where one of the two fermions --- either the quark, or the antiquark
--- has a much smaller longitudinal momentum fraction than the other. This limit is responsible for the logarithmic
enhancement in $F_{2D}^{q\bar q g}$, but is less interesting for jet production; see Sect.~\ref{sec:qqgFD} below.}.
By itself, the existence of a form of  collinear factorisation for $F_{2D}^{q\bar q g}$ at large $Q^2$ is of course
not a surprise. What is remarkable though, is the fact that such a factorisation emerges from the colour dipole 
picture ({\it a priori} valid in a different kinematical region than the collinear factorisation) and that it offers an explicit result
for this diffractive structure function, from first principles.

Albeit previously used for computing $F_{2D}^{q\bar q g}$  \cite{GolecBiernat:1999qd}, 
the Pomeron UGD has not been systematically studied in the literature. In particular, until the recent studies  in \cite{Iancu:2021rup,Hatta:2022lzj}, one failed to notice its
peculiar $K_\perp$--dependence: unlike the  Weiszäcker-Williams UGD (the gluon occupation
number in the target  \cite{Iancu:2003xm}), which enters the TMD factorisation
for inclusive dijets \cite{Dominguez:2011wm} and shows a slowly-decaying tail $\propto 1/K_\perp^2$ 
at large momenta $K_\perp\gg Q_s$, 
the Pomeron UGD shows a rather sharp transition around $K_\perp \sim Q_s$:
it takes values of order one at lower momenta $K_\perp \lesssim Q_s$ (where it is controlled by  
the black disk limit $\mathcal{T}_{gg}(R)\simeq 1$), but it decays very fast, like $1/K_\perp^4$,
in the colour-transparency regime at $K_\perp\gg Q_s$ (see Sect.~\ref{sect:pomeron} for details).  
This in turn implies that the bulk of the diffractive (2+1)-jet events have a transverse momentum imbalance 
$K_\perp \sim Q_s$ between the two hard jets
--- unlike the inclusive dijets, where this distribution is shifted towards larger values $K_\perp\gg Q_s$.

The last point suggests that one should be able to directly measure $Q_s$ from the transverse momentum
imbalance between the hard jets in a diffractive event. This possibility is indeed tantalising and should not be underestimated, 
but one should temper the optimism by recalling that the $K_\perp$--distribution can be strongly modified by the
Sudakov effect (the radiation in the final state).  This effect has not yet been computed for the diffractive process
at hand, but is presumably important. That said, this process has one additional virtue, which is absent
for inclusive dijets: the strong sensitivity to saturation persists after integrating the cross-section over $K_\perp$
(since the integral is dominated by values $K_\perp \sim Q_s$). This operation replaces
the {\it unintegrated} gluon distribution of the Pomeron by its {\it integrated} version and at the same time eliminates
the Sudakov effect. 

Thus, remarkably, even the would-be standard calculation of the diffractive hard dijet production
in the collinear factorisation is in fact {\it controlled} by gluon saturation (at small Bjorken-$x$ and small Pomeron-$x$,
of course). In the usual applications of collinear factorisation, this is obscured by the fact that the Pomeron gluon distribution
(a.k.a. the gluon DPDF) is not really computed from first principles, but merely parametrised to provide the
initial condition for the DGLAP evolution \cite{Gribov:1972ri,Altarelli:1977zs,Dokshitzer:1977sg}
at some moderately hard scale $Q_0^2$. But for sufficiently high energies
($\xbj\ll 1$) and sufficiently large values for the nuclear mass $A$ and/or the rapidity gap $Y_{\mathbb{P}}$ (such that
$Q_s^2(A,Y_{\mathbb{P}})\gg \Lambda^2$), our present calculation provides a leading-order estimate for the gluon DPDF
{\it from first principles}, which is strongly sensitive to saturation and to the high-energy evolution
with increasing $Y_{\mathbb{P}}$: the solution to the B-JIMWLK equations 
\cite{Balitsky:1995ub,JalilianMarian:1997jx,JalilianMarian:1997gr,Kovner:2000pt,Iancu:2000hn,Iancu:2001ad,Ferreiro:2001qy} 
(which for the present purposes can be replaced by the Balitsky-Kovchegov (BK) equation at large $N_c$ \cite{Kovchegov:1999yj}) 
acts as an initial
condition (more precisely, a source term, see Sect.~\ref{sect:dglap}) for the DGLAP evolution.
To our knowledge, this is a unique situation where the high-energy pQCD evolution and the evolution with increasing
virtuality can be matched with each other without further approximations and without any ambiguity. Our numerical
solutions to the combined BK+DGLAP equations demonstrate that the sensitivity of this diffractive process
to gluon saturation --- in particular, the strong peak of the  $K_\perp$--distribution at $K_\perp \sim Q_s$ --- 
is preserved by the ensemble of the parton evolution in pQCD.

This paper is structured as follows. In Sect.~\ref{sec:kin} we specify the kinematics and identify the interesting
physical regimes. In Sect.~\ref{sec:eik} we study the $q\bar q g$ component of the light-cone wavefunction of
the virtual photon in the limit where the gluon emission is sufficiently soft to be treated in the eikonal approximation.
In terms of diffraction, this corresponds to the case of a large diffractive mass, or  small  $\beta\ll 1$.
We thus unveil the effective gluon dipole picture and the associated factorisation, both at the amplitude level and at
the level of the cross-section for diffractive 2+1 jet production. In Sect.~\ref{sect:factx} we extend the previous construction
beyond the eikonal approximation, in such a way to cover the regime of moderate diffractive mass, 
or (moderately) large $\beta\sim\order{1}$. We then formulate the TMD factorisation for the diffractive 2+1 jet cross-section,
from which we deduce the respective collinear factorisation by integrating out the transverse momentum imbalance $K_\perp$ between
the two hard jets. In Sect.~\ref{sect:pomeron} we provide a detailed study  of the gluon distribution of the Pomeron, for
both the ``unintegrated'' version (the gluon diffractive TMD) and the ``integrated'' one (the gluon DPDF). We use the 
McLerran-Venugopalan (MV) model  \cite{McLerran:1993ni,McLerran:1994vd}, and sometimes the GBW model 
\cite{GolecBiernat:1999qd}, as a tree-level approximation and include both the high-energy evolution over the rapidity gap
--- i.e. the evolution of the gluon dipole amplitude described by the BK equation --- and the DGLAP evolution of
the Pomeron gluon distribution --- i.e. the emission of additional gluons\footnote{We ignore dynamical quarks in the DGLAP
evolution, for simplicity.} whose transverse momenta $k_\perp$ are strongly ordered within the range $Q_s^2\ll
k_\perp^2\ll P_\perp^2$. The final section \ref{sec:conc} summarises our main conclusions, proposes a few observables 
associated with 2+1 diffractive jet production, and presents the case for a similar process in the context of UPCs.

\section{Diffractive trijets in the correlation limit: the general picture}
\label{sec:kin}
 
As explained in the Introduction, we are interested in the diffractive production of 3 jets --- 2 relatively 
hard jets plus a semi-hard one --- 
in electron-nucleus ($eA$) deep inelastic scattering (DIS) at relatively high energy, or 
small Bjorken $\xbj\ll 1$. By ``diffractive production'' we more precisely mean a {\it coherent}
process where the scattering is elastic and the hadronic target does not break in the final state.

Let us start by specifying the kinematics. We work in a frame where both the virtual photon and the nuclear target are ultrarelativistic. The photon is a right mover,
with 4-momentum $q^\mu= (q^+, -Q^2/2q^+,\bm{0}_\perp )$  (in light-cone notations) and space-like
virtuality $q^\mu q_\mu=-Q^2$. The nucleus is a left mover,
with 4-momentum $P^\mu_N = (0, P^-_N,\bm{0}_\perp)$ per nucleon (we neglect the nucleon mass).
The Bjorken variable is defined as $\xbj\equiv Q^2/(2 q\cdot P_N)= Q^2/(2 q^+P_N^-)$.

For this kinematics, it becomes appropriate to describe DIS within the {\it colour dipole picture}:
the virtual photon decays into a quark-antiquark ($q\bar q$) pair --- a ``colour dipole'' --- which has a relatively large 
coherence time, much larger than the longitudinal extent of the target:
 $\tau_q=2q^+/Q^2\gg R_A/\gamma$,  with $R_A\simeq A^{1/3}R_N$  the nuclear radius and $\gamma \gg 1$ 
 the Lorentz factor for the boosted nucleus. One roughly has $R_N/\gamma \simeq 1/P_N^-$, so the above condition on 
 the $q\bar q$ coherence time amounts to  $\xbj \ll  A^{-1/3}$ and is indeed well satisfied when $\xbj\le 10^{-2}$.
 In this picture, the QCD scattering refers to the collision between the $q\bar q$ pair and the nucleus.

To produce a 3-jet final state, the quark or the antiquark must also radiate a gluon. This $q\bar q g$ parton system
(``three jets'' at leading order in perturbative QCD) is put on shell by the scattering.
We denote the 4-momenta of the produced partons as $k_i^\mu=(k_i^+,k_i^-,\bk_i)$, with $k_i^-=k_{i\perp}^2/2k_i^+$,
where $i=1,2,3$ refers to the quark, the antiquark, and the gluon, respectively. We shall mostly work with the longitudinal
fractions $\vartheta_i=k_i^+/q^+$, with $\vartheta_1+\vartheta_2+\vartheta_3=1$, and we shall denote
$\xi\equiv\vartheta_3$ for the gluon. We anticipate that the interesting situation is such that $\xi\ll 1$,
whereas $\vartheta_1$ and  $\vartheta_2$ take generic values, with $\vartheta_1+\vartheta_2\simeq 1$. 
In what follows, we shall often use the condition $\xi\ll 1$ to simplify the kinematics.

A distinguished experimental signature of coherent diffraction is the existence of a relatively large
{\it rapidity gap} in the final state ---  a rapidity interval $Y_{\mathbb{P}}$ 
between the target  and the produced jets (the ``diffractive system'') which is void of other particles (hadrons or jets).
This is the interval in rapidity covered by the elastic exchange between the target and the diffractive system
--- a colourless exchange a.k.a. the {\it Pomeron}. Within perturbative QCD, the Pomeron starts as a pair of two gluons.
For sufficiently large values of $Y_{\mathbb{P}}$, such that $\abar Y_{\mathbb{P}}\gtrsim 1$ (with $\abar\equiv
\alpha_s N_c/\pi$), the Pomeron ``evolves'' via the emission of additional, soft, gluons (i.e. gluons which carry
small longitudinal momentum fractions $x=k^-/P_N^-\ll 1$). When the gluon occupation number inside the Pomeron 
becomes sufficiently large (of order $1/\abar$), non-linear effects like gluon saturation start to be important,
and the high-energy evolution with increasing  $Y_{\mathbb{P}}$ is governed by the non-linear B-JIMWLK equation
\cite{Balitsky:1995ub,JalilianMarian:1997jx,JalilianMarian:1997gr,Kovner:2000pt,Iancu:2000hn,Iancu:2001ad,Ferreiro:2001qy}.
The gluon density  in a nucleus is larger (roughly, by $ A^{1/3}$) than in a proton, hence saturation
effects start at somewhat lower values of $Y_{\mathbb{P}}$ when the hadronic target is a large nucleus ($A\gg 1$).
The importance of the non-linear effects for diffraction is controlled by $Q_s(A, Y_{\mathbb{P}})$ ---
the target saturation momentum at the rapidity gap. This is a semi-hard (few GeV) transverse momentum scale,
 whose square is proportional to the gluon density per unit transverse area \cite{Iancu:2002xk,Iancu:2003xm,Gelis:2010nm,Kovchegov:2012mbw}.
 
Yet another hallmark of coherent diffraction is the fact that the total transverse momentum
$\bm{\Delta}$ transferred from the target to the diffractive system is {\it soft}, of the order the inverse size of the nucleus,
$\Delta_\perp\sim 1/R_A$, which is as soft as the QCD confinement scale $\Lambda$. This is indeed necessary for
the nucleus not to break up.  By transverse momentum conservation, we have  $\bk_1+\bk_2+\bk_3=\bm{\Delta}$.
The three jets to be considered in what follows are all much harder, $k_{i\perp}\gg \Lambda$ for $i=1,2,3$, hence one can
neglect $\Delta_\perp$ for the present purposes: $\bk_1+\bk_2+\bk_3\simeq 0$. Formally, this is tantamount to considering
a homogeneous target --- a disk with large transverse area $S_\perp$.
This also implies that the transverse momentum imbalance $\bK\equiv \bk_1+\bk_2$  between the $q\bar q$ jets
is fixed by the respective momentum of the gluon jet:  $\bK\simeq -\bk_3$. As we shall shortly argue, the interesting
situation is such that $k_{1\perp}\simeq k_{2\perp}\gg k_{3\perp}\sim Q_s(A, Y_{\mathbb{P}})$:
 the $q\bar q$  jets are hard and nearly back-to-back in the transverse plane, whereas the gluon jet is semi-hard.

The rapidity gap $Y_{\mathbb{P}}$ is related to the kinematics of DIS and of the final state via
 $Y_{\mathbb{P}}=\ln (1/x_{\mathbb{P}})$, where  $x_{\mathbb{P}}$ is the fraction of the target 
 longitudinal momentum $P_N^-$ which is ``carried by the Pomeron'', i.e. which is transferred from a nucleon
 to the diffractive system. Notice that, within the colour dipole picture, the
 ``minus'' component of the 4-momentum refers to the light-cone (LC) energies
 of  the produced jets. The respective conservation law
 reads
 \beq\label{xgdef}
 x_{\mathbb{P}} P_N^-=\frac{1}{2q^+}\left(\frac{k^2_{1\perp}}{\vartheta_1} +\frac{k^2_{2\perp}}{\vartheta_2} 
 +\frac{k^2_{3\perp}}{\vartheta_3} +Q^2\right),\eeq
which immediately implies
 \begin{align}\label{xP}
x_{\mathbb{P}}= \,\frac{Q^2+M_{q\bar q g}^2}{2q\cdot P_N}\,,
\end{align}
where $M_{q\bar q g}^2 $is the ``diffractive mass'' (the invariant mass squared of the three final jets):
\begin{align}\label{MX}
 M_{q\bar q g}^2   \equiv (k_1+k_2+k_3)^2 =\, \frac{k^2_{1\perp}}{\vartheta_1} +\frac{k^2_{2\perp}}{\vartheta_2} 
 +\frac{k^2_{3\perp}}{\vartheta_3} -(\bk_1+\bk_2+\bk_3)^2.
\end{align}
When characterising diffractive events, it is also customary
to specify the complementary rapidity difference $\ln(1/\beta)\equiv \Ybj-Y_{\mathbb{P}}$, with $\Ybj=\ln(1/\xbj)$. 
This is the rapidity difference between the Pomeron and the virtual photon, hence the longitudinal phase-space for
the quantum evolution of the projectile. One clearly has 
\begin{align}\label{beta}
\beta\equiv \,\frac{Q^2}{Q^2+M_{q\bar q g}^2}\,.
\end{align}

\begin{figure}[t] \centerline{
\centerline{\includegraphics[width=0.6\textwidth]{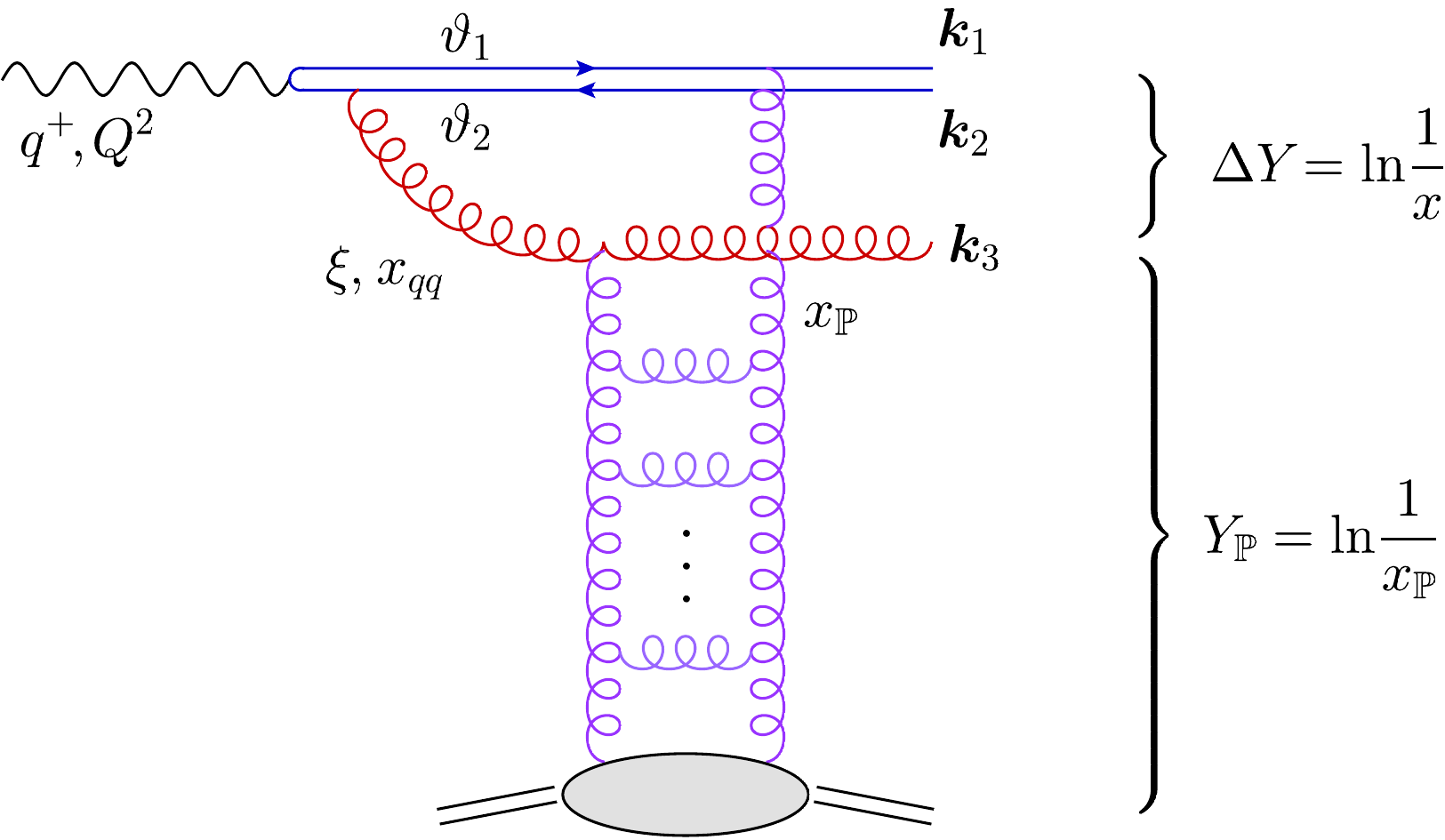}}}
\caption{\small Schematic representation of a Feynman graph contributing to diffractive trijet production. The 
colourless, ``Pomeron'', exchange  is represented by the gluon ladder.}
 \label{fig-diff}
\end{figure}

Another rapidity scale of interest is the rapidity difference $\Delta Y$ between the Pomeron and the
hard $q\bar q$ dijets. This is computed as $\Delta Y=\ln(x_{\mathbb{P}}/ x_{q\bar q})$, with $ x_{q\bar q}$ the ``minus'' longitudinal momentum fraction transferred to the $q\bar q$ jets. By analogy with \eqn{xgdef}
one can write
 \beq\label{xqqdef}
 x_{q\bar q} =\frac{1}{2q^+P_N^-}\left(\frac{k^2_{1\perp}}{\vartheta_1} +\frac{k^2_{2\perp}}{\vartheta_2} 
  +Q^2\right)\,=\, \frac{Q^2+M_{q\bar q}^2+K_{\perp}^2}{2P_N\cdot q}\,,
\eeq
where $M_{q\bar q}^2 $ refers to the $q\bar q$ system alone (the approximate equality below is valid when 
$\vartheta_1+\vartheta_2\simeq 1$):
\begin{align}\label{Mqq}
 M_{q\bar q}^2   \equiv (k_1+k_2)^2 \simeq\, \frac{k^2_{1\perp}}{\vartheta_1} +\frac{k^2_{2\perp}}{\vartheta_2}  -(\bk_1+\bk_2)^2,
\end{align}
and we recall that $\bk_1+\bk_2=\bK$. The following ratio
\beq\label{xdef}
x\equiv  \frac{x_{q\bar q}}{x_{\mathbb{P}}} \,\simeq\,\frac{Q^2+M_{q\bar q}^2+K_{\perp}^2}{Q^2+M_{q\bar q g}^2}
\,=\,\frac{Q^2+M_{q\bar q}^2+K_{\perp}^2}{Q^2+M_{q\bar q}^2+K_{\perp}^2 +{k^2_{3\perp}}/{\xi}}
\,,
\eeq
will play an important role in what follows: this is the splitting fraction of the gluon with respect to the Pomeron. 
Implicit in this interpretation, there is a change of perspective: the semi-hard gluon
is now viewed as being emitted by the Pomeron, and not by the $q\bar q$ pair.
This alternative viewpoint will naturally emerge when the results of our calculation 
will be seen to take the form of a {\it collinear factorisation}. 
From this new perspective, the rapidity difference $\Delta Y= \ln(1/x)$ is the longitudinal phase-space
for the quantum evolution of the target between the Pomeron and the hard dijets.

The various transverse momentum and rapidity scales are summarised in Fig.~\ref{fig-diff}.
After these kinematical considerations, let us briefly turn to the dynamics and anticipate some important conditions which will
orient our subsequent analysis. We would like to argue that the most interesting trijet configurations for a study of gluon saturation
are such that the $q\bar q$ dijets are {\it relatively hard}, with transverse momenta
 $k_{1\perp}^2,\, k_{2\perp}^2\gg Q_s^2$ and sizeable longitudinal momenta, $\vartheta_1,
  \vartheta_2\sim\order{1}$, whereas the gluon jet is {\it semi-hard}, 
$k_{3\perp}^2 \sim  Q_s^2$, and also {\it soft}: $\xi \lesssim Q_s^2/Q^2 \ll 1$.
(We use the simpler notation $Q_s^2\equiv  Q_s^2(A, Y_{\mathbb{P}})$.)
For brevity, and also by analogy with the terminology used for inclusive dijets \cite{Dominguez:2011wm}, we shall refer to this special 
kinematics as the ``correlation limit''. 

Albeit special, such 2+1 jet configurations are not rare: they are expected to dominate over
exclusive dijet production in the high virtuality regime at $Q^2\gg Q_s^2$.
Indeed, as well known (see also Sect.~\ref{sec:eik} below), the  photon virtuality
limits the transverse size $r$ of the $q\bar q$ fluctuation to values $r\lesssim 1/\bar Q$, 
with $\bar Q^2\equiv \vartheta_1 \vartheta_2 Q^2$. By the uncertainty principle, the two quarks are
produced with transverse momenta $k_{1\perp}\sim k_{2\perp}\sim \bar Q$, which are hard
($\bar Q^2\gg Q_s^2$) when  $\vartheta_1,  \vartheta_2\sim\order{1}$.
However, a small $q\bar q$ dipole with size $r\ll 1/Q_s$ interacts only weakly,   due to colour transparency:
its scattering amplitude can be estimated as $\mathcal{T}_{q\bar q}(r)\sim r^2Q_s^2 \ll 1$. This suppression is important
for the exclusive dijet production, which proceeds via the elastic scattering of the small $q\bar q$ dipole: the
respective cross-section involves the square $|\mathcal{T}_{q\bar q}(r)|^2$ of the dipole amplitude
(see e.g.  \cite{Salazar:2019ncp} and also App.~\ref{sec:el} below). But is does not penalise the 
diffractive production of 2+1 jets, because the emission of the semi-hard gluon opens up the colour space
and yields a large-size ($R\sim 1/Q_s$) partonic configuration which can strongly interact with the target,
despite the $q\bar q$ pair being small.

 Indeed, the constraint due to virtuality is less important for soft gluon emissions: a gluon 
 with longitudinal fraction $\xi\ll 1$ can have a large transverse separation
  $R\sim 1/(Q\sqrt{\xi})\gg 1/Q$ w.r.t. its sources, as suggested by a formation time argument (see Sect.~\ref{sect:factx}
for actual calculations). When the gluon formation time $\tau_3 = 2\xi q^+/ k_{3\perp}^2$ is comparable
to the coherence time $\tau_q = 2q^+/Q^2$ of the virtual photon, the gluon  transverse  momentum is relatively soft,
$ k_{3\perp}^2\sim \xi Q^2$, hence its transverse separation $R\sim 1/k_{3\perp}$ from its sources is quite large,
$R^2\sim 1/(\xi Q^2)$. So, the transverse scale of the gluon fluctuation can be dynamically adjusted, by varying $\xi$.
 In the context of elastic scattering, it is natural for this scale to be $Q_s$: when $R\gtrsim 1/Q_s$ the scattering is as strong
as possible and there is no suppression due to colour transparency.

In view of the above, our subsequent analysis of diffractive jets will focus on 2+1 jet production in the correlation limit.
For the sake of pedagogy, we will start our analysis (in Sect.~\ref{sec:eik})
with the case where the gluon is very soft, $\xi\ll Q_s^2/Q^2\ll 1$, so the virtuality plays no role  
for its emission: the formation time obeys $\tau_3 \ll \tau_q$, meaning that the gluon is emitted very close
to the time of scattering with the nuclear shockwave. Accordingly, one can treat its emission in the eikonal approximation,
which drastically simplifies the problem: the gluon emission quasi-trivially factorises from the remaining part of the 
amplitude and an effective gluon dipole picture immediately emerges.

But despite being simple and also quite instructive, 
this eikonal limit is not the most interesting case for phenomenology. Indeed, when
$k_{3\perp}^2 \sim Q_s^2$ and $\xi\ll Q_s^2/Q^2$, the diffractive mass \eqref{MX} is controlled by the gluon LC energy and is
much larger than both the virtuality $Q^2$ and the invariant mass $ M_{q\bar q}^2$ of the hard dijets:
\beq
M_{q\bar q g}^2\simeq \frac{k_{3\perp}^2}{\xi} \,\gg\, Q^2 \sim M_{q\bar q}^2.
\eeq
This in turn implies that both $\beta$ and $x$ are small\footnote{Notice that, when $k_{1\perp}^2,\, k_{2\perp}^2\sim Q^2$,
we also have $Q^2 \sim M_{q\bar q}^2$, hence $\beta$ and $x$ are of the same order of magnitude.}
(with $x>\beta$, of course): parametrically,
$\beta\sim x\sim \xi Q^2/Q_s^2\ll 1$. Therefore the rapidity interval $\ln(1/\beta)$ for the
evolution of the projectile is quite large, whereas the (complementary) rapidity gap $Y_{\mathbb{P}}$ is 
correspondingly reduced. This situation does not suit our general objective, which is to have
the largest possible value for $Y_{\mathbb{P}}$ in order to maximise the high-energy
evolution of the gluon distribution of the Pomeron.

Vice-versa, this argument shows that the most interesting regime for a study of gluon saturation in the target 
is when  $\beta$ and $x$ are of order one. Via Eqs.~\eqref{MX}--\eqref{xdef}, this requires ${k_{3\perp}^2}/{\xi}\sim Q^2$,
or $\xi\sim Q_s^2/Q^2$. This is the regime to which we shall devote most of our subsequent analysis. In particular, in
Sect.~\ref{sect:factx}, we will demonstrate that the gluon dipole picture applies to this large-$\beta$ regime  as well.
Once again, this will lead to a TMD factorisation for the diffractive trijet production, albeit the way how this works
in practice is more subtle than in the eikonal approximation.

\section{Small $\beta$,  or large diffractive mass}
\label{sec:eik}

\begin{figure}[t] 
\centerline{
\includegraphics[width=.85\textwidth]{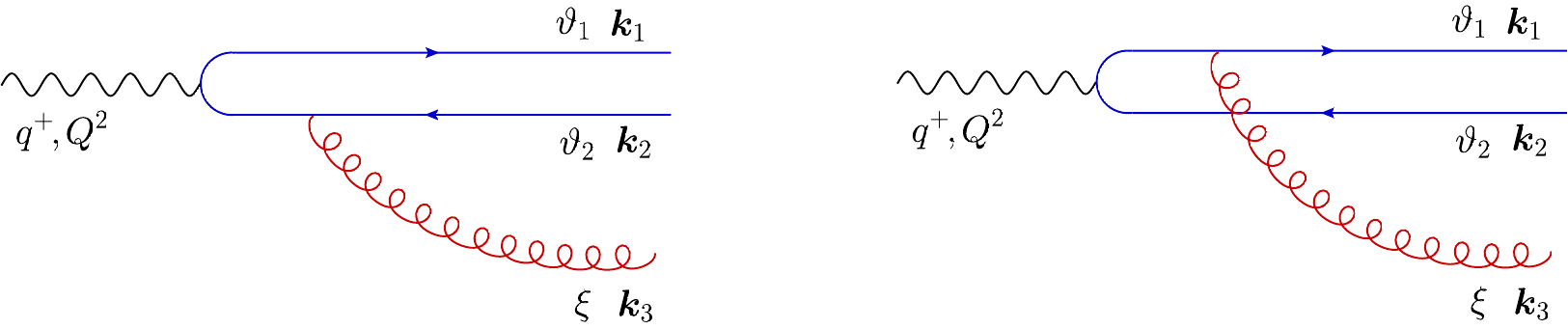}}
\caption{\small Feynman graphs in light-cone perturbation theory describing the quark-antiquark-gluon fluctuation
of the virtual photon.}
\label{fig:qqg}
\end{figure}

In order to compute the diffractive tri-parton production, we shall rely on the light-cone wavefunction (LCWF) formalism, that is,
we shall build the quark-antiquark-gluon ($q\bar q g$) Fock space component of the wavefunction of the virtual photon.
The virtual photon first decays into a $q\bar q$ pair and subsequently a gluon is emitted by either the quark, or the antiquark
(see Fig.~\ref{fig:qqg}). The ensuing tri-parton system scatters off the nuclear target. To compute the partonic
decays and the scattering, it is convenient to work in the projectile light-cone gauge $A^+=0$. 

In this gauge and Lorentz frame, 
the parton structure of the hadronic target is not manifest, so one cannot directly study gluon saturation.
The target is rather depicted as a dense collection of ``valence'' colour sources
which interact with the projectile via Coulomb exchanges.
Yet, the final result for the cross-section being gauge and boost invariant,
the effects of saturation are properly taken into account, in the form of multiple scattering. In turn, multiple scattering
at high energy can be efficiently resummed within the eikonal approximation. This exploits the fact that the transverse coordinates of the partons from the projectile are not modified by the collision. The only effect of the latter
is a rotation of the parton colour state, as described by a Wilson line in the appropriate representation of the colour group SU$(N_c)$.

As announced, we start our analysis with the case where the gluon is very soft: $\xi\ll Q_s^2/Q^2$. In this case, the gluon emission can be computed in the eikonal approximation, which leads to important simplifications.
Incidentally, the results that we shall obtain here can also be used to deduce the Kovchegov-Levin (KL) equation \cite{Kovchegov:1999ji}
describing the small--$\beta$ evolution of the diffractive cross-section --- that is, the evolution via the emission of soft gluons in
the wavefunction of the $q\bar q$ projectile, within the rapidity interval $ \Ybj-Y_{\mathbb{P}}=\ln(1/\beta)$ complementary
to the gap. The soft gluon emission that we will explicitly compute here can be seen as the first step in that evolution (see also
\cite{Hatta:2006hs} for an alternative derivation of the KL equation, 
based on the dipole picture \cite{Mueller:1993rr} of the high-energy evolution).

\subsection{The $q\bar q g$ Fock state in the eikonal approximation}
\label{sec:qqg}

When computed in the eikonal approximation, the gluon emission factorises from the remaining part of the amplitude, which 
describes the decay of the virtual photon into a $q\bar q$ pair. So it is convenient to first display the LCWF representing
the $q\bar q$ fluctuation alone. Before we proceed, let us first summarise our  conventions and notations. Concerning the normalisation of the single-particle
(bare) states, we shall follow the conventions in  Refs.~\cite{Iancu:2018hwa,Iancu:2020mos}.  In particular we shall work with
helicity states,  $\lambda_{1,2}=\pm 1/2$, for the quark and the antiquark, and with 
linear polarisation states for the transverse (virtual) photon and for the emitted gluon. The polarisation vectors for an on-shell gluon
with transverse momentum $\bk$ and longitudinal momentum $k^+$ take the form
\beq\label{pol}
\epsilon^\mu(k, \lambda=1,2)=\left(0,\,\frac{\bm{\epsilon}_\lambda\cdot\bk}{k^+}\,,\,\bm{\epsilon}_\lambda\right),\quad
\bm{\epsilon}_1=(1,0),\ \ \bm{\epsilon}_2=(0,1)\,, 
\eeq
and obey $k\cdot \epsilon(k, \lambda)=0$ and $\epsilon^2(k, \lambda)=-1$.
Notice that $ {\epsilon}_\lambda^i=\delta_{\lambda}^ i$, so in what follows we shall indicate a linear polarisation state
by the 2-dimensional vector index which indicates the direction of polarisation in the transverse plane: 
$\lambda\to i$. Similarly, the polarisation vectors for the virtual photon read
\beq
\epsilon^\mu_T(q, i=1,2)=(0,0, \bm{\epsilon}_i),\qquad
\epsilon^\mu_L(q)=(q^+/Q,Q/2q^+,\bm{0}_\perp),
\eeq
and obey $q\cdot \epsilon_T=q\cdot \epsilon_L=0$, $ \epsilon_T^2=-1$, and $ \epsilon_L^2=1$.
We will show intermediate results only for the case of a virtual photon with transverse polarisation ($\gamma_{T}^{i}$),
but extend the final results to a longitudinal photon ($\gamma_L$)  as well.
We shall  first consider the virtual photon LCWF in the absence of the scattering with the nuclear target
and display the respective  $q\bar q$  and  $q\bar q g$ components. The effects of the scattering will be then added
in the next subsection.

We start with the $q\bar q$ component. This is most conveniently written in momentum space and reads
\begin{align}\label{qqLCWF}
\left|\gamma_{T}^{i}(q)\right\rangle_{q\bar{q}}\, 
=\sum_{\lambda_{1,2}=\pm 1/2}\,\sum_{\alpha,\beta=1}^{N_c}\,\delta_{\alpha\beta} &
\int_{0}^{1}\rmd\vartheta_1 \rmd\vartheta_2 \,\delta(1-\vartheta_1 -\vartheta_2)
\int \rmd^{2}\bm{k}_1  \rmd^{2}\bm{k}_{2}\,\delta^{(2)}(\bm{k}_1 +\bm{k}_2)
\nn
&\,\times\,\psi^i_{\lambda_{1}\lambda_{2}}(\vartheta_1,  \bk_1)\,
\left|{q}_{\lambda_{1}}^{\alpha}(\vartheta_1, \bk_1)\bar{q}_{\lambda_{2}}^{\beta}(\vartheta_2, \bk_2)
\right\rangle .
\end{align}
Here $i=1,\,2$ is the polarisation index for the transverse photon, $\alpha,\beta=1\dots N_c$ are colour indices in the fundamental
representation, and the other notations were already explained.
The various $\delta$--functions in \eqn{qqLCWF} express the conservation of colour and (longitudinal and transverse) momentum.
This LCWF involves the amplitude
\begin{align}\label{psiqq}
\psi^i_{\lambda_1\lambda_{2}}(\vartheta,  \bk)\,=\,\sqrt{\frac{q^+}{2}}\,\frac{ee_{f}}{(2\pi)^3}\,
\frac{\varphi_{\lambda_{1}\lambda_{2}}^{il}(\vartheta)\,k^{l}}{\bm{k}^{2}+ 
\vartheta(1-\vartheta) Q^{2}}\,,
\end{align}
where $e_f$ is the electric charge of the quark flavour under consideration, the function
\begin{equation}\label{phidef}
\varphi_{\lambda_1\lambda_{2}}^{il}(\vartheta)\,\equiv\,
\delta_{\lambda_{1}\lambda_{2}}\left[(2\vartheta-1)\delta^{il}+2i\varepsilon^{il}\lambda_{1}\right],
\end{equation}
encodes the helicity structure of the photon decay vertex, and the expression in the denominator is the remnant of the original,
{\it energy}, denominator (recall that $\vartheta_1+\vartheta_2=1$ and $\bk_1+\bk_2=0$)
\begin{align}
	\label{EDqq0}
	E_{q\bar{q}} - E_{\gamma} = \frac{1}{2q^+}
	\left(
	\frac{k_{1 \perp}^2}{\vartheta_1} + 
	\frac{k_{1 \perp}^2}{\vartheta_2} + {Q^2} 
	\right)
	=\frac{1}{2 q^+\vartheta_1\vartheta_2}
	\left({k_{1 \perp}^2} +{\vartheta_1  \vartheta_2} Q^2
	\right),
	\end{align}
which describes the off-shellness of the $q\bar q$ fluctuation of the virtual photon. In what follows,
we shall often use the shorthand notation $\bar Q^2\equiv {\vartheta_1  \vartheta_2} Q^2$.
Note the symmetry property
\beq
\psi^i_{\lambda_1\lambda_{2}}(\vartheta,  \bk)\,=\,\psi^{i\,*}_{\lambda_2\lambda_{1}}(1-\vartheta, - \bk),
\eeq
which expresses the invariance of the amplitude under charge conjugation, i.e. under the exchange of the quark
and the antiquark.

We are now prepared to exhibit the $q\bar q g$ component  of the LCWF, with the gluon emission as
computed in the eikonal approximation --- that is, by using the condition $\xi\equiv  \vartheta_3\ll 1$ in order to simply
both the emission vertex and the corresponding energy denominator. This reads
\begin{align}\label{qqgLCWF} 
\left|\gamma_{T}^{i}(q)\right\rangle_{q\bar{q}g}\, 
= & \int_{0}^{1}\rmd\vartheta_1 \rmd\vartheta_2\,\delta(1-\vartheta_1 -\vartheta_2)\int_0^1  \rmd\xi
\int \rmd^{2}\bm{k}_1  \rmd^{2}\bm{k}_{2}\rmd^{2}\bm{k}_{3}\,\delta^{(2)}(\bm{k}_1 +\bm{k}_2+\bk_3)
\nn
&\,\times\,
\Psi^{ij}_{\lambda_{1}\lambda_{2}}(\vartheta_1,  \bk_1, \vartheta_2, \bk_2, \xi, \bk_3)
\,t^a_{\alpha\beta} \,
\left|{q}_{\lambda_{1}}^{\alpha}(\vartheta_1, \bk_1)\bar{q}_{\lambda_{2}}^{\beta}(\vartheta_2, \bk_2) g_j^a(\xi, \bk_3)
\right\rangle .
\end{align}
Here, the sums over repeated discrete indices are kept implicit;
 $a=1\dots N_g$, with $N_g=N_c^2-1$, is a colour index for the gluon, while $j=1,2$ 
is the gluon polarisation state. Notice that the gluon longitudinal momentum has been neglected in
the respective conservation law, which still enforces $\vartheta_1+\vartheta_2=1$, as in the absence of the gluon emission.

The $q\bar q g$ amplitude takes a factorised form, as anticipated:
\begin{align}\label{qqgpsi}
	\Psi^{ij}_{\lambda_1\lambda_2}&\,=\,\sqrt{\frac{q^+}{2}}
	\frac{g}{(2\pi)^3}\,\frac{1}{\sqrt{\xi}}
	\frac{2k_3^j}{k_{3\perp}^2}\left[\psi^i_{\lambda_1\lambda_{2}}(\vartheta_1+\xi,  \bk_1+\bk_3)-
	\psi^i_{\lambda_1\lambda_{2}}(\vartheta_1,  \bk_1)\right].
\end{align}
The first term inside the square brackets refers to gluon emission by the quark: prior to this emission,
the quark had a longitudinal momentum $\vartheta_1+\xi\simeq \vartheta_1$ and a transverse momentum
$\bk_1+\bk_3= -\bk_2$, and this is  the kinematics which matters for the photon decay vertex.
The second term inside the square brackets refers to gluon emission by the anti-quark; it comes with a minus
sign due to the opposite colour charge of the emitter.

 The  factor in front of the square brackets in \eqn{qqgpsi} describes
gluon emission in the eikonal approximation. It can be understood as follows: the factor
\beq
\frac{2k_3^j}{k_{3\perp}^2}\,=\,\frac{\bm{\epsilon}_j\cdot\bk_3}{k^+_3}\,\frac{2k^+_3}{k_{3\perp}^2}\,\eeq
has been generated as the product between the minus component $\epsilon^-_j$ of the gluon polarisation vector (cf. \eqn{pol}), which
controls the vertex for soft gluon emission\footnote{This minus component dominates the vertex
$\gamma_\mu \epsilon^\mu_j\simeq \gamma^+ \epsilon^-_j$ for soft gluon emission because it is proportional to $1/k^+_3$.}, and the
inverse $1/(E_{q\bar{q}g} - E_{\gamma})$
of the energy denominator for the $q\bar q g$ state, which for sufficiently small $\xi$ is dominated by the gluon energy:
\begin{align}\label{EDqqg0}
	E_{q\bar{q}g} - E_{\gamma} = \frac{1}{2q^+}
	\left(
	\frac{k_{1\perp}^2}{ \vartheta_1} + 
	\frac{k_{2\perp}^2}{\vartheta_2} +
	\frac{k_{3\perp}^2}{\xi} +{Q^2}
	\right)\simeq \frac{k_{3\perp}^2}{2k_3^+}\,. 	\end{align}
The other non-trivial ingredients  in \eqn{qqgpsi} are the factor of $g$ from the gluon emission vertex and the
 factor $1/\sqrt{\xi}$  from the normalisation of the gluon field. Recalling the expression \eqref{psiqq}
 for the $q\bar q$ amplitude, one obtains our final result for the $q\bar q g$ amplitude in the eikonal approximation and in
 the absence of scattering:
\begin{align}\label{Psieik}
	\Psi^{ij}_{\lambda_1\lambda_2}\, = \,-\frac{e e_f g  q^+}{(2\pi)^6}\,\frac{1}{\sqrt{\xi}}\,
	\varphi_{\lambda_{1}\lambda_2}^{il}(\vartheta_1)\left[\frac{k_1^l}{k_{1 \perp}^2+\bar Q^2}+\frac{k_2^l}{k_{2\perp}^2+\bar Q^2}\right]
	\frac{k_3^j}{k_{3\perp}^2}.
\end{align}
Notice that the relative sign between the two terms inside the brackets has changed as compared to the original equation
 \eqref{qqgpsi}. Yet, one should
remember that in the most interest kinematics, the hard momenta $\bk_1$ and $\bk_2$ are nearly back-to-back, so
 \eqn{Psieik} truly involves the {\it difference} between gluon emissions by the quark
and by the antiquark, as expected.

One may understand the limitations of the eikonal approximation by inspection of  \eqn{EDqqg0}:  for the gluon LC energy to control the
energy denominator, $\xi$ must be much smaller than the ratio between the gluon transverse momentum
$k_{3\perp}^2$ and any of the hard scales in the problem: the photon virtuality $Q^2$ and the transverse momenta 
$k_{1\perp}^2$ and $k_{2\perp}^2$ of the two quarks. In the most interesting regime,
where the hard scales are comparable with each other, $k_{1\perp}^2 \sim k_{2\perp}^2 \sim Q^2\gg Q_s^2$, whereas the gluon momentum
is semi-hard, $k_{3\perp}^2 \sim Q_s^2$, one needs the rather stringent condition $\xi\ll Q_s^2/Q^2\ll 1$.
This condition will be relaxed in Sect.~\ref{sect:factx}, where we will study the sub-eikonal (or finite--$\xi$) corrections to both the 
gluon emission vertex and the energy denominator. 

\subsection{Adding multiple scattering off the target}

We shall now complete our $q\bar q g$ state by adding the effects of multiple scattering off the nuclear target in the eikonal approximation.
As already mentioned, this approximation is most conveniently formulated in the transverse coordinate space, so we start by changing
the representation in \eqn{qqgLCWF}. Using
\beq\hspace*{-0.5cm}
\left|{q}_{\lambda_{1}}^{\alpha}(\vartheta_1, \bk_1)\bar{q}_{\lambda_{2}}^{\beta}(\vartheta_2, \bk_2) g_j^a(\xi, \bk_3)
\right\rangle = \int \rmd^2\bx\, \rmd^2\by\, \rmd^2\bz\,
	e^{-i \bk_1\cdot \bx - i \bk_2\cdot \by - i \bk_3\cdot \bz}\,
\left|{q}_{\lambda_{1}}^{\alpha}(\vartheta_1, \bx)\bar{q}_{\lambda_{2}}^{\beta}(\vartheta_2, \by) g_j^a(\xi, \bz)
\right\rangle,\eeq
one immediately finds
\begin{align}\label{qqgLCWF} 
\left|\gamma_{T}^{i}(q)\right\rangle_{q\bar{q}g}\, 
= &
\int_{0}^{1}\rmd\vartheta_1 \rmd\vartheta_2\,\delta(1-\vartheta_1 -\vartheta_2)\int_0^1  \rmd\xi
\int \rmd^{2}\bm{x}\,  \rmd^{2}\bm{y}\,\rmd^{2}\bm{z}
\nn
&\,\times\,\Psi^{ij}_{\lambda_{1}\lambda_{2}}(\vartheta_1,  \bx, \vartheta_2, \by, \xi, \bz)\,t^a_{\alpha\beta} \,
\left|{q}_{\lambda_{1}}^{\alpha}(\vartheta_1, \bx)\bar{q}_{\lambda_{2}}^{\beta}(\vartheta_2, \by) g_j^a(\xi, \bz)
\right\rangle,
\end{align}
with the amplitude in the coordinate-representation defined as:
\begin{align}
\Psi^{ij}_{\lambda_{1}\lambda_{2}}(\vartheta_1,  \bx, \vartheta_2, \by, \xi, \bz)\,=\,&
\int \rmd^{2}\bm{k}_1  \rmd^{2}\bm{k}_{2}\rmd^{2}\bm{k}_{3}\,\delta^{(2)}(\bm{k}_1 +\bm{k}_2+\bk_3)
\nn
&\,\times\,e^{-i \bk_1\cdot \bx - i \bk_2\cdot \by - i \bk_3\cdot \bz}\,\Psi^{ij}_{\lambda_{1}\lambda_{2}}(\vartheta_1,  \bk_1, \vartheta_2, \bk_2, \xi, \bk_3)\,.
\end{align}
Consider e.g. the first piece in \eqn{Psieik}, which describes the gluon emission by the antiquark (see Fig.~\ref{fig:qqg} left).
This piece is independent of the antiquark momentum $\bk_2$, hence the relevant Fourier transform reads
\begin{align}\hspace*{-0.5cm}
\int \rmd^{2}\bm{k}_1  \rmd^{2}\bm{k}_{3}\,e^{-i \bk_1\cdot (\bx- \by) + i \bk_3\cdot (\by-\bz)}\,
\frac{k_1^l}{k_{1 \perp}^2+\bar Q^2}\,	\frac{k_3^j}{k_{3\perp}^2}= (2\pi)^2 \frac{(\by-\bz)^j}{(\by-\bz)^2}
\frac{(\bx-\by)^l}{|\bx-\by|}\,\bar{Q} K_1(\bar{Q}|\bx-\by|)\,.
\end{align}
After similarly treating the second piece in \eqn{Psieik}, i.e. the emission by the quark, one finds
\begin{align}\label{Psix0}
\Psi^{ij}_{\lambda_{1}\lambda_{2}}(\vartheta_1,  \bx, \vartheta_2, \by, \xi, \bz)\,=\,&
\frac{e e_f g  q^+}{(2\pi)^4}\,\frac{1}{\sqrt{\xi}}\,
	\varphi_{\lambda_{1}\lambda_2}^{il}(\vartheta_1)\,\frac{r^l}{r}\,
	\bar{Q} K_1(\bar{Q}r)\left[
	\frac{(\bx-\bz)^j}{(\bx-\bz)^2}-
	\frac{(\by-\bz)^j}{(\by-\bz)^2}
	\right],
\end{align}
where $\bm{r} = \bm{x}-\bm{y}$ is the transverse separation between the quark and the antiquark.
The $r$-dependent factor which also involves the modified Bessel function $K_1(\bar{Q}r)$ describes the decay
of the transverse virtual photon into a $q\bar q$ pair with size $r$. The two terms in the square brackets are
the Weiszäcker-Williams (WW) kernels for gluon emissions by the quark and by the antiquark. 
In these terms, a difference like $\bx-\bz$ should be
understood as the transverse separation between the gluon and its emitter  {\it after} the emission.
Hence, according to \eqn{Psix0}, the quark and the antiquark have exactly the same transverse positions ($\bx$ and
respectively $\by$) both {\it before}, and {\it after}, emitting the gluon. This is a consequence of 
the eikonal approximation, valid at  small $\xi\ll 1$. In general, i.e. for generic values of $\xi$, 
the transverse coordinate of the emitter could be modified by the gluon recoil (see also the discussion
in Sect.~\ref{sect:factx}, notably around \eqn{relrecoil}).

This discussion confirms that the eikonal approximation used for computing a soft gluon emission inside the projectile wavefunction
is indeed equivalent to that used for the scattering: in both cases the
transverse coordinate of a parton which interacts with a soft gluon is a ``good quantum number''.

\begin{figure}[t] 
\centerline{
\includegraphics[width=.85\textwidth]{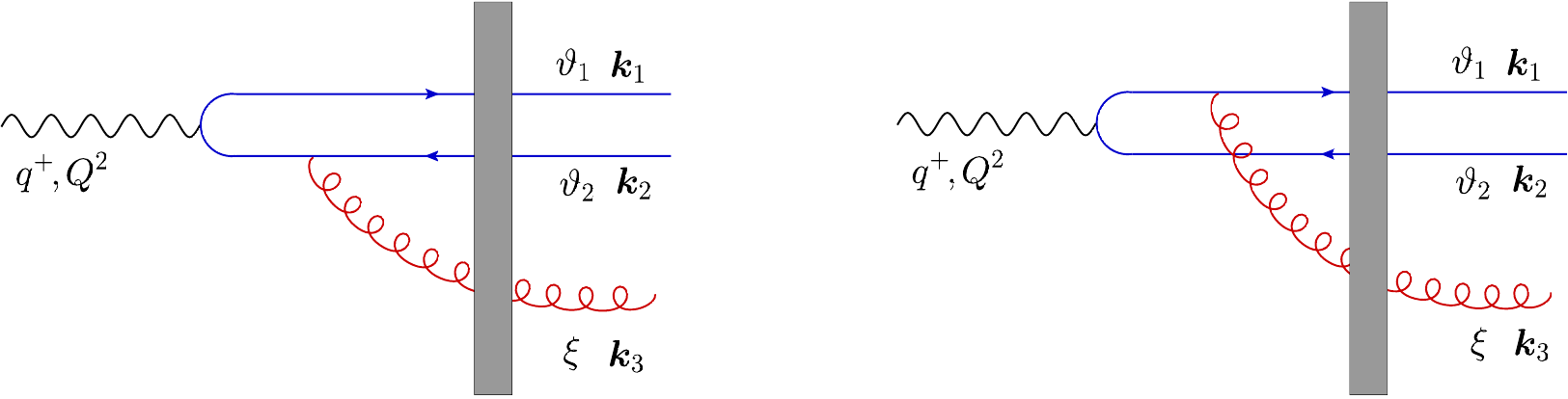}}
\caption{\small Feynman graphs representing the amplitude for the scattering between the $q\bar q g$ system 
and the nuclear target (represented as a shockwave) in the case where the gluon emission occurs before the
scattering.}
\label{fig:qqgSW}
\end{figure}

\begin{figure}[t] 
\centerline{
\includegraphics[width=.85\textwidth]{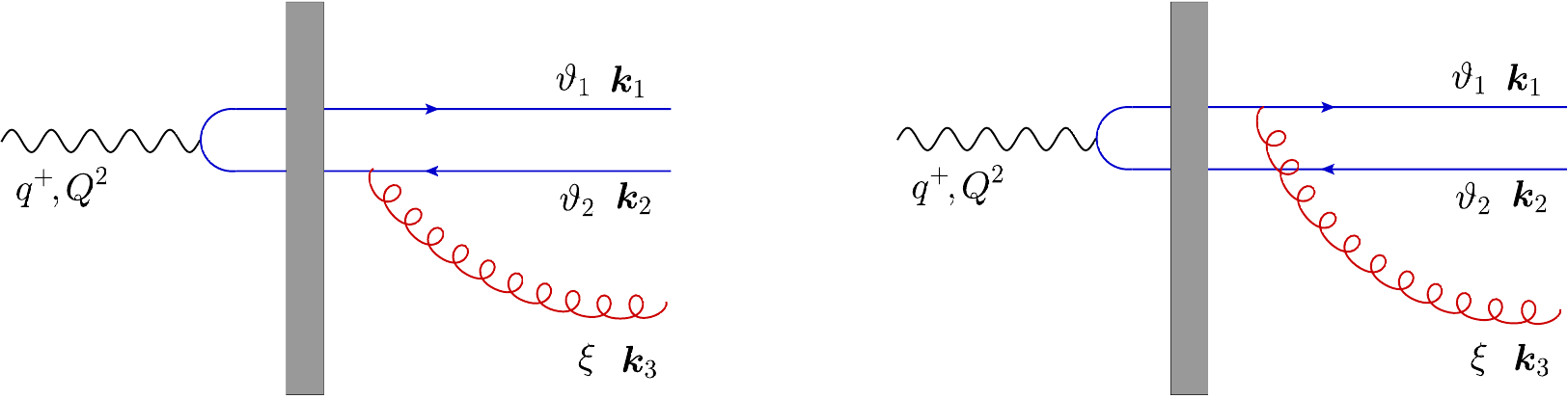}}
\caption{\small The same as in Fig~\ref{fig:qqgSW}, but for the case where the gluon is emitted after the
collision. }
\label{fig:qqgSW2}
\end{figure}

At this level, it is straightforward to add the effects of the collisions. In the kinematics of interest, the
nuclear target can be treated as a shockwave, due to Lorentz contraction. This shockwave can be inserted 
either in the final state (see Fig.~\ref{fig:qqgSW}), meaning that all the three partons will interact with it, or in the intermediate $q\bar q$ state (prior
to the gluon emission, see Fig.~\ref{fig:qqgSW2}), so only this $q\bar q$ pair will scatter. In both cases, the effect of
the scattering consists in multiplying with a product of Wilson lines --- one for each parton. These are colour matrices, so they 
also multiply the original matrix $t^a$ from the gluon emission vertex. 
It is easy to see that the final result takes the form
\begin{align}\label{Psix}
\Psi^{ij, \,\alpha\beta}_{\lambda_{1}\lambda_{2}}(\vartheta_1,  \bx, \vartheta_2, \by, \xi, \bz)\,=\,&\,
\frac{e e_f g  q^+}{(2\pi)^4}\,\frac{1}{\sqrt{\xi}}\,
	\varphi_{\lambda_{1}\lambda_2}^{il}(\vartheta_1)\,\frac{r^l}{r}\,
	\bar{Q} K_1(\bar{Q}r)
	\nonumber \\*[0.2cm]
&\times \Bigg\{\left[
	\frac{(\bx-\bz)^j}{(\bx-\bz)^2}-
	\frac{(\by-\bz)^j}{(\by-\bz)^2}\right] U^{ab}(\bm{z})V(\bm{x})t^{b}V^{\dagger}(\bm{y})
\nonumber \\*[0.2cm]
  & \ \ \  - \left[\frac{(\bx-\bz)^j}{(\bx-\bz)^2}\, t^{a}V(\bm{x})V^{\dagger}(\bm{y})
  - \frac{(\by-\bz)^j}{(\by-\bz)^2}\,V(\bm{x})V^{\dagger}(\bm{y})t^{a}\right] 
\Bigg\}_{\alpha\beta}.
\end{align}
$V(\bm{x})$ and $V^{\dagger}(\bm{y})$ are Wilson lines in the fundamental representation and
refer to the quark and the antiquark, respectively, whereas $U(\bm{x})$ belongs to the adjoint representation
and refers to the gluon:
\beq\label{SA}
 V(\bm{x})=\,{\rm T}\exp\left\{ ig\int dx^{+}\, t^{a}A^{-}_a(x^{+},\bm{x})\right\}, \qquad   
U(\bm{x})=\,{\rm T}\exp\left\{ ig\int dx^{+}\,T^{a}A_{a}^{-}(x^{+},\bm{x})\right\}.
    \eeq
The target field $A^{-}_a$  represents Coulomb exchanges and must be treated as a random quantity, to be averaged
over when computing the cross-section (see below). The first line insides the accolades, where the colour structure also
involves the gluon Wilson lines, corresponds to the shockwave insertion in the final state. The second line refers
to the intermediate state, with the first (second) term corresponding to gluon emission by the quark (antiquark).

\eqn{Psix} represents the most general 3-parton scattering state that can be produced via (generally inelastic) DIS in
the approximations of interest, but this is not quite the result we need. This amplitude would enter the total cross-section
for $q\bar q g$ production, but here we are rather interested in the respective {\it diffractive} process. By ``diffractive'' we more
precisely mean the coherent process where both the target and the 3-parton system undergo elastic scattering. 
This in particular implies that the projectile emerges as a colour singlet in the final state. The elastic amplitude 
is therefore obtained as the colour-singlet projection of the general amplitude \eqref{Psix}.  

To understand this projection,
notice that the colour structure of the $q\bar q g$ state is of the generic type (we keep only colour indices)
$\mathcal{O}^a_{\alpha\beta}\left|{q}^{\alpha}\bar{q}^{\beta}g^a \right\rangle$, where $\mathcal{O}^a$ is
a colour matrix in the fundamental representation and carries an adjoint colour index, e.g.
$\mathcal{O}^a=U^{ab}(\bm{z})V(\bm{x})t^{b}V^{\dagger}(\bm{y})$. The corresponding colour-singlet state
corresponds to $\mathcal{O}^a\to t^a$. So, the relevant projection operator is
\begin{align}\label{projD}
\mathbb{P}_D\, \mathcal{O}^a_{\alpha\beta}\,\equiv\,
\frac{1}{C_FN_c} \tr\big[t^c \mathcal{O}^c]\,t^a_{\alpha\beta}\,.
\end{align}
Its action on the colour structures in \eqn{Psix} amounts to
\begin{align}\label{proj1}
\Big[V(\bm{x})V^{\dagger}(\bm{y}^\prime)t^{a}
\Big]_{\alpha\beta}
 \,\to\,{S}_{q\bar{q}}(\bm{x}, \bm{y})\, t^a_{\alpha\beta},\qquad
 \Big[U^{ab}(\bm{z})V(\bm{x})t^{b}V^{\dagger}(\bm{y})\Big]_{\alpha\beta}
\,\to\,S_{q\bar{q}g}(\bm{x}, \bm{y}, \bm{z}) \,t^a_{\alpha\beta},
\end{align}
where we introduced the  elastic $S$-matrices for a  $q\bar q$ dipole and for a $q\bar qg$ system (in a colour singlet
state), respectively
\begin{align}\label{proj1}
{S}_{q\bar{q}}(\bm{x}, \bm{y})&\,\equiv\,\frac{1}{N_c} \,\tr\Big[V(\bm{x})V^{\dagger}(\bm{y})\Big],
\nonumber\\*[0.2cm]
S_{q\bar{q}g}(\bm{x}, \bm{y}, \bm{z})&\,\equiv\,\frac{1}{C_FN_c} U^{cb}(\bm{z}) \tr\left[t^cV(\bm{x})t^{b}V^{\dagger}(\bm{y})\right].
\end{align}

After also implementing this colour projection, we arrive at our main result in this section: an expression for the LCWF which controls the diffractive
production of a $q\bar q g$ system in the limit where the gluon is sufficiently soft, $\xi\ll Q_s^2/Q^2$, for its emission
to be treated in the eikonal approximation:
\begin{align}\label{qqgDWF} 
\left|\gamma_{T}^{i}(q)\right\rangle_{q\bar{q}g}^{\rm D}\, 
= &
\int_{0}^{1}\rmd\vartheta_1 \rmd\vartheta_2\,\delta(1-\vartheta_1 -\vartheta_2)\int_0^1  \rmd\xi
\int \rmd^{2}\bm{x}\,  \rmd^{2}\bm{y}\,\rmd^{2}\bm{z}
\nn
&\,\times\,\Psi^{ij,\,{\rm D}}_{\lambda_{1}\lambda_{2}}(\vartheta_1,  \bx, \vartheta_2, \by, \xi, \bz)\,t^a_{\alpha\beta} \,
\left|{q}_{\lambda_{1}}^{\alpha}(\vartheta_1, \bx)\bar{q}_{\lambda_{2}}^{\beta}(\vartheta_2, \by) g_j^a(\xi, \bz)
\right\rangle,
\end{align}
with the diffractive amplitude
\begin{align}\label{PsixD}
\Psi^{ij, \,{\rm D}}_{\lambda_{1}\lambda_{2}}(\vartheta_1,  \bx, \vartheta_2, \by, \xi, \bz)\,=\,&\,
 \frac{e e_f g  q^+}{(2\pi)^4}\,\frac{1}{\sqrt{\xi}}\,
	\varphi_{\lambda_{1}\lambda_2}^{il}(\vartheta_1)\,\frac{r^l}{r}\,\bar{Q} K_1(\bar{Q}r)
	\nonumber \\*[0.2cm]
&\times \left[
	\frac{(\bx-\bz)^j}{(\bx-\bz)^2}-
	\frac{(\by-\bz)^j}{(\by-\bz)^2}\right]\left[
	{S}_{q\bar{q}g}(\bx,\by,\bz) - 
        {S}_{q\bar{q}}(\bx,\by)
	\right].
\end{align}

\subsection{The diffractive trijet cross-section}
\label{sect:3jet}

Given the previous result for the diffractive LCWF,  cf.~Eqs.~\eqref{qqgDWF}--\eqref{PsixD}, it is straightforward to compute
the cross-section for diffractive trijet production for sufficiently small $\xi\ll Q_s^2/Q^2$, or large diffractive mass $M_{q\bar q g}^2\gg Q^2$.
This is obtained by ``counting'' the number of partons in the final state, or, more precisely, by computing the average of
the parton number densities:
\begin{align}\label{3jeteik}
 \frac{\rmd \sigma_{\rm D}^{\gamma_T^* A
  \rightarrow q\bar q g A}}
  {\rmd k_1^+
  \rmd k_2^+
  \rmd k_3^+
  \rmd^{2}\bm{k}_{1}
  \rmd^{2}\bm{k}_{2}
  \rmd^{2}\bm{k}_{3}} &
\,(2\pi)\,\delta(q^{+}-k_1^+-k_2^+-k_3^+)\nn
&\,=\frac{1}{2}\!\!\!
{\phantom{\big\langle}}_{\,\,q\bar{q}g}^{\,\,\,\,\,\rm D}\!\left\langle \gamma_{T}^{i}(q)\right|\,\mathcal{N}_{q}(k_{1})\,\mathcal{N}_{\bar{q}}(k_{2})\,\mathcal{N}_{g}(k_{3})\,\left|\gamma_{T}^{i}(q)\right\rangle_{q\bar{q}g}^{\rm D},
\end{align}
where $\mathcal{N}_{q}(k_{1})$, $\mathcal{N}_{\bar{q}}(k_{2})$, and $\mathcal{N}_{g}(k_{3})$ are particle number density operators
for quarks, antiquarks, and gluons, and the overall factor 1/2 comes from the average over the 2 transverse polarisations. 
Once again, we follow the conventions in Refs.~\cite{Iancu:2018hwa,Iancu:2020mos}. A straightforward calculation yileds
  \begin{align}
  \label{sigma3jet}
  \frac{\rmd \sigma_{\rm D}^{\gamma_T^* A
  \rightarrow q\bar q g A}}
  {\rmd k_1^+
  \rmd k_2^+
  \rmd k_3^+
  \rmd^{2}\bm{k}_{1}
  \rmd^{2}\bm{k}_{2}
  \rmd^{2}\bm{k}_{3}} 
  =\frac{\alpha_{em}\,N_c}{2 \pi^4 q^+}
  \left(\sum e_{f}^{2}\right)
  \left(\vartheta_1^2+\vartheta_2^2\right)
  \delta(q^{+}-k_1^+-k_2^+)
  \frac{\alpha_s C_{F}}{\xi q^+}
  \big|\tilde{\mcal{A}}_{q\bar{q} g}^{lj} \big|^2, 
\end{align}
where $\alpha_{em}=e^2/4\pi$, $\alpha_{s}=g^2/4\pi$, a summation over the indices $l$ and $j$ is implicitly assumed, and the 
reduced amplitude $\tilde{A}_{q\bar{q} g}^{lj}$ reads (we recall that $\bm{r} = \bm{x}-\bm{y}$)
\begin{align}
	\label{tildeA}
	\tilde{\mcal{A}}_{q\bar{q} g}^{lj} = 
	&\int\frac{\rmd^2\bx}{2\pi}
	\int\frac{\rmd^2\by}{2\pi}\,
	e^{-i \bk_1\cdot \bx - i \bk_2\cdot \by}\,
	\frac{r^l}{r}\,
	\bar{Q} K_1(\bar{Q}r)
	\nn 
	&\times\int\frac{\rmd^2\bz}{2\pi}\,
	e^{-i \bk_3\cdot \bz}
	\left[
	\frac{(\bx-\bz)^j}{(\bx-\bz)^2}-
	\frac{(\by-\bz)^j}{(\by-\bz)^2}
	\right]
	\left[
	\mcal{S}_{q\bar{q}g}(\bx,\by,\bz) - 
	\mcal{S}_{q\bar{q}}(\bx,\by)
	\right].
\end{align}
This is not exactly the same as the original amplitude \eqref{PsixD}, since we have 
factorised out the information about the distribution of the longitudinal momentum fractions 
$\vartheta_1,\,\vartheta_2$ and $\xi$ at the two emission vertices. This information has been used to 
construct the  splitting functions explicit in the cross-section \eqref{sigma3jet}:
$1/\xi$ for the soft gluon and  $(\vartheta_1^2+\vartheta_2^2)$ for the photon decay; 
the latter has been obtained via the identity
\beq
\varphi_{\lambda_{1}\lambda_{2}}^{ij}(\vartheta)\,
\varphi_{\lambda_{1}\lambda_{2}}^{il\,*}(\vartheta)\,
=\,2\delta^{jl}\left[1+(1-2\vartheta)^2\right]\,
=\,4\delta^{jl}\left[\vartheta^2+(1-\vartheta)^2\right].
\eeq
Finally, the ($q\bar qg$ and $q\bar q$) elastic $S$-matrices in \eqn{tildeA} are now denoted with calligraphic letters, to emphasise
the fact that they include the average over the colour fields in the target; e.g.
\begin{align}
	\label{Sqqg}
	\mcal{S}_{q\bar{q}g}(\bx,\by,\bz)
	\equiv
	\frac{1}{C_{F} N_{c}}
	\left\langle 
	U^{ba}(\bm{z})\,
	\mathrm{tr}
	\big[V(\bm{x})t^{a}
	V^{\dagger}(\bm{y})t^{b}
	\big]
	\right\rangle_{Y_{\mathbb P}}.
\end{align}
As indicated by our notation, this expectation value should be evaluated with the CGC weight function evolved to the 
rapidity scale $Y_{\mathbb P}$.
Notice that the target averaging is performed already at the level of the amplitude, due to our restriction to coherent scattering.  
This has an important consequence for our analysis: the total transverse momentum transferred between the target
and the 3-parton projectile is set by the scale $\sim 1/R_A$ for transverse inhomogeneity in the target --- a soft scale, of order $\Lambda$.
(This is {\it a priori} obvious by translational symmetry in the transverse plane and will be explicitly verified in the next subsection.)
Accordingly, the imbalance $ \bm{k}_1+ \bm{k}_2$ between the hard jets is still controlled by the semi-hard momentum
of the gluon jet, like in the absence of scattering: $|\bm{k}_1+ \bm{k}_2|\simeq k_{3\perp}$.

\subsection{The gluon dipole picture}
\label{sect:corrlimit}

The amplitude in Eq.~\eqref{tildeA} has a suggestive structure, characteristic of the dipole picture:
it is the convolution between an amplitude describing the decay $\gamma_T^*\to q\bar q$ of the virtual photon
(the same as appearing e.g. in the inclusive DIS cross-section) times a factor which describes both
the gluon emission ($q\bar q\to q\bar q g$) and the elastic scattering of the ensuing system of partons.
This convolution is not factorising because the second factor depends in a non-trivial way on the $q\bar q$ dipole size $\bm{r}$. 

Yet, as we shall demonstrate in what follows, genuine factorisation can be recovered after exploiting the 
separation of  transverse scales inherent in the correlation limit: besides being soft ($\xi\ll 1$), the gluon also has a relatively
low transverse momentum: $k_{3\perp}\sim Q_s\ll k_{1\perp},\, k_{2\perp}\sim Q$. This leads to important simplifications
which become transparent when using appropriate variables. For the $q \bar{q}$ pair we replace $\bk_1$ and $\bk_2$ by the relative 
momentum $\bm{P}$ and the total one $\bm{K}$, defined as
\begin{align}
	\label{PandK}
	\bm{P} \equiv
	\frac{\vartheta_2 \bm{k}_1 -
	\vartheta_1 \bm{k}_2}{\vartheta_1 +\vartheta_2}
	\,,\qquad
	\bm{K} \equiv \bm{k}_1 + \bm{k}_2,
\end{align}
with the inverse relations given by 
\begin{align}
	\label{PandKinv}
	\bm{k}_1 =\bm{P} +
	\frac{ \vartheta_1}{\vartheta_1 +
	\vartheta_2}\,\bm{K}
	\,,\qquad
	\bm{k}_2=-\bm{P} +
	\frac{ \vartheta_2}{\vartheta_1 +
	\vartheta_2}\,\bm{K}.
\end{align}
In coordinate space, we replace the positions $\bx$ and $\by$  by the transverse separation $\br$ and
the centre of energy $\bm{b}$:
\begin{align}
	\label{bdef}
	\br = \bx-\by,\qquad
	\bm{b}\equiv
	\frac{\vartheta_1\bx + \vartheta_2\by}
	{\vartheta_1 +\vartheta_2}\
\end{align} 
so that the inverse relations read
\begin{align}
	\label{bandrinv}
	\bx = \bb + \frac{\vartheta_2}{\vartheta_1+\vartheta_2}\,\br
	\,,\qquad
	\by = \bb - \frac{\vartheta_1}{\vartheta_1+\vartheta_2}\,\br.
\end{align}
Notice that for the purposes of this paper, one may let $\vartheta_1 + \vartheta_2 \simeq 1$ in all the above expressions. By further defining the separation $\bm{R} = \bz - \bb$ between the gluon and the $q\bar{q}$ pair, we can finally rewrite the combined phase in the triple Fourier transform in Eq.~\eqref{tildeA} as
\begin{align}
	\label{phase}
	\bm{k}_{1}\cdot\bm{x}+ 
	\bm{k}_{2}\cdot\bm{y}+
	\bm{k}_{3}\cdot\bm{z}=
	\bm{P}\cdot\bm{r}+ 
	(\bm{K}+\bm{k}_{3})\cdot\bm{b} + 
	\bm{k}_{3}\cdot \bm{R}\,.
\end{align}
Even though it is not critical for the existence of the correlation limit, we shall further assume that the target is homogeneous in the transverse plane. 
Then an elastic $S$-matrix like $\mcal{S}_{q\bar{q}g}(\bx,\by,\bz)$
 depends only upon differences between pairs of coordinates and thus it is independent of $\bm{b}$. The only 
$\bm{b}$-dependence that remains is the one in the phase, and the respective integrations in the 
direct amplitude and the complex conjugate amplitude combine to give
\begin{align}
	\label{intb}
	\int \frac{\rmd^2 \bar{\bb}}{2\pi}\,
	e^{i (\bm{K}+\bm{k}_{3})\cdot\bar{\bm{b}}} 
	\int \frac{\rmd^2 \bb}{2\pi}\,
	e^{-i (\bm{K}+\bm{k}_{3})\cdot\bm{b}} 
	=\int \rmd^2 \bar{\bb} \,
	\delta^{(2)}(\bm{K}+\bk_3)=
	S_{\perp} \delta^{(2)}(\bm{K}+\bk_3),
\end{align}
with $S_\perp$ the transverse area of the nucleus. This confirms that, for the case of coherent diffraction, the transverse imbalance of the $q\bar{q}$ pair
is minus the transverse momentum of the gluon.  Accordingly, the correlation limit can be reformulated as $P_{\perp} \sim Q \gg K_{\perp} = k_{3\perp} \sim Q_s(Y_{\mathbb P})$.

This hierarchy between the transverse momenta has implications for the transverse coordinates conjugated to them. 
As clear from \eqn{phase}, 
the hard scale $P_{\perp}$ controls the size $r$ of the $q\bar{q}$ pair, while the semi-hard momentum $k_{3\perp}$ regulates the gluon separation
$R=|\bm{R}|$ from its sources. So, the correlation limit implies the condition $r \ll R$ in coordinate space. Together with the flow of the colour, this condition
implies that the $q\bar{q}g$ projectile effectively acts as a {\it gluon-gluon dipole}. One leg of this dipole is made with the $q\bar{q}$ pair 
(which turns into a colour octet state after the gluon emission) and the other leg is the gluon itself. This becomes manifest by inspection of
the respective $S$-matrix: to the order of accuracy one can approximate 
$\bx \simeq \by \simeq \bb$  in Eq.~\eqref{Sqqg}. After recalling the SU($N_c$) relations $U^{\dagger ab} t^b= V t^{a}V^{\dagger} $
and $2 C_F N_c=N_c^2-1$, we readily find
\begin{align}
	\label{Sqqg-gd}
	\hspace{-0.1cm}
	\mcal{S}_{q\bar{q}g}(\bx,\by,\bz)\simeq \mcal{S}_{q\bar{q}g}(\bb,\bb,\bm{R}+\bb) 
     = 	\frac{1}{N_c^2-1}
	\big\langle
	\mathrm{Tr}
	\big[U(\bm{R}+\bb) 
	U^{\dagger}(\bb) \big]
	\big\rangle
	=\mcal{S}_g(\bm{R}),
\end{align}
with $\mcal{S}_g$ the average $S$-matrix for the elastic scattering of a $gg$ dipole of transverse separation $\bm{R}$. In the same limit, we  
have $\mcal{S}_{q\bar{q}}(\bm{x},\bm{y}) \simeq \mcal{S}_{q\bar{q}}(\bm{b},\bm{b}) = 1$, hence the $S$-matrices combination appearing in Eq.~\eqref{tildeA} reduces to
\begin{align}
	\label{SminusS}
	\mcal{S}_{q\bar{q}g}(\bb,\bb,\bm{R}+\bb) - 
	\mcal{S}_{q\bar{q}}(\bb,\bb)
	\simeq
	\mcal{S}_g(\bm{R}) -1 = -\mcal{T}_g(\bm{R}),
\end{align}
which is independent upon the size $\br$ of the small $q\bar q$ pair.

It remains to simplify the structure of the WW kernels for the gluon emission. Their difference 
in the second line of Eq.~\eqref{tildeA} vanishes when $r\to 0$ and can be expanded to linear order in $r/R$ to yield
\begin{align}
	\label{WWdiff}
	\frac{(\bx-\bz)^j}{(\bx-\bz)^2}-
	\frac{(\by-\bz)^j}{(\by-\bz)^2}
	\simeq
	\frac{r^j}{R^2}- 
	\frac{2 R^j (\br \cdot \bm{R})}{R^4} = 
	\frac{r^i}{R^2}
	\left(
	\delta^{ij} - \frac{2 R^i R^j}{R^2}
	\right).
\end{align}

After these simplifications,  the integrations over $\br$ and over $\bm{R}$ factorise from each other and 
the diffractive trijet cross section in Eqs.~\eqref{sigma3jet} and \eqref{tildeA} reduces to
\begin{align}
  \label{sigma3corr}
  \hspace{-0.9cm}
  \frac{\rmd \sigma_{\rm D}^{\gamma_T^* A
  \rightarrow q\bar q g A}}
  {\rmd \vartheta_1
  \rmd \vartheta_2
  \rmd \xi
  \rmd^{2}\!\bm{P}
  \rmd^{2}\!\bm{K}
  \rmd^{2}\bm{k}_{3}}=
  \frac{S_{\perp}\alpha_{em}N_c}{2 \pi^4}
  \left(\sum e_{f}^{2}\right)\!
  \left(\vartheta_1^2 \!+\! \vartheta_2^2\right)
  \delta(1\!-\!\vartheta_1 \!-\! \vartheta_2)\,
  \delta^{(2)}(\bm{K}\!+\!\bk_3)
  \frac{\alpha_s C_{F}}{\xi}
  \big|\mcal{A}_{q\bar{q} g}^{lj} \big|^2\!, 
\end{align}
where the new amplitude $\mcal{A}_{q\bar{q} g}^{lj}$ (defined without the trivial factor arising from the integration over $\bm{b}$) 
has a \emph{factorized} structure, as anticipated: 
 \begin{align}
	\label{Afact}
	\mcal{A}_{q\bar{q} g}^{lj} = 
	\mcal{H}^{li}(\bP,\bar{Q})
	\mcal{G}^{ij}(\bK,Y_{\mathbb P}).
\end{align} 

The {\it hard factor} is given by
\begin{align}
	\label{hardamp}
	\mcal{H}^{li}(\bP,\bar{Q}) =
	\int 
	\frac{\rmd^2 \br}{2\pi}\,
	e^{-i \bP \cdot \br}
	\frac{r^l r^i}{r}\,
	\bar{Q} K_1(\bar{Q}r) =
	\frac{1}{P_\perp^2 + \bar{Q}^2}
	\left(
	\delta^{li} - 
	\frac{2 P^l P^i}{P_{\perp}^2+\bar{Q}^2}
	\right)
\end{align}
and encompasses the amplitude for the decay of the transverse photon with polarization $l$ into a $q\bar q$ state 
and the vertex $\propto r^i$ for the subsequent emission of a transverse gluon with polarization $i$.  This hard factor 
is in fact identical to that occurring in the TMD factorization for inclusive dijet production\footnote{Note
however a difference in the physical interpretation: unlike in the current study, where the factor $r^i$ in the integrand 
in Eq.~\eqref{hardamp} corresponds to a gluon emission by the $q\bar q$ pair, in the case of inclusive dijets, it rather
describes the absorption of a gluon from the target.} \cite{Dominguez:2011wm}.
 It becomes traceless, as expected, in the limit $\bar{Q}^2=0$, i.e.~when the transverse photon is real.

The {\it semi-hard} tensorial distribution
\begin{align}
	\label{Gij}
	\mcal{G}^{ij}(\bK,Y_{\mathbb P}) = 
	\int \frac{\rmd^2 \bm{R}}{2 \pi}\,
	e^{i \bK \cdot \bm{R}}
	\left(
	\delta^{ij} - \frac{2 R^i R^j}{R^2}
	\right)
	\frac{\mcal{T}_g(R,Y_{\mathbb P})}{R^2}
	= \left(\frac{K^i K^j}{K_{\perp}^2} - 
	\frac{\delta^{ij}}{2} \right)
	\mcal{G}(K_{\perp},Y_{\mathbb P}),
\end{align}
which is real, symmetric, traceless and dimensionless, encodes the spatial distribution of the gluon emission by the small $q\bar{q}$ pair together with the scattering between the effective $gg$ dipole and the target. In order to arrive at the second equality we have assumed that the scattering amplitude $\mcal{T}_g$ depends only on the magnitude of the vector $\bm{R}$. By 
contracting Eq.~\eqref{Gij} with $K^i K^j/K_{\perp}^2$, one can find the following integral representation
for the scalar quantity $\mcal{G}(K_{\perp},Y_{\mathbb P})$:
\begin{align}
	\label{Gscalar}
	\mcal{G}(K_{\perp},Y_{\mathbb P}) = 
	2 \int \frac{\rmd^2 \bm{R}}{2 \pi}\,
	e^{i \bK \cdot \bm{R}}
	\left[1 - \frac{2 (\bK \cdot \bm{R})^2}{K_{\perp}^2 R^2}
	\right]
	\frac{\mcal{T}_g(R,Y_{\mathbb P})}{R^2} =
	2 \int_0^{\infty} \frac{\rmd R}{R}\,
	J_2(K_{\perp} R)\,
	\mcal{T}_g(R,Y_{\mathbb P}). 
\end{align}
We shall systematically study this quantity later on, after first generalising the above results to larger 
values $\xi\sim k^2_{3\perp}/Q^2$, which require going beyond the eikonal approximation for the emitted gluon.

The square of the (real) amplitude in Eq.~\eqref{Afact}, which is the quantity appearing in the cross section 
\eqref{sigma3corr}, simplifies even further. The tensor decomposition in Eq.~\eqref{Gij} implies 
\begin{align}
	\label{GijGkj}
	\mcal{G}^{ij}(\bK,Y_{\mathbb P})\,
	\mcal{G}^{kj}(\bK,Y_{\mathbb P}) = 
	\frac{\delta^{ik}}{4}\,[\mcal{G}(K_\perp,Y_{\mathbb P})]^2,
\end{align}  
meaning that we  only  need the following contraction  for the hard factor:
\begin{align}
    \label{hardsquare}
    [\mcal{H}^{li}(\bP,\bar{Q})]^2 = 
	2\,\frac{P_{\perp}^4 + \bar{Q}^4}{(P_{\perp}^2 + \bar{Q}^2)^4},
\end{align}
So, we finally arrive at the relatively simple expression for the amplitude squared
\begin{align}
	\label{Aljsquare}
	\big|\mcal{A}_{q\bar{q} g}^{lj} \big|^2 =
	\frac{1}{2}\,
	\frac{P_{\perp}^4 + \bar{Q}^4}{(P_{\perp}^2 + \bar{Q}^2)^4}\,
	[\mcal{G}(K_\perp,Y_{\mathbb P})]^2. 
\end{align}

To conclude this section, let us briefly describe the corresponding results for a virtual photon with longitudinal polarisation. They
are easily obtained by making the replacements 
$\vartheta_1^2+\vartheta_2^2 \to 4 \vartheta_1 \vartheta_2$ in Eq.~\eqref{sigma3jet} 
and $(r^l/r)K_1(\bar{Q} r) \to K_0(\bar{Q} r) $ in Eq.~\eqref{tildeA}. The latter eventually leads to the 
replacement $P_{\perp}^4 + \bar{Q}^4 \to 2 P_{\perp}^2 \bar{Q}^2$ in the numerator of 
Eqs.~\eqref{hardsquare} and \eqref{Aljsquare}. Obviously, the  
longitudinal cross section is of the same order with the transverse one when 
$\bar{Q} \sim P_{\perp}$, but it is negligible in cases where the photon becomes 
quasi-real, i.e.~when $Q^2 \to 0$, like in ultraperipheral heavy ion collisions. 

\section{Gluon dipole picture and TMD factorisation  for generic values of $\beta$}
\label{sect:factx}

Our construction of the gluon dipole picture in the previous section was limited to very soft gluon emissions, with
$\xi\ll  Q_s^2/Q^2\ll 1$, for which the eikonal approximation applies. Yet, as explained in Sect.~\ref{sec:kin}, this
regime  is not the most interesting ones for the phenomenology: it corresponds to relatively low values
$\beta\sim x\sim \xi Q^2/Q_s^2\ll 1$, hence to a rapidity gap $Y_{\mathbb P}$ which is significantly smaller than its
maximal possible value\footnote{Strictly speaking, the maximal value $Y_{\rm{max}}$ of
the rapidity gap for a given kinematics of the hard $q\bar q$ dijets truly is $Y_{\rm max}=\ln(1/x_{q\bar q})$; in practice
though $x_{q\bar q}$ is comparable to $\xbj$ (since $M_{q\bar q}^2\sim Q^2$ for the kinematics at hand),  
hence one indeed has $Y_{\rm max}\simeq \Ybj$.} $\Ybj$.

In this section, we shall demonstrate that the gluon dipole picture and the associated factorisation for diffractive
trijet production also holds for larger values $\xi \sim Q_s^2/Q^2 \ll 1$ (corresponding to $\beta,\,x\sim\order{1}$ and
hence $Y_{\mathbb P}\simeq \Ybj$), although the respective argument is considerably more subtle.
As before, we shall assume the ``correlation limit''
$k_{1\perp}^2,\, k_{2\perp}^2\sim Q^2\gg k_{3\perp}^2 \sim Q_s^2(A, Y_{\mathbb{P}})$ --- which is the typical
configuration for diffractive trijets in DIS at large $Q^2$, as also explained in Sect.~\ref{sec:kin}.

Before we proceed, it is convenient to rewrite the diffractive longitudinal fractions $\beta$ and $x$ in terms of the relative
and total transverse momenta of the hard dijets, as introduced in \eqn{PandK}. Using $P_{\perp}\sim Q\gg K_\perp= k_{3\perp}$
and $\xi\ll 1$, it is easy to check that $M_{q\bar q}^2\simeq P_{\perp}^2/(\vartheta_1 \vartheta_2)$ and therefore
 \begin{align}
	\label{xapprox}
	\beta \,\simeq \,
	\frac{\bar{Q}^2}{\bar{Q}^2 + 
	P_{\perp}^2  + \vartheta_1\vartheta_2
	K_{\perp}^2/\xi}, \qquad x \,\simeq \,
	\frac{\bar{Q}^2+P_{\perp}^2 }
	{\bar{Q}^2+P_{\perp}^2 + \vartheta_1 \vartheta_2 K_{\perp}^2/\xi}.
\end{align}
We recall that $\bar{Q}^2= \vartheta_1 \vartheta_2 Q^2$, and
$ \vartheta_1$ and  $\vartheta_2$ take generic values with  $\vartheta_1 +\vartheta_2 \simeq 1$; hence, all the hard scales
are parametrically of the same order: $\bar Q\sim Q\sim P_\perp$.  \eqn{xapprox} confirms that $\beta$ and $x$ become of order one
when  $\xi\sim K_{\perp}^2/Q^2$. It furthermore shows that in order to approach the limit $x\to 1$, one needs to consider even
larger values $\xi\gg K_{\perp}^2/Q^2$, with $\xi < 1$ of course. 

Our subsequent approximations will apply for all values of $\xi$ satisfying
$\xi\ll K_\perp/Q\ll 1$. Such values are still small enough to allow for simplifications like $1-\xi\simeq 1$ or
$ \vartheta_1+\xi\simeq  \vartheta_1$ (such simplifications have already been used in deriving the expressions for $\beta$ and $x$ in 
\eqn{xapprox}). But they can be large enough to cover not only the physically most interesting regime at $\xi\sim K_{\perp}^2/Q^2$,
but also (at least to some extent) the limit $x\to 1$ --- namely, we will approach that limit by studying $\xi$ values within
the range $K_{\perp}^2/Q^2\ll \xi \ll K_\perp/Q$.

For the physical interpretation of the subsequent results, it is useful to keep in mind that relatively hard gluons with 
$\xi\gtrsim k_{3\perp}^2/Q^2$ have large formation times, which are comparable to, or even larger than, the coherence time
of the virtual photon: $\tau_3= 2\xi q^+/k_{3\perp}^2 \gtrsim \tau_\gamma= 2q^+/Q^2$. Accordingly, such gluon emissions
may not be completely formed at the time of scattering with the nuclear shockwave. Besides, theses emissions are sensitive to the photon
virtuality $Q^2$, as clear from the fact all the terms in the energy denominator \eqref{EDqqg0} (or in the denominators
in \eqn{xapprox}) become comparable. We shall return to such physical considerations after some
technical developments.

 \subsection{Sub-eikonal corrections to the $q\bar qg$ wavefunction}
 \label{sect:noneik}
 
 As in Sect.~\ref{sec:eik}, we shall study the $q\bar q g$ Fock space component of the light-cone wavefunction of the virtual photon.  However, unlike in Sect.~\ref{sec:eik}, we shall now keep the sub-eikonal corrections 
 (the terms of first order in $\xi$) in both the energy
 denominator and the gluon emission vertices. Indeed, in these particular cases, 
 the contributions linear in $\xi$ are enhanced by large ratios
 of transverse momenta (see below).  For the same reason, one also needs to keep the instantaneous contributions,
  i.e.~contributions to the LCWF where the propagator of the intermediate quark or antiquark is instantaneous.
 On the other hand, we shall neglect $\xi$ next to unity, or next to the ``large'' longitudinal fractions
 $ \vartheta_1$ and  $\vartheta_2$.

Still as in  Sect.~\ref{sec:eik}, we shall first consider the LCWF in the absence of the scattering and in momentum space.
(The effects of the collision will be added later, after a suitable excursion through the coordinate representation.)
For the moment we shall omit all the factors which are the same as in the eikonal approximation 
and we will reinsert them at the very end, when computing the cross-section.

Let us begin with the regular terms, i.e.~those involving the non-instantaneous piece of the intermediate (anti)quark propagator
(see Fig.~\ref{fig:qqg}). The relevant energy denominators have already been presented in Sect.~\ref{sec:qqg}.
For the $q\bar q$ intermediate state, the energy difference $E_{q\bar{q}} - E_{\gamma} $ is given by \eqn{EDqq0} 
when the  gluon is emitted by the antiquark, and by a similar expression, but with $k_{1\perp}^2\to k_{2\perp}^2$,
for a gluon emission by the quark. The energy difference $E_{q\bar{q}g} - E_{\gamma}$ corresponding to the
$q\bar q g$ state is the same for both cases and is shown in the first equality in \eqn{EDqqg0}.
 Of course, for the present purposes, one must keep all the terms in that expression: when $\xi\gtrsim k_{3\perp}^2/Q^2$
 and $k_{1\perp}^2,\, k_{2\perp}^2\sim Q^2$,  all these terms are comparable with each other.
 
 The gluon emission vertices for generic values of $\xi$ are also well known in the literature --- they can be found e.g. in
 the calculations of next-to-leading order corrections to inclusive jets, or inclusive dijets (or hadrons), in either DIS, or
 in proton-nucleus collisions. Here we shall simply adapt these vertices from the literature and explain the origin
 of their $\xi$-dependence. 
 
 Up to an overall factor, which includes the coupling constants and the helicity structure of the photon decay vertex,
 and which can be simply read from  \eqn{Psieik}, the generalisation of \eqn{Psieik} including sub-eikonal corrections reads as follows
 \begin{align}
	\label{psi0}
	\Psi^{lj}_{\rm reg} = -
	\left[ \frac{k_1^l
	\left(k_3^j + 
	\frac{\xi}{1-\vartheta_1}\,k_1^j\right)}
	{k_{1\perp}^2 + \bar{Q}^2} + 
	\frac{k_2^l
	\left(k_3^j+
	\frac{\xi}{1-\vartheta_2}\,k_2^j\right)}
	{k_{2\perp}^2 + \bar{Q}^2}
	\right]
	\frac{1}{k_{3\perp}^2 + \mcal{M}^2},
\end{align}
where the subscript ``reg'' reminds that this is solely the contribution of the non-instantaneous graphs.
The scale $\mcal{M}^2$, which plays the role of an {\it effective gluon virtuality}\footnote{This ``virtuality'' is meant
in the sense of LC perturbation theory, that is, as a difference between on-shell LC energies at an emission
vertex, cf. \eqn{EDqqg0}; in other terms,  $\mcal{M}^2$ plays the same role for the $q\bar q g$ state as $\bar{Q}^2$ for the
$q\bar q$ state.}
encompasses the other terms in the energy difference \eqref{EDqqg0}, besides the gluon LC energy:
\begin{align}
	\label{Msquared}
	\mcal{M}^2\equiv
	\xi \left(
	\frac{k_{1\perp}^2}{ \vartheta_1} + 
	\frac{k_{2\perp}^2}{\vartheta_2} 
	+Q^2\right) \simeq 
	\frac{\xi}{\vartheta_1 \vartheta_2}
	\left(
	P_{\perp}^2 + \bar Q^2 
	\right).
\end{align}
The second equality above follows from (cf. \eqn{PandKinv})
\begin{align}
    \frac{k_{1\perp}^2}{\vartheta_1} + 
	\frac{k_{2\perp}^2}{\vartheta_2} =
	\frac{\vartheta_1 +\vartheta_2}
	{\vartheta_1 \vartheta_2}\, 
	P_\perp^2 +	
	\frac{K_{\perp}^2}{\vartheta_1 +\vartheta_2} 
	\,\simeq\, 
	\frac{P_\perp^2}{\vartheta_1\vartheta_2}\,.
\end{align}
To understand the emission vertices, consider e.g. the first term within the square braces in \eqn{psi0}, where the
gluon is emitted by the antiquark. First, the transverse photon decays into a $q\bar q$ pair with momenta $(\vartheta_1, \,\bk_1)$ 
and $(1-\vartheta_1, \,-\bk_1)$, so the respective vertex is proportional to $k_1^l$. Next, the antiquark $(1-\vartheta_1, \,-\bk_1)$,
emits a gluon ($\xi, \bk_3$) and the respective vertex is proportional to the relative transverse velocity 
$\bk_3/\xi+\bk_1/(1-\vartheta_1)$. We point out that Eq.~\eqref{psi0} contains the {\it exact} dependence of the $q\bar q g$ amplitude upon the various transverse momenta in the absence of scattering.
	
The gluon emission vertex from the antiquark can be simplified as 
\begin{align}
	\label{evsub}
	k_3^j+
	\frac{\xi}{1-\vartheta_1}k_1^j
	\,\simeq\, 
	k_3^j+\frac{\xi}{\vartheta_2}P^j,
\end{align}
and similarly for the emission from the quark, since the omitted terms of the order of $\mcal{O}(\xi^2 P_{\perp},\xi K_{\perp}, \dots)$ are small in comparison to (at least one of) the terms being kept. The term proportional to $\xi$ in Eq.~\eqref{evsub} represents sub-eikonal corrections and is strictly speaking small compared to the leading one since, as anticipated, we  work under the assumption 
that $\xi \ll k_{3\perp}/P_\perp$.

Let us now open a parenthesis and explain the motivation for this particular constraint. As observed in Sect.~\ref{sec:eik},
when the gluon emission is computed beyond the eikonal approximation, the transverse coordinate of the emitter can change,
due to recoil. Consider an emission by the antiquark.
 If $\bm{y}'$ and $\by$ are the corresponding positions before and after the gluon emission, one has
\begin{align}
	\label{deltay}
	\bm{y}' = \frac{\vartheta_2 \bm{y} + \xi \bz}{\vartheta_2 + \xi}
	\,\Rightarrow\,
	\Delta \by \equiv \by - \by' = 
	\frac{\xi (\by - \bz)}{\vartheta_2 + \xi}
	\simeq
	-\frac{\xi \bm{R}}{\vartheta_2}. 
\end{align}
Thus, in the region of interest, the relative change in the position of the antiquark quark reads
\begin{align}
	\label{relrecoil}
	\frac{|\Delta \by|}{r}
	\simeq
	\frac{\xi}{\vartheta_2}\,\frac{R}{r}
	\sim 
	\frac{\xi}{\vartheta_2}\,\frac{P_{\perp}}{k_{3\perp}} 
	\,\Rightarrow\,
	\frac{|\Delta \by|}{r}
	\ll 1
	\quad \mathrm{when} \quad
	\xi \ll k_{3\perp}/P_{\perp},
\end{align}
where we have assumed that sizes are correlated to momenta via the respective Fourier transforms, more precisely $R \sim 1/k_{3\perp}$ and $r \sim 1/P_{\perp}$ (cf.  Eq.~\eqref{phase}).
Thus, the inequality $\xi \ll k_{3\perp}/P_{\perp}$ guarantees that the transverse recoil of the quark and antiquark due to the gluon emission is negligible, so in particular, the $q\bar q$ pair is (comparatively) small both before, and after the emission.
In other terms, the hierarchy between the final transverse momenta unambiguously translates into a corresponding
hierarchy among transverse sizes {\it at all the stages of the process}, including the intermediate $q\bar q$ state.
In turn this ensures that our whole description of the process in coordinate space is indeed valid.

Although small, the correction of the order of $\xi$ in the vertices is important for what follows since the dominant effects cancel between the emissions by the quark and the antiquark. Now we shall expand the LCWF in Eq.~\eqref{psi0} in powers of $k_{3\perp}/P_\perp$ and to the order of interest we rewrite it as
\begin{align}\label{psireg1}
	\Psi^{lj}_{\rm reg} \simeq -
	\left( 
	\frac{k_1^l}{k_{1\perp}^2 + \bar{Q}^2} + 
	\frac{k_2^l}{k_{2\perp}^2 + \bar{Q}^2}
	\right)
	\frac{k_3^j}{k_{3\perp}^2 + \mcal{M}^2}\,-\, 
	\frac{\xi }{\vartheta_1 \vartheta_2} 
	\frac{P^l P^j}{P_{\perp}^2 + \bar{Q}^2}
	\frac{1}{k_{3\perp}^2 + \mcal{M}^2}\,.
\end{align}
In the first term we have kept only the dominant piece $k_3^j$ in the gluon emission vertex (cf. \eqn{evsub}), whereas in the second term we have similarly kept just the leading pieces, $k_1^l \simeq -k_2^l \simeq P^l$, in the photon decay vertex (cf. \eqn{PandKinv}).  The difference within the parenthesis in the first term can be further expanded as
\begin{align}
	\frac{k_1^l}{k_{1\perp}^2 + \bar{Q}^2} + 
	\frac{k_2^l}{k_{2\perp}^2 + \bar{Q}^2} 
	\simeq
	\frac{1}{P_{\perp}^2 +\bar{Q}^2}
	\left( 
	\delta^{li} - \frac{2 P^l P^i}{P_{\perp}^2 +\bar{Q}^2}
	\right)
	K^{i}
	= -\mcal{H}^{li}(\bP) k_{3}^i,
\end{align}
where we have recognised the hard factor $\mcal{H}^{li}(\bP)$ defined in Eq.~\eqref{hardamp} and also recalled that $\bm{K}=-\bk_3$. Using Eq.~\eqref{Msquared} to eliminate the explicit factor of $\xi$ in the second term of \eqn{psireg1}, we deduce our final result
for this regular piece of the LCWF:
\begin{align}
	\label{psireg}
	\Psi^{lj}_{\rm reg}(\bP,\bk_3) =
	\mcal{H}^{li}(\bP)\,
	\frac{k_3^i\, k_3^j}
	{k_{3\perp}^2 +\mcal{M}^2} -
	\frac{P^l P^j}
	{(P_{\perp}^2 + \bar{Q}^2)^2}
	\frac{\mcal{M}^2}{k_{3\perp}^2 + \mcal{M}^2}\,.
\end{align}
As a consistency check, the two terms in Eq.~\eqref{psireg} are of the same order when $\xi \sim k_{3\perp}^2/P_{\perp}^2$. 

The full LCWF involves an additional piece, that was ignored so far, which is generated by the instantaneous piece of the quark propagator and reads \cite{Beuf:2017bpd} (see App.~\ref{sect:lcwf} for more details)
 \begin{align}
 	\label{psiinst}
 	\Psi_{\rm inst}^{lj}(\bP,\bk_3)= 
 	\frac{\delta^{lj}}{2}\frac{\xi }
 	{\vartheta_1 \vartheta_2} 
 	\frac{1}{{k_{3\perp}^2 +\mcal{M}^2}} = 
 	\frac{\delta^{lj}}{2}
 	\frac{1}{P_{\perp}^2 + \bar{Q}^2}\,
 	\frac{\mcal{M}^2}{k_{3\perp}^2 +\mcal{M}^2}\,.
 \end{align}
 The second term of $\psi^{lj}_{\rm reg}$ in Eq.~\eqref{psireg} combines with $\psi^{lj}_{\rm inst}$ in \eqref{psiinst} to give a contribution proportional to the hard factor $\mcal{H}^{li}(\bP)$, so that the complete virtual photon LCWF takes the \emph{factorized} form (unlike the regular piece alone)
\begin{align}
	\label{psitotal}
	\Psi_{\rm tot}^{lj}(\bP,\bk_3) = 
	\Psi_{\rm reg}^{lj}(\bP,\bk_3) +
	\Psi_{\rm inst}^{lj}(\bP,\bk_3)=
	\mcal{H}^{li}(\bP)\,
	\frac{k_3^i\, k_3^j+\delta^{ij} \mcal{M}^2/2}
	{k_{3\perp}^2 +\mcal{M}^2}\,.
 \end{align}
At this point, it is instructive to decompose the total LCWF into a traceless and a diagonal part (so long as the indices $i$ and $j$ are concerned), namely
\begin{align}
	\label{psisplit}
	\Psi^{lj}_{\rm tot}(\bP,\bk_3)=
	\Psi_{\mathbb{P}}^{lj}(\bP,\bk_3)+ 
	\Psi_{\rm diag}^{lj}(\bP,\bk_3),
\end{align} 
where the two pieces are given by
\begin{align}
	\label{psiPd}
	\Psi_{\mathbb{P}}^{lj}(\bP,\bk_3) \equiv 
	\mcal{H}^{li}(\bP)\,
	\frac{k_3^i\, k_3^j- \delta^{ij} k_{3\perp}^2/2}
	{k_{3\perp}^2 +\mcal{M}^2}
	\quad\,
	\mathrm{and}
	\quad\,
	\Psi_{\rm diag}^{lj}(\bP,\bk_3) \equiv 
	\frac{\mcal{H}^{lj}(\bP)}{2}\,.
\end{align}	
We shall eventually keep only the traceless, ``Pomeron'', piece, since it will be clear in a while that the diagonal piece does not contribute to the scattering process.
 
 Here comes a crucial point in our analysis. In principle, all the momenta $\bm{P}$, $\bm{K}$ and $\bk_3$ can be modified due to the scattering by contributions of the order of $Q_s$. Since however we are interested in hard dijets with $P_\perp^2\gg Q_s^2$, we can neglect the change in the relative momentum $\bm{P}$. This is consistent with the fact that, in the gluon dipole picture, the Wilson lines from the $gg$ dipole $S$--matrix do not involve the transverse separation $\br$ of the $q\bar q$ pair (the variable conjugated to $\bm{P}$, cf.~\eqref{phase}). So we shall still work at fixed $\bm{P}$, but we shall take the Fourier transform w.r.t.~$\bk_3$ (the variable conjugate to $\bm{R}$), in order to include multiple scattering in the eikonal approximation. It is straightforward to find that the LCWF in this mixed space becomes 
\begin{align}
	\label{psimixed}
	\Psi_{\mathbb{P}}^{lj}(\bP,\bm{R}) = 
	\int \frac{\rmd^2 \bk_3}{(2\pi)^2}\, 
	e^{i \bk_3 \cdot \bm{R}}\,
	\psi_{\mathbb{P}}^{lj}(\bP,\bk_3)
	=
	\frac{1}{2\pi}\,\mcal{H}^{li}(\bP)
	\left( \frac{\delta^{ij}}{2} - 
	\frac{R^i R^j}{R^2} \right)
	\mcal{M}^2 K_2(\mcal{M}R),
	\end{align}
with $K_2(\mcal{M}R)$ the modified Bessel function of second rank.
Now it suffices to multiply the above $\psi_{\mathbb{P}}^{lj}(\bP,\bm{R})$ with the $gg$ dipole amplitude $\mcal{T}_g(R)$ and then perform the inverse Fourier transform in order fix the \emph{final} gluon momentum. The latter is still equal to minus the imbalance of the $q\bar{q}$ (recall that we consider elastic scattering off a homogeneous nucleus) for which we shall still use the notation $-\bK$ in order to avoid introducing additional notation. 

To summarise, the full amplitude to the accuracy of interest and in the presence of scattering can be expressed as
\begin{align}
	\label{Aljnew}
	\mcal{A}^{lj}_{q\bar{q}g} 
	= \mcal{H}^{li}(\bP) 
	\mcal{G}^{ij}(\bK,\mcal{M}, Y_{\mathbb P}),
\end{align}
with  
\begin{align}
	\label{Gijnew}
	\hspace{-0.8cm}
	\mcal{G}^{ij}(\bK,\mcal{M}, Y_{\mathbb P}) &\,=\! 
	\int\! \frac{\rmd^2 \bm{R}}{2\pi}\,
	e^{i \bK \cdot \bm{R}}
	\left( 
	\frac{\delta^{ij}}{2} \!-\! 
	\frac{R^i R^j}{R^2} 
	\right)\! 
	\mcal{M}^2 K_2(\mcal{M}R)\, 
	\mcal{T}_g(R,Y_{\mathbb P}) \nn &\, = 
	\left( 
	\frac{K^i K^j}{K_{\perp}^2} \!-\! 
	\frac{\delta^{ij}}{2} 
	\right)
	\mcal{G}(K_{\perp},\mcal{M} ,Y_{\mathbb P}),
\end{align}
where, as in Eq.~\eqref{Gij} we have made it explicit that the tensorial distribution $\mcal{G}^{ij}(\bK,Y_{\mathbb P})$ is traceless. The scalar distribution, after also performing the angular integration, reads
\begin{align}
	\label{Gscalarnew}
	\mcal{G}(K_{\perp},\mcal{M} ,Y_{\mathbb P})=
	\mcal{M}^2 \int_0^\infty 
	\rmd R\, R\, 
	J_2(K_{\perp} R) 
	K_2(\mcal{M} R) \mcal{T}_g(R,Y_{\mathbb P}).
\end{align}
Regarding the diagonal piece of the LCWF in Eq.~\eqref{psiPd}, we immediately observe that it is independent of the gluon transverse momentum $\bk_3$, thus its Fourier transform to the mixed space ($\bP$,$\bm{R}$) generates a $\delta$-function $\delta^{(2)}(\bm{R})$. As anticipated, this gives a vanishing result after multiplication with the gluon dipole amplitude $\mcal{T}_g(R,Y_{\mathbb P})$, due to colour transparency.

\eqn{Aljnew} is clearly a generalisation of the previous result in \eqn{Afact}, to which it reduces for sufficiently small $\xi$: 
indeed, when $\xi\ll K_\perp^2/Q^2$, we also have $\mcal{M}^2/K_\perp^2\ll 1$ (recall  \eqn{Msquared}), 
so one can take the limit  $\mcal{M} R\ll 1$ inside the integrand of \eqn{Gscalarnew}. By also using
$\mcal{M}^2 K_2(\mcal{M}R)\simeq 2/R^2$ for $\mcal{M}R\ll 1$, one recovers the result of the
eikonal approximation in Eq.~\eqref{Gscalar}. 

Yet, at a first sight, the presence of the virtuality scale $\mcal{M}^2$ in the 
distribution \eqref{Gijnew} seems to violate the factorisation between the hard and the semi-hard sectors: indeed,
as clear from \eqn{Msquared}, this scale is built with the hard scales $Q^2$ and $P_\perp^2$. However, \eqn{Msquared} also
involves the ``plus'' longitudinal momentum fraction of the gluon $\xi=k_3^+/q^+$, which is not the right variable for 
the collinear factorisation: in the latter, the ``semi-hard TMD'' should be a target distribution, hence it must depend upon the
``minus'' longitudinal fraction $x$ --- the fraction  of the Pomeron longitudinal momentum $x_{\mathbb{P}}P_N^-$ which
is taken by the gluon exchanged between the Pomeron and the hard dijets.
Eqs.~\eqref{xapprox} and \eqref{Msquared} trivially  
lead to
\begin{align}
	\label{Msquaredx}
	\mcal{M}^2 = 
	\frac{x}{1-x} K_{\perp}^2
\end{align}
and the TMD-like factorization of the diffractive amplitude $\mcal{A}_{q\bar{q}g}^{lj}$ in Eq.~\eqref{Aljnew} is now manifest. 
From now on, we shall replace $\mcal{M}\to x$ in the arguments of the semi-hard distribution; e.g. we shall write
$\mcal{G}(K_{\perp},x,Y_{\mathbb P})$.

Eq.~\eqref{Gscalarnew} features two Bessel functions. The function $J_2(K_{\perp}R)$ has emerged due to the particular tensor structure of $G^{ij}(\bK,x,Y_{\mathbb P})$, while the hyperbolic one $K_2(\mcal{M}R)$ reflects the gluon virtuality and reduces the available phase space when $x$ is not small. 
Since the traceless property of the tensor $G^{ij}(\bK,x,Y_{\mathbb P})$ is independent of the value of $x$, Eqs.~\eqref{GijGkj} and \eqref{Aljsquare} which are necessary for proceeding to calculate the cross section remain valid.

\subsection{TMD factorisation for diffractive trijets}
\label{sec:fact}

To summarise the analysis in the previous subsection, the  cross section for diffractive trijet production 
 in the correlation limit $P_\perp^2\sim Q^2\gg k_{3\perp}^2 \sim Q_s^2(A, Y_{\mathbb{P}})$ and for 
 any $\xi\lesssim  Q_s^2/Q^2$ is still given by Eq.~\eqref{sigma3corr}, but with the more general amplitude displayed in 
 Eqs.~\eqref{Aljnew}--\eqref{Gscalarnew}. As already explained, $\xi$ is not the right variable to exhibit collinear factorisation,
 so it is convenient to make a change of variable from $\xi$ to either  $x$, or $x_{\mathbb P}$. In practice, we shall use
 $x_{\mathbb P}$, since this is directly related to the rapidity gap $Y_{\mathbb P}$ in the final state. 
 From Eqs.~\eqref{xP} and \eqref{xdef},
 one easily finds
  \begin{align}
	\label{xitoxp}
	\frac{x_{\mathbb{P}}-x_{q\bar q}} {x_{q\bar q}}=\frac{K_{\perp}^2/\xi}{Q^2+M_{q\bar q}^2+K_{\perp}^2}
	\ \Longrightarrow\ 
		\frac{\rmd\xi}{\xi}= 
	\frac{\rmd x_{\mathbb{P}}}
	{x_{\mathbb{P}}-x_{q\bar q}} =
	\frac{\rmd x_{\mathbb{P}}}
	{x_{\mathbb{P}}}\,\frac{1}{1-x}\,. 
\end{align}
It is furthermore trivial to integrate over the transverse momentum $\bk_3$ of the gluon jet by using the respective
$\delta$-function  Eq.~\eqref{sigma3corr}, which fixes $\bk_3=-\bK$.

To summarise, the differential cross-section for
producing a pair of hard dijets with relative transverse momentum $\bP$, total transverse momentum $\bK$, 
longitudinal momentum fractions $\vartheta_1$ and $\vartheta_2$, and such that there is a rapidity gap
$Y_{\mathbb P}$ between the diffractive system (which also includes the unmeasured, gluon, jet) and the target,
takes the following, {\it factorised}, form:
\begin{align}
	\label{3jetsD1}
	\frac{\rmd \sigma_{\rm D}^{\gamma_{T,L}^* A
	\rightarrow q\bar q g A}}
  	{\rmd \vartheta_1
  	\rmd \vartheta_2
  	\rmd^{2}\!\bm{P}
  	\rmd^{2}\!\bm{K}
  	\rmd Y_{\mathbb P}} = 
  	 H_{T,L}(\vartheta_1,\vartheta_2, {Q}^2, P_{\perp}^2)\,
  	 \frac{\dif xG_{\mathbb{P}}(x, x_{\mathbb{P}}, K_\perp^2)}
  	 {\dif^2\bm{K}}.
 \end{align}
 This generalises Eq.~\eqref{sigma3corr} to generic values $x < 1$.
 The ``hard factors'' have been defined as 
 \begin{align}
 	\label{HardT}
	H_T(\vartheta_1,\vartheta_2, {Q}^2, P_{\perp}^2)
	\equiv 
	{\alpha_{em}\alpha_s}
	\Big(\sum e_{f}^{2}\Big) \,
	\delta(1-\vartheta_1-\vartheta_2)  
	\left(\vartheta_1^{2} + 
	\vartheta_2^{2}\right)
	\frac{P_{\perp}^4 + \bar{Q}^4}
	{(P_{\perp}^2 + \bar{Q}^2)^4}\,
 \end{align}
 for a photon with transverse polarisation and,  respectively,
\begin{align}
	\label{HardL}
	H_L(\vartheta_1,\vartheta_2,{Q}^2, P_{\perp}^2)
	\equiv 
	{\alpha_{em}\alpha_s}\Big(\sum e_{f}^{2}\Big)\,
	\delta(1-\vartheta_1-\vartheta_2)  
	\vartheta_1 \vartheta_2 \,
	\frac{8P_{\perp}^2 \bar{Q}^2}{(P_{\perp}^2 + \bar{Q}^2)^4}\,
 \end{align}
 for the case of a longitudinal photon.
 
 Furthermore, the ``semi-hard factor'' in \eqn{3jetsD1} is defined as
 \begin{align}
	\label{pomugddef}	
	\frac{\rmd xG_{\mathbb{P}}
	(x, x_{\mathbb{P}}, K_{\perp}^2)}
	{\rmd^2 \bK} \equiv 
	\frac{S_\perp (N_c^2-1)}{4\pi^3}\,  \frac{
	[\mcal{G}(K_{\perp},x,Y_{\mathbb P})]^2}{2\pi(1-x)}, 
 \end{align}
 where the factor $1-x$ in the denominator arises from the Jacobian for the change of variables in Eq.~\eqref{xitoxp}.
 It is here understood that  $x$ is not an independent variable; rather its value is fixed by the kinematics of DIS and
  that of the diffractive final  state, as follows (see also \eqn{xapprox})
 \beq
 x= \frac{x_{q\bar q}}{x_{\mathbb P}}\,=\,\beta\, \frac{x_{q\bar q}}{\xbj}
 \,\simeq\,\beta\, \frac{\bar Q^2+P_\perp^2}{\bar Q^2}\,.\eeq
  
\begin{figure}[t] 
\centerline{
\includegraphics[width=.83\textwidth]{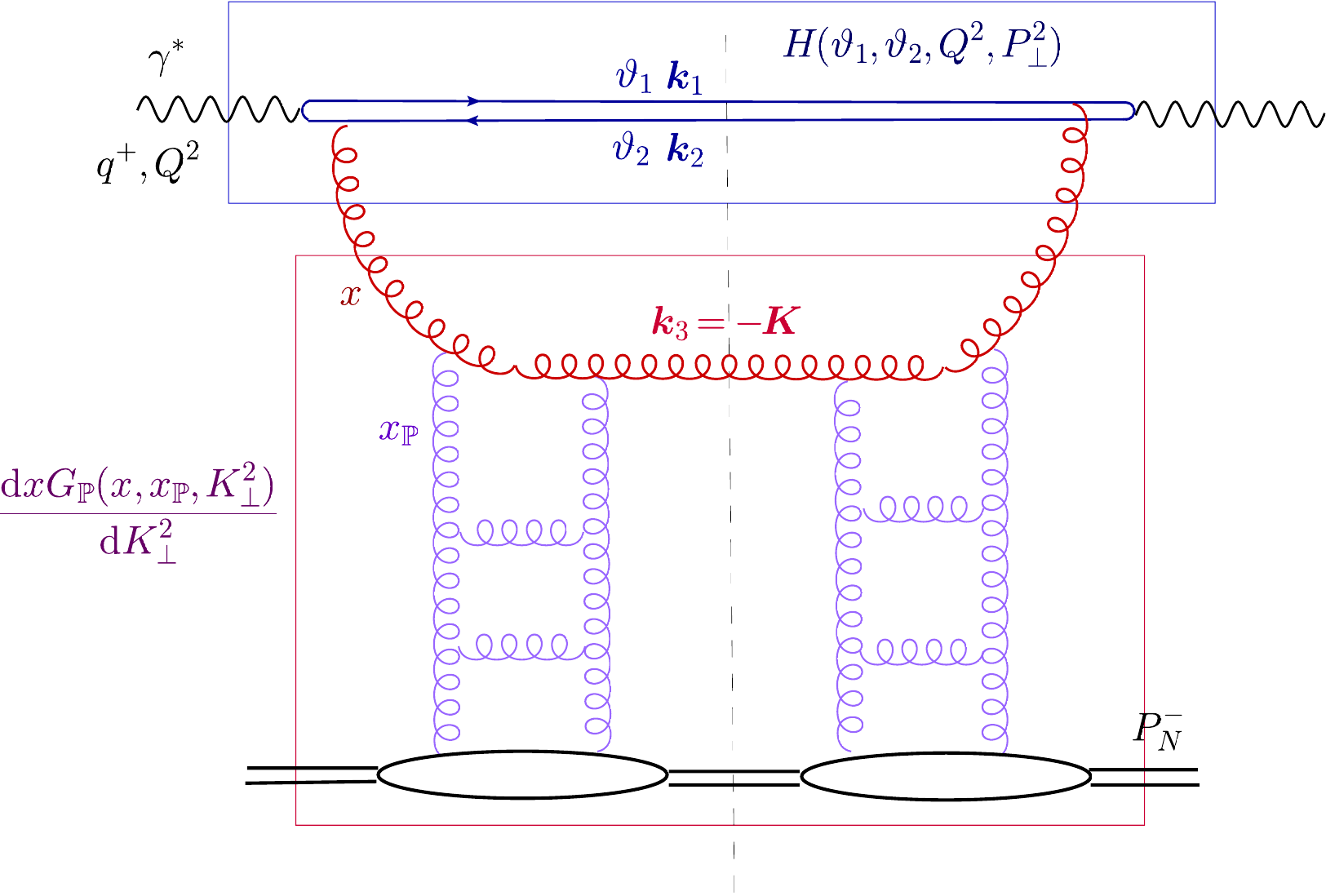}}
\caption{\small Graphical representation of the TMD factorisation for diffractive 2+1 jet production.}
\label{fig:TMDfact}
\end{figure}

The factorised structure of 
 Eq.~\eqref{3jetsD1} is recognised as a {\it diffractive version of TMD factorisation} (see also Fig.~\ref{fig:TMDfact}).
The cross-section is the product of a 
  hard factor describing the formation and the scattering of a small projectile (here, the $q\bar q$ dipole) and a semi-hard factor
that can be interpreted as the unintegrated  (or transverse-momentum dependent) gluon distribution of the
``Pomeron'' --- a colourless excitation of the hadronic target, which carries a (minus) 
 longitudinal momentum fraction equal to $x_{\mathbb{P}}$. This interpretation motivates our notation for the semi-hard
 factor (cf. the l.h.s. of \eqref{pomugddef}): $xG_{\mathbb{P}}(x, x_{\mathbb{P}}, K_{\perp}^2)$ denotes the
 gluon distribution of the Pomeron, for gluons with longitudinal momentum fraction $x$ (w.r.t. to the Pomeron) 
 and any transverse momentum
 $k_\perp \le K_\perp$; hence,  its derivative w.r.t. $K_\perp^2$ yields the respective {\it unintegrated} distribution
 --- a diffractive TMD.
 
 Underlying this interpretation, there is a new perspective over this diffractive process.
 This would be the actual physical picture if 
 the process was studied  in the target infinite momentum frame and the light-cone gauge $A^-_a=0$. 
 In this new picture, the ``projectile'' refers to the hard $q\bar q$ dijets alone, whereas the  gluon emission is
 associated with the  target LCWF. More precisely, the Pomeron emits {\it two} gluons in a overall colour singlet
 state (a $gg$ dipole): one gluon with splitting fraction $x$ and transverse momentum $\bm{K}$, 
 which is then absorbed by the projectile ---  this is the gluon  ``measured'' by the diffractive TMD --- and another one 
  with splitting fraction $1-x$ and transverse momentum $-\bm{K}$, which is released in the final state. 
  Both gluons represent the unique gluon from the original, ``colour dipole'', picture, but at different stages:
  before and respectively after its scattering with the target.

 The TMD  factorisation in \eqn{3jetsD1} is very similar to that for inclusive dijet production in the 
 correlation limit \cite{Dominguez:2011wm,Metz:2011wb,Dominguez:2011br}. The hard factors are {\it exactly} the 
 same\footnote{In fact, the normalisation of the hard factors in Eqs.~\eqref{HardT}--\eqref{HardL} has been precisely
 chosen such as to match the respective results for inclusive dijets. This choice, together with our current calculation
 of diffractive dijets, uniquely fixes the normalisation of the Pomeron UGD in \eqn{pomugddef}.}, 
 whereas the Weiszäcker-Williams UGD (the gluon occupation number in the target)
 is  replaced by the diffractive TMD \eqref{pomugddef}, which expresses the gluon occupation number in the
 Pomeron. The precise definition for the latter with our present conventions reads as follows:
  \begin{align}
	\label{occup}	
		\Phi_{\mathbb P}(x,x_{\mathbb P},K_{\perp}^2) \equiv 
		\frac{(2\pi)^3}{S_\perp 2(N_c^2-1)}\, \frac{\rmd xG_{\mathbb{P}}
	(x, x_{\mathbb{P}}, K_{\perp}^2)}{\rmd^2 \bK}
		= \frac{
	[\mcal{G}(K_{\perp},x,Y_{\mathbb P})]^2}{2\pi(1-x)}\,.
 \end{align}

From the general discussion in Sect.~\ref{sec:kin}, we expect the bulk of the Pomeron UGD to be localised at transverse
momenta $K_\perp$ of the order of the saturation momentum $Q_s(Y_{\mathbb{P}})$. This will be confirmed by
the analysis in the next section, where we shall evaluate the integral in \eqn{Gscalarnew} with various approximations
for the gluon dipole amplitude. Yet, this strong correlation between the target saturation momentum and 
the transverse momentum imbalance  $K_\perp$ between the two hard jets can be spoilt in practice 
by the {\it Sudakov effect} --- i.e. by final-state radiation. The quark and the antiquark are generally
off-shell after the collision, hence they can radiate gluons with transverse momenta $p_\perp\lesssim P_\perp$.
The recoil associated with such emissions can affect the $K_\perp$--distribution of the final dijets and thus hinder the
effects of gluon saturation. This final-state radiation is a higher-order correction, yet its effects can be numerically 
large, since enhanced by the Sudakov double-logarithm  $\ln^2(P_{\perp}^2/K_{\perp}^2)$: when computing the cross-section
for a given imbalance $K_\perp$, one must forbid real gluon emissions with much larger momenta
$K_\perp \ll p_\perp \ll P_\perp$, whereas there is no similar constraint on the virtual gluon emissions.
This mismatch between the phase-spaces for real and respectively virtual emissions is the origin of the Sudakov double-logs
 \cite{Mueller:2013wwa}.  Whereas an explicit calculation of the Sudakov effect for this particular process 
  (diffractive 2+1 jets production) is undoubtedly important, this is left for 
 further work.
 
 Following an original proposal in \cite{Iancu:2021rup}, here we shall use an alternative strategy
 to circumvent the Sudakov effect, which consists in integrating the cross-section \eqref{3jetsD1} 
 over the hard dijet imbalance $\bm{K}$. This integration is formally restricted to $K_\perp\le P_{\perp}$, to preserve
 the scale hierarchy in the problem, yet this restriction has little consequences in
 practice, since the integral is controlled by values
 $K_\perp\sim Q_s(Y_{\mathbb{P}})$, as we shall see. This manipulation leads
 to the expected version of collinear factorisation for (non-exclusive) diffractive dijet production,
that is,
\begin{align}
	\label{gluondipColl}
	\frac{\rmd \sigma_{\rm D}^{\gamma_{T,L}^* A
	\rightarrow q\bar q g A}}
  	{\rmd \vartheta_1
  	\rmd \vartheta_2
  	\rmd^{2}\!\bm{P}
  	\rmd Y_{\mathbb P}} = 
  	 H_{T,L}(\bar{Q}^2, P_{\perp}^2)\,
  	 xG_{\mathbb{P}}(x, x_{\mathbb{P}}, P_\perp^2),
 \end{align}
 which features the standard gluon distribution of the Pomeron, a.k.a. the gluon DPDF:
\begin{align}
	\label{xGP}	
	xG_{\mathbb{P}}(x, x_{\mathbb{P}}, P_{\perp}^2)
	= \int_0^{P_{\perp}}\!
	\dif^2\bK\,
	\frac{\rmd xG_{\mathbb{P}}(x, x_{\mathbb{P}}, K_{\perp}^2)}
	{\rmd^2 \bK} 
	= 
	\frac{S_\perp (N_c^2-1)}{4\pi^3} 
	\int_0^{P_{\perp}}\!
	\dif^2\bK\, 
	\Phi_{\mathbb P}(x,x_{\mathbb P},K_{\perp}^2).
 \end{align}
 The integration over $\bm{K}$ removes the restriction on the final state and hence allows for the standard cancellations
 between real and virtual emissions, thus eliminating the Sudakov double-logs, as expected. On the other hand,
 these cancellations still allow for single logarithmic corrections due to the final-state radiation, which express
 the DGLAP evolution of the DPDF. This evolution becomes important when $\alpha_s\ln (P_{\perp}^2/Q_s^2)\gtrsim 1$
 and will be indeed included in our approach, in Sect.~\ref{sect:dglap}  below. 
 
 The above discussion also implies that,  in general, the measured pair of hard dijets can be accompanied 
 in the final state by more than just one semi-hard gluon: to leading order
 accuracy, these are the gluons generated by the DGLAP evolution, hence they are strongly ordered in 
 transverse momenta,  within the range $Q_s^2(Y_{\mathbb{P}})\sim k_{3\perp}^2\ll k_{4\perp}^2\ll k_{5\perp}^2\ll \cdots
 k_{n\perp}^2 \ll P_\perp^2$.
(The first gluon, with $ k_{3\perp}\sim Q_s(Y_{\mathbb{P}})$, defines the tree-level process, with 2+1 jets, that we have
originally considered; the harder gluons  $k_{4\perp},\, k_{5\perp}\cdots $ are generated by the DGLAP evolution.)
The general features characterising such a $(2+n)$--jets, diffractive, final state
 are \texttt{(i)} the presence of two hard jets, which are nearly back to back
($ k_{1\perp}\simeq k_{2\perp} \simeq P_\perp$) and carry a fraction $ x_{q\bar q}$ of the target longitudinal momentum $P_N^-$, 
and \texttt{(ii)} the existence of a rapidity gap $Y_{\mathbb{P}}=\ln (1/x_{\mathbb{P}})$ 
between the hard dijets and the elastically scattered target,
with $x_{\mathbb{P}} > x_{q\bar q}$. The most interesting situation is such that $x_{\mathbb{P}} \sim x_{q\bar q}$,
or $x\sim\order{1}$, and will be studied at length in what follows.

 \section{The gluon distribution of the Pomeron}
 \label{sect:pomeron}
 
 In this section, we will perform systematic studies, analytic and numerical, for the gluon distribution of the Pomeron,
 as introduced in the previous section. We will consider both the unintegrated version of this distribution, i.e. the
 gluon occupation number, and its integrated version, i.e. the total number of gluons in the structure of the Pomeron.
The essential ingredient of this distribution is the scattering amplitude $\mcal{T}_g(R,Y_{\mathbb P})$
 of a gluon-gluon dipole, for which we shall consider two cases:  intermediate energies (moderate values 
 of $Y_{\mathbb P}$, such that $\abar Y_{\mathbb P}\ll 1$), where one can rely on
the MV model \cite{McLerran:1993ni,McLerran:1994vd}, and
 higher energy/$Y_{\mathbb P}$, where $\mcal{T}_g(R,Y_{\mathbb P})$ is
 obtained from numerical solutions to the BK equation  \cite{Balitsky:1995ub,Kovchegov:1999yj} with an initial
 condition given again by the MV model. In the case of the integrated Pomeron gluon distribution (a.k.a. the
 gluon diffractive parton distribution function), we shall also study the effects of the DGLAP evolution 
 \cite{Gribov:1972ri,Altarelli:1977zs,Dokshitzer:1977sg} between the Pomeron and the hard $q\bar q$ dijets.

\subsection{The unintegrated  distribution}
\label{sect:pomugd}

We start with the unintegrated distribution defined in Eq.~\eqref{pomugddef} with
the function $\mcal{G}(K_{\perp}, x,Y_{\mathbb P})$ given by Eq.~\eqref{Gscalarnew}. From the derivation of the latter,
it should be quite clear that the integration variable $R$ in Eq.~\eqref{Gscalarnew}
represents the gluon dipole size at the time of the collision with the nuclear shockwave.
In general, this integral is controlled by 3 momentum scales: the gluon momentum $K_{\perp}$ (the same as
the dijet imbalance), the gluon virtuality  $\mcal{M}$, and the target saturation momentum $Q_s(A, Y_{\mathbb{P}})$
(as encoded in the structure of the  scattering amplitude $\mcal{T}_g(R,Y_{\mathbb P})$).

The first two among these scales restrict the integration over $R$, in the sense of introducing an effective upper limit $R_{\rm max}$.
The Bessel function $J_2(K_{\perp} R)$ is rapidly oscillating when its argument becomes larger than unity, whereas the
modified Bessel function $K_2(\mcal{M} R)$ is exponentially decaying when $\mcal{M} R> 1$. So, clearly, 
$R_{\rm max}={\rm min}\{1/K_{\perp},\, 1/\mcal{M}\}$. We can distinguish several cases, depending upon the value of $x$.

\texttt{(i)} When $x\ll 1$, \eqn{Msquaredx} shows that $\mcal{M}\ll K_\perp$,
so clearly $R_{\rm max}=1/K_{\perp}$  and the virtuality plays no role; in fact, in this case $\mcal{M} R\ll 1$ for any
$R\le R_{\rm max}$, so one can replace $\mcal{M}^2 K_2(\mcal{M}R)\simeq 2/R^2$ (meaning that we return to the result
 \eqref{Gscalar} of the eikonal approximation).  
 
 \texttt{(ii)} When $x$ is very close to one, such that $1-x\ll 1$, we have the opposite
 situation:  $\mcal{M}\gg K_\perp$, hence $R_{\rm max}=1/\mcal{M}$ and we can replace $J_2(K_{\perp} R)\simeq (K_{\perp} R)^2/8$
 for analytic estimates.

 \texttt{(iii)} When both $x$ and $1-x$ are of order one, $K_{\perp}$ and  $\mcal{M}$ are commensurable with each other.
 
\begin{figure}[t] 
\centerline{
\includegraphics[width=.65\textwidth]{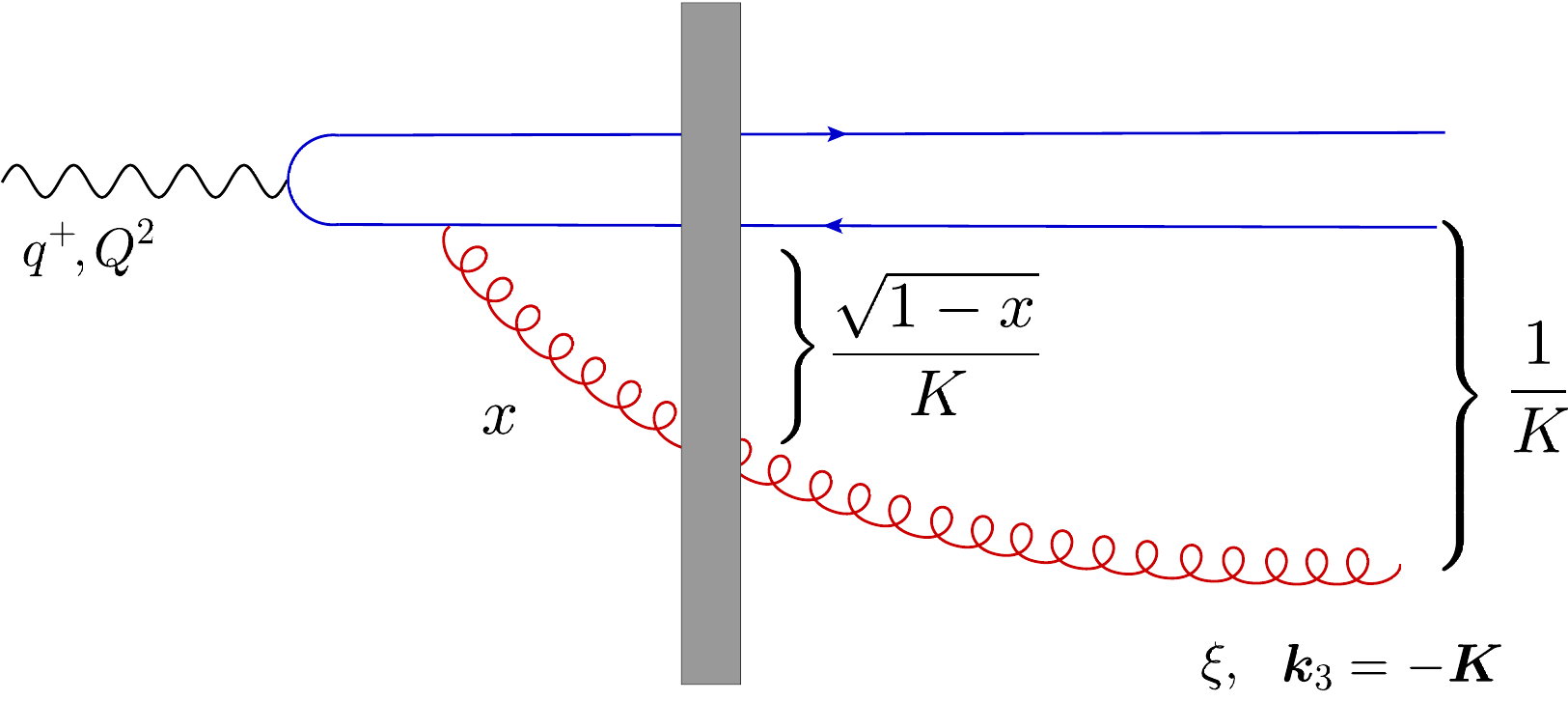}}
\caption{\small  Pictorial illustration of the $q\bar q g$ Fock state in the regime where the gluon formation time
is relatively large, of the order of the coherence time of the virtual photon.}
\label{fig:Rmax}
\end{figure}

 So, in practice, we are always in a two-scale problem, $R_{\rm max}$ and $1/Q_s(A,Y_{\mathbb{P}})$, but such that
 the value of  $R_{\rm max}$ depends upon $x$. By inspection of the previous arguments, it becomes clear that, parametrically,
 \beq\label{Rmax}
 R_{\rm max}^2(x)\,\sim\,\frac{1-x}{K_\perp^2}\,.\eeq
This upper limit on the gluon dipole size has a simple physical interpretation \cite{Iancu:2021rup} (see also
Fig.~\ref{fig:Rmax}): the transverse separation
between the gluon and its emitter grows via diffusion, $R^2(\tau)\sim \tau/k_3^+$,
where $\tau$ is the interval in $x^+$ from the photon decay vertex. This distance becomes of order $1/k_{3\perp}$ when 
$\tau$ is of the order of the formation time $\tau_3=2k_3^+/k_{3\perp}^2$. When $x$ is near one, however, this formation
time becomes larger than the coherence time $\tau_\gamma= 2q^+/Q^2$
of the virtual photon (the time at which the $q\bar q$ pair crosses the nuclear shockwave).
 Indeed, $1-x\ll 1$ corresponds to $\xi \gg k_{3\perp}^2/Q^2$ (cf. Eq.~\eqref{xapprox}), which in turn implies
$\tau_3\gg \tau_\gamma$, as explained at the beginning of Sect.~\ref{sect:factx}.
Accordingly, the transverse separation $R(\tau_\gamma)$ at the time of scattering is smaller than $1/k_{3\perp}$.
Recalling that $k_{3\perp}=K_\perp$ and $k_3^+=\xi q^+$, one immediately finds 
$R^2(\tau_\gamma)\sim 1/(\xi Q^2)$, which is indeed the same as $R_{\rm max}^2$ in \eqn{Rmax}.

Furthermore for any value of $x$, the gluon dipole amplitude $\mcal{T}_g(R,Y_{\mathbb P})$ determines what are the
values of $R$ which control the integration in Eq.~\eqref{Gscalarnew}. Indeed, so long as the scattering is weak,
meaning for small dipole sizes $R\ll 1/Q_s$, this amplitude is rapidly increasing with $R$, roughly like 
$\mcal{T}_g(R)\propto R^2Q_s^2$, and then the whole integrand is rapidly increasing as well. This rise saturates
for larger dipole sizes $R\gtrsim 1/Q_s$, where the amplitude reaches the unitarity limit $\mcal{T}_g(R)=1$. So, clearly,
there are two situations: (a) when $R_{\rm max}(x)\ll  1/Q_s$, the integral is dominated by its (effective) upper limit
$R_{\rm max}(x)$; in that case, the scattering is weak; (b) when $R_{\rm max}(x)\gtrsim  1/Q_s$, the integral is rather
controlled by $R\sim 1/Q_s$ since the integrand is strongly peaked at that value; in that case, the scattering is
predominantly strong. Recalling \eqn{Rmax}, one deduces that there are two physical regimes depending upon the
value of the momentum imbalance $K_\perp$: a weak scattering regime at $K_\perp\ll \tilde Q_s(x)$ and a
strong scattering regime at $K_\perp\gtrsim \tilde Q_s(x)$. Here, $\tilde Q_s^2(x)\equiv
\tilde{Q}_s^2(x,A, Y_{\mathbb{P}})$ is an {\it effective saturation momentum}, 
defined as
      \begin{align}
	\label{Qstilde}
	\tilde{Q}_s^2(x,A, Y_{\mathbb{P}}) \equiv (1-x) Q_s^2(A, Y_{\mathbb{P}}).
\end{align}
which controls the process of diffractive trijet production for generic values of $x$. At small $x\ll 1$, this reduces, of course,
to the standard saturation scale of the target. But when $x\sim 1$, \eqn{Qstilde} takes into account the effects of
the virtuality on the gluon emission.

Incidentally, the previous discussion also shows that, for generic $x\sim\order{1}$, the transverse momentum condition
underlying the correlation limit should be generalised to
\beq\label{PvsKx}
P_\perp^2\,\gg\,\frac{K_\perp^2}{1-x}\,.\eeq
Indeed, one needs to ensure that the size $r\sim 1/P_\perp$ of the $q\bar q$ pair is much smaller than the 
gluon transverse separation  $R_{\rm max}$ at the time of scattering.

Having understood the relevant scales at work, let us now study the properties of the Pomeron UGD in more detail.
We start with the strong scattering regime, where we can replace the gluon dipole amplitude by its black disk limit
($\mcal{T}_g(R)\to 1$), as just explained. This yields (we will omit the rapidity variables $Y_{\mathbb{P}}$ or $_{\mathbb{P}}$ 
whenever unessential, to simplify writing)
\begin{align}
\label{pomugdlow}
	\mcal{G}(K_{\perp}, x)=
	\mcal{M}^2 \int_0^\infty 
	\rmd R\, R\, 
	J_2(K_{\perp} R) 
	K_2(\mcal{M} R) =
	\frac{K_{\perp}^2}{K_{\perp}^2 +\mcal{M}^2} = 
	1-x
	\quad
	\mathrm{for} 
	\quad
	K_{\perp} \ll \tilde{Q}_s(x),
\end{align}
where we have also used  \eqn{Msquaredx}. Together with \eqn{occup}, this result implies that 
 the gluon occupation number $\Phi_{\mathbb P}(x,K_{\perp})$ inside the Pomeron
 reaches a limiting value $(1-x)/2\pi$ at low $K_{\perp} \ll \tilde{Q}_s$.

Remarkably, the result in Eq.~\eqref{pomugdlow} turns out to be identical with the respective prediction of the LCWF
in the absence of scattering, that is, $\Psi_{\mathbb{P}}^{lj}(\bP,\bK)$ from \eqn{psiPd}; indeed, the scalar projection of
the semihard, tensorial, component of $\Psi_{\mathbb{P}}^{lj}(\bP,\bK)$ reads ${K_{\perp}^2}/({K_{\perp}^2 +\mcal{M}^2})$,
in agreement with the r.h.s. of Eq.~\eqref{pomugdlow}. So, strong scattering appears to be equivalent to no scattering at all, 
which might look surprising at a first sight. In fact, this is a general characteristic of the black disk limit, that in the present
context can be more explicitly understood as follows. 

From the general discussion in Sect.~\ref{sec:eik}, we recall 
that the gluon dipole amplitude $\mcal{T}_g(R)$ has been constructed as the difference of 2 distinct contributions
(cf.  Eqs.~\eqref{tildeA}, \eqref{SminusS} and Figs.~\ref{fig:qqgSW} and \ref{fig:qqgSW2}): \texttt{(i)}
a piece where the gluon emission occurs prior to the collision, which yields the 3-parton ($q\bar q g$) $S$-matrix in 
Eq.~\eqref{tildeA} --- equivalently, the gluon dipole $S$-matrix $\mcal{S}_g(\bm{R}) $ in  \eqref{SminusS},
and  \texttt{(ii)}
a piece where the gluon is emitted after the scattering, which yields the $q\bar q$ $S$-matrix
in Eq.~\eqref{tildeA} and the unit term in the r.h.s. of \eqn{SminusS}.  
In the ``strong scattering'' regime under consideration, the kinematics is such that the 3-parton system
is completely absorbed, meaning $\mcal{S}_g(\bm{R}) \to 0$, but the intermediate $q\bar q$ pair is still at
colour transparency, that is, $\mcal{S}_{q\bar q}(\br)=1$. We thus conclude that the black disk limit for the
trijet production is in fact controlled by configurations where the semi-hard, gluon, jet has been emitted only after
the scattering and hence it did not participate in the collision.
 
 Consider now the weak scattering regime at $K_{\perp} \gg \tilde{Q}_s(x)$. In this case, we need a specific model for 
 the gluon dipole amplitude (at least, for relatively small dipoles) in order to proceed. To start with, we use 
 the MV model, which is a reasonable approximation so long as $Y_{\mathbb{P}}$ is low enough ($\alpha_s
 Y_{\mathbb{P}}\ll 1$) for the high-energy evolution of the Pomeron to be negligible. In this model, 
  the elastic amplitude for a $gg$ dipole is given by
\begin{align}
	\label{MVgg}
	\mcal{T}_g(R) = 1 -
	\exp \left( 
	-\frac{R^2 Q_{gA}^2}{4}\,\ln\frac{4}{R^2 \Lambda^2}
	\right),
\end{align}
where the scale $Q_{gA}^2$ is a measure of the density of valence colour sources per unit transverse
area weighted with the Casimir factor $C_A=N_c$ for gluons (since the projectile is made with a $gg$ pair).
Accordingly, this scale $Q_{gA}^2$ is roughly twice as large as the respective scale for a $q\bar{q}$ dipole:
$Q_{gA}^2 = (N_c/C_F) Q_{A}^2$. The saturation scale in this model is conventionally defined as the value of $2/R$
for which the exponent is one, that is
\begin{align}
	\label{MVQs}
	Q_s^2 = Q_{gA}^2 \ln \frac{Q_s^2}{\Lambda^2}.
\end{align}

When evaluating Eq.~\eqref{Gscalarnew} for  $K_{\perp} \gg \tilde{Q}_s(x)$, one can replace the dipole amplitude in Eq.~\eqref{MVgg} 
with its single scattering approximation,
\begin{align}
	\label{MVgg1scatt}
	\mcal{T}_g(R)  \simeq \frac{R^2 Q_{gA}^2}{4}\,\ln\frac{4}{R^2 \Lambda^2}
	\quad
	\mathrm{for} 
	\quad
	R \ll  1/{Q}_s.
\end{align}
The ensuing integral over $R$ is controlled by the effective upper cutoff $R_{\rm max}$, as
introduced by (one or the other of) the Bessel functions. One finds
\begin{align}
\label{pomugdhigh}
	\mcal{G}(K_{\perp},x) &=
	\frac{\mcal{M}^2 Q_{gA}^2}{4} \int_0^\infty 
	\rmd R\, R^3 J_2(K_{\perp} R) K_2(\mcal{M} R)
	\ln \frac{4}{R^2 \Lambda^2} 
	\nn
	&\simeq (1-x)^2 (1+2x)\, \frac{Q_{gA}^2}{K_{\perp}^2}
	\ln \frac{K_{\perp}^2}{(1-x)\Lambda^2}
	\nn
	&= (1-x) (1+2x) \frac{\tilde{Q}_s^2(x)}{K_{\perp}^2}
	\left(
	1+ \frac{\ln K_{\perp}^2/\tilde{Q}_s^2(x)}{\ln Q_s^2/\Lambda^2}
	\right)	
	\quad
	\mathrm{for} 
	\quad
	K_{\perp} \gg \tilde{Q}_s.
\end{align}
The second line is the ``natural form'' in the absence of saturation and reduces to the already known result in the limit $x \to 0$ \cite{Iancu:2021rup}, while in the third line we have expressed the answer in terms of the saturation scale $\tilde{Q}_s^2$ (cf.~the definitions in Eqs.~\eqref{Qstilde} and \eqref{MVQs}). The dominant $K_\perp$-dependence as encoded in the factor
$1/K_{\perp}^2$ can be recognised as the bremsstrahlung spectrum for the gluon emission by either the $q\bar q$ pair,
or by the Pomeron --- depending upon the perspective that we have over this process.

For more clarity, it is convenient to summarise now the limiting behaviours previously found for the Pomeron UGD,
at very low and very large transverse momenta, respectively
\begin{align}\label{Philimits}
\Phi_{\mathbb{P}}(x, K_\perp^2)\,\simeq \frac{1-x}{2\pi}
    \begin{cases}    \displaystyle{\ 1}
         &
        \text{for \ $K_\perp\!\ll \tilde Q_s(x)$}
                  \\*[0.4cm]
               \displaystyle{\ \frac{\tilde Q_s^4(x)}{K_\perp^4}} &
        \text{for \ $K_\perp\!\gg \tilde Q_s(x)$}\,,
         \end{cases}
\end{align}
where the second line strictly holds within the MV model, cf. \eqn{pomugdhigh}, and
 we have ignored slowly varying functions of $x$ or $K_\perp^2$,  to emphasise  the dominant respective
dependences. These results show several interesting features:

{\it (a)} a strong suppression  $\propto 1/K_\perp^4$ for
 large values of the dijet momentum imbalance $K_\perp\!\gg \tilde Q_s(x)$. This suppression is a consequence of the
 elastic nature of the scattering. It confirms that the bulk of the events are characterised by $K_\perp\sim \tilde Q_s(x)$,
 that is, they are {\it controlled by the physics of saturation};
 
 {\it (b)} a rather strong  suppression when $x\to 1$; from the viewpoint of projectile evolution, this is naturally understood
 as the suppression of gluon emissions with very large formation times $\tau_3\gg \tau_\gamma$; alternatively, when the
 gluon is viewed as a part of the LCWF of the Pomeron, this suppression is due to the reduction in phase-space for
 emissions at large $x$. This ultimately justifies our emphasis in this study on gluon emissions with lower formation times
 $\tau_3\lesssim \tau_\gamma$ and hence relatively soft longitudinal momenta $\xi\lesssim K_\perp^2/Q^2\ll 1$.

 {\it (c)} geometric scaling:  within the approximations  underlying \eqn{Philimits}, 
 the function $\Phi_{\mathbb{P}}(x, K_\perp^2)/(1-x)$
 depends upon its arguments only via the dimensionless variable $K_\perp/\tilde Q_s(x)$. This scaling is only
 approximate (e.g. it is violated by the factor $1+2x$ visible in \eqn{pomugdhigh}) and will be further studied
 in what follows, both analytically and numerically.
 
 It is furthermore interesting to compare the above estimates for the gluon occupation number in
 the Pomeron to the standard (``Weiszäcker-Williams'') occupation number for small-$x$ gluons 
 in a dense hadronic target  \cite{Iancu:2002xk,Iancu:2003xm,Gelis:2010nm,Kovchegov:2012mbw}  --- a quantity which in particular enters 
 the TMD factorisation for inclusive dijet production in DIS at small Bjorken $x$ \cite{Dominguez:2011wm}.
 This quantity can be explicitly computed in the MV model (see e.g. \cite{Iancu:2002xk,Iancu:2003xm}), and one
 finds the following limiting behaviours:
 \begin{align}\label{PhiWW}
\Phi_{WW}(K_\perp^2)\,\simeq \frac{1}{\alpha_s N_c}
    \begin{cases}    \displaystyle{\ \ln\frac{Q_{s}^2}{K_\perp^2}}
         &
        \text{for \ $K_\perp\!\ll  Q_{s}$}
                  \\*[0.4cm]
               \displaystyle{\ \frac{ Q_{gA}^2}{K_\perp^2}} &
        \text{for \ $K_\perp\!\gg Q_{s}$}\,,
         \end{cases}
\end{align}
where the scales $Q_{gA}$ and $Q_s$ have the same meaning as in Eqs.~\eqref{MVgg}--\eqref{MVQs}.
Note that there is no $x$-dependence in these formulae\footnote{In the context of inclusive dijets production,
the relevant value of $x$ is $x_{q\bar q}$, i.e. the energy fraction taken by the produced $q\bar q$ pair, cf.
\eqn{xqqdef}, which is indeed small, $x_{q\bar q}\ll1 $, for DIS at high energy.}: 
they strictly hold at small (but not {\it too} small) values $x\ll 1$, 
and an explicit $x$-dependence would occur only after taking into account the high-energy BK/JIMWLK evolution (which
goes beyond the MV model).

Comparing Eqs.~\eqref{Philimits} and \eqref{PhiWW}, one can notice two important differences:

\texttt{(I)} In the strong scattering regime at low momenta $K_\perp\!\lesssim  Q_{s}$, the WW gluon occupation number
is of $\order{1/\alpha_s}$ (up to a slowly varying logarithm) and hence it is parametrically larger
than for gluons in the Pomeron in the corresponding regime at $K_\perp\!\lesssim \tilde Q_s(x)$. This difference
can be understood as follows. The low-momentum gluons from the target are genuinely {\it saturated}: they are densely packed,
 with occupation numbers are large as possible. A small external projectile, like the $q\bar q$ pair in the context
 of inclusive dijet production, directly couples to one of these saturated gluons\footnote{Whenever we speak about ``saturated
 gluons'', we implicitly have in mind the description of the scattering process in the target  LC gauge $A^-_a=0$.
 In the alternative description using the projectile LC gauge $A^+_a=0$, the ``saturated gluons'' are replaced by
 Coulomb exchanges responsible for strong scattering.}.
  On the contrary,  the gluon distribution of the Pomeron remains
  {\it dilute}, with occupation numbers of $\order{1}$, even in the  strong scattering regime at 
  $K_\perp\!\lesssim \tilde Q_s(x)$. In that case, the phenomenon of saturation does not refer to the gluon
  exchanged between the Pomeron and the $q\bar q$ projectile, but to its  {\it sources}: the wee gluons
  composing the Pomeron.   Within the MV model, these gluons are generated via radiation from the valence quarks;
 in general, additional sources are produced by the high-energy evolution over the rapidity interval $Y_{\mathbb{P}}$.
 A gluon occupation number of order one is naturally
obtained as the product between the density of saturated sources, of $\order{1/\alpha_s}$, and the factor  $\alpha_s$ 
from the vertex describing  the emission of the $gg$ pair from these sources.

\texttt{(II)} At large transverse momenta $K_\perp\!\gg Q_{s}$, the  WW UGD in \eqn{PhiWW}
decreases like $1/K_\perp^2$, which is the bremsstrahlung spectrum for  a single gluon emission.
This decrease is considerably slower than that observed for the Pomeron UGD at $K_\perp\!\gg \tilde Q_s(x)$,
cf. Eq.~\eqref{Philimits}. 
This means that most of the inclusive dijet events lie in the tail of the $K_\perp$--distribution at
$Q_s^2\ll K_\perp^2\ll Q^2$ (they are logarithmically distributed within the range $Q_s^2\ll K_\perp^2\ll Q^2$;
see also Sect.~\ref{sect:pomgd} below) and thus are insensitive to gluon saturation, at 
variance with the typical diffractive events.
 
These considerations demonstrate that the diffractive production of hard dijets is a more sensible probe of gluon saturation 
than the inclusive dijet production in the correlation limit.

\subsection{More refined estimates for the UGD}

After these general considerations, let us deduce some more refined estimates for the Pomeron UGD, first within the
MV model and then by also including the high-energy evolution of the Pomeron (i.e. of the gluon dipole amplitude   in \eqn{Gscalarnew}).

First, we will use the MV model to construct a global approximation for the function $\mcal{G}(x,K_{\perp})$ valid throughout
the strong scattering regime, i.e. for all
momenta $K_{\perp} \lesssim \tilde{Q}_s(x)$. (The previous respective estimates in Eq.~\eqref{pomugdlow} or Eq.~\eqref{Philimits}
refer only to the low-momentum  limit $K_\perp \ll \tilde Q_s(x)$.)  For any such value of $K_\perp$,
the integral over $R$ in \eqn{Gscalarnew} is controlled by relatively large dipole sizes
 $R\gtrsim 1/Q_s$, for which the dipole amplitude in \eqn{MVgg} can be approximated by a Gaussian\footnote{As a matter
 of fact,  it is possible to obtain quasi-exact analytic expressions for any $K_\perp$ (in the two limiting cases of small $x$ and small $1-x$) 
 even for the full MV model --- that is, beyond the Gaussian (or GBW) approximation.
 To this end, we replace $R$ inside the logarithms by the value that renders the argument of the respective Bessel function of the order of one. Both cases can be encompassed by letting $R \to 2 \sqrt{1-x}/K_{\perp}$, which eventually means that the final results
  in Eqs.~\eqref{pomugdsat0} and \eqref{pomugdsat1} remain valid for the full MV model and any $K_\perp$ provided one makes
 the replacement $\tilde{Q}_s^2 \to Q_A^2(1-x) \ln [K_{\perp}^2/(1-x)\Lambda^2]$. In the case $x\ll 1$, this procedure
  leads to Eq.~(16) in Ref.~\cite{Iancu:2021rup}.}:
   $\mcal{T}_g(R) \simeq 1 - e^{-R^2 Q_s^2/4}$. (In other terms,
 the MV model is effectively replaced by the GBW model  \cite{GolecBiernat:1999qd}.)
    In order to get analytic results, we focus on the two limiting cases
 $x \to 0$ and $x\to 1$, but our subsequent analysis will also shed light on the behaviour for generic values of $x$.
 
 When $x\ll 1$, the gluon virtuality plays no role and we are left with a simple integral  \begin{align}
\label{pomugdsat0}
	\mcal{G}(K_{\perp},x)
	&\, =
	2 \int_0^\infty \!
	\frac{\rmd R}{R}\, 
	J_2(K_{\perp} R)
	\left( 1 - e^{-R^2 Q_s^2/4} \right) \nn &\, = 
	\frac{Q_s^2}{K_{\perp}^2}\!
	\left( 1 - e^{-K_{\perp}^2/Q_s^2} \right)
	\quad
	\mathrm{for} 
	\quad
	K_{\perp} \lesssim Q_s\,,\
	x\ll1.
\end{align}
(We have also used  $\tilde Q_s(x)\simeq Q_s$ at small $x$.)
This equation describes the change in the spectrum of the produced gluon via multiple scattering off the saturated gluons.
The Gaussian broadening is clearly visible for $K_{\perp}$ around the saturation momentum. But the Gaussian also 
controls the behaviour at low $K_\perp$, via its  small-argument expansion:
 by expanding to second order, one finds 
\beq
\mcal{G}(K_{\perp},x)\simeq 1 - \frac{K_{\perp}^2}{2Q_s^2}\quad
	\mathrm{for} 
	\quad
	K_{\perp} \ll Q_s\,.\ 
	x\ll1.
\eeq
This estimate describes the deviation from the ``black disk limit'' in  Eq.~\eqref{pomugdlow} with increasing $K_{\perp}/Q_s$.

When $1-x\ll 1$, we can approximate $J_2(K_{\perp} R) \simeq K_{\perp}^2R^2/8$ and we find
\begin{align}
\label{pomugdsat1}
	\hspace*{-0.8cm}
	\mcal{G}(K_{\perp},x)
	&=
	\frac{\mcal{M}^2 K_{\perp}^2}{8} 
	\int_0^\infty 
	{\rmd R}\, R^3\, K_2(\mcal{M} R)
	\left( 1 - e^{-R^2 Q_s^2/4} \right)
	\nn
	&\hspace*{-0.5cm}= (1-x)
	\left[1- 
	\frac{K_{\perp}^2}{2\tilde{Q}_s^2}
	+
	\frac{K_{\perp}^4}{2\tilde{Q}_s^4}
	-
	\frac{K_{\perp}^6}{2\tilde{Q}_s^6}\,
	e^{K_{\perp}^2/\tilde{Q}_s^2}
	E_1(K_{\perp}^2/\tilde{Q}_s^2)
	\right]
	\quad
	\mathrm{for} 
	\quad
	K_{\perp} \lesssim \tilde{Q}_s
	\,,\,
	1-x \ll 1,
\end{align}
where we used $\mcal{M}^2 \simeq K_{\perp}^2/(1-x)$ and  $E_1(z) = \int_z^{\infty} {\rm d}t\, e^{-t}/t$ is
 the exponential integral function. 
It is interesting to compare Eqs.~\eqref{pomugdsat0} and \eqref{pomugdsat1}: when expanding for $K_{\perp} \ll \tilde{Q}_s$, the two expansions agree with each other to order $(K_{\perp}/\tilde{Q}_s)^2$ and the discrepancy starts only at the fourth order. Thus, when $\mcal{G}(x,K_{\perp})/(1-x)$ is viewed as a function of $K_{\perp}/\tilde{Q}_s$ for various values of $x$, one expects to see a much better scaling behavior below the saturation scale than above it (cf.~the additional $1+2x$ factor in Eq.~\eqref{pomugdhigh}), i.e.~we expect
\begin{align}
	\label{scaling}
	\frac{\mcal{G}(K_{\perp},x)}{1-x}
	\simeq F(K_\perp/\tilde{Q}_s)
	\quad
	\mathrm{for} 
	\quad
	K_{\perp} \lesssim \tilde{Q}_s.
\end{align}
We should stress again that the scaling is only approximate and it becomes even less accurate above the saturation scale. Employing again Eqs.~\eqref{pomugdsat0} and \eqref{pomugdsat1} we can easily estimate the maximum scaling violation in the transition region around $\tilde{Q}_s$.  We find
\begin{align}
	\label{ratioatQs}
	\frac
	{[\mcal{G}(K_{\perp}=\tilde{Q}_s(x), x)/(1-x)]_{x\to 1}}
	{[\mcal{G}(K_{\perp}=\tilde{Q}_s(x), x)/(1-x)]_{x\to 0}}
	=\frac{1 - e E_1(1)/2}{1 - e^{-1}}\simeq1.11,	
\end{align}
which implies a scaling violation of $\sim$10\% for $\mcal{G}(K_{\perp},x)$ and thus $\sim$20\% for its square (the Pomeron UGD).

\begin{figure}[t] 
	\centerline{\includegraphics[width=0.8\columnwidth]{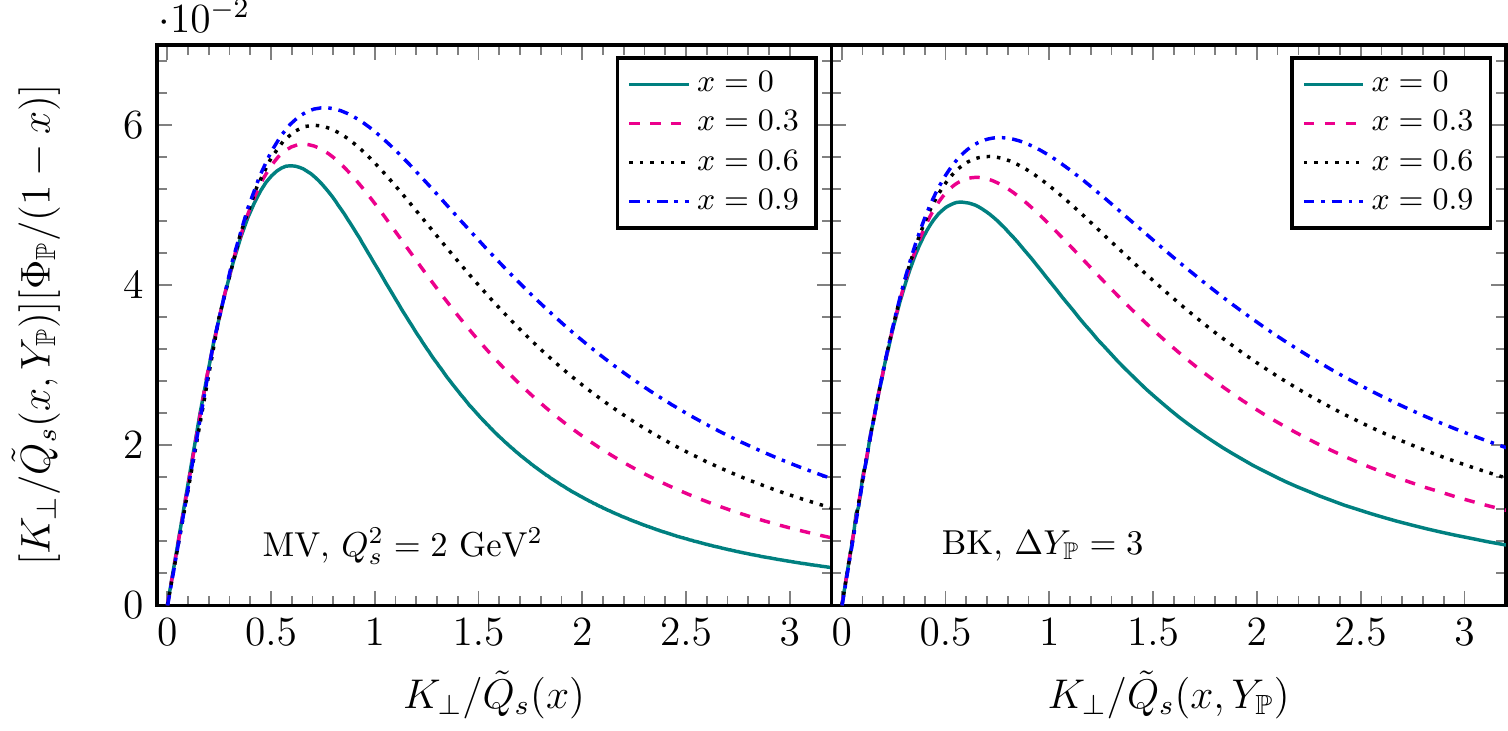}}
	\caption{\small The gluon occupation number for the Pomeron
 	as obtained from Eqs.~\eqref{Gscalarnew} and \eqref{occup}, scaled with $[K_{\perp}/\tilde{Q}_s(x,Y_{\mathbb P})]/(1-x)$ and plotted as a function of $K_\perp/\tilde Q_s(x,Y_{\mathbb P})$, for various values of $x$ and $Y_{\mathbb P}$. 
 	Left panel:  $\mathcal{T}_{g}({R})$ is given by the MV model. 
 	Right panel:  $\mathcal{T}_{g}({R,Y_{\mathbb P}})$ is obtained from the solution to the BK equation with the initial condition at 
	$\Delta Y_{\mathbb P}=0$ (i.e. $Y_{\mathbb P} = Y_{0}$) given by the MV model.}
\label{fig:dxGP}
\end{figure}

All the features of the Pomeron distribution discussed above can be verified by making a direct numerical calculation of Eq.~\eqref{Gscalarnew}, with the results exhibited in the left panel of Fig.~\ref{fig:dxGP}. The slope in the origin verifies Eq.~\eqref{pomugdlow}, the good scaling in the regime $K_{\perp} \lesssim \tilde{Q}_s$ is clearly visible, the values around the saturation scale confirm the $\sim$20\% difference given in Eq.~\eqref{ratioatQs}, and, finally, the scaling violation becomes stronger in the high momentum tail.

So far, we have exclusively relied on the MV model to evaluate the gluon dipole scattering amplitude 
$\mcal{T}_g(R,Y_{\mathbb P})$, meaning
that we have implicitly considered relatively small values for the diffractive gap $Y_{\mathbb P}$, such that $\alpha_s Y_{\mathbb P}
\ll 1$. Yet, the most interesting regime for a study of gluon saturation is when the diffractive gap is large enough,
$\alpha_s Y_{\mathbb P}\gtrsim 1$, for the effects of the high-energy evolution to become important. This evolution amplifies
the distribution of colours sources (gluons with longitudinal momentum fractions $x\ge x_{\mathbb P}$) within the Pomeron and
thus enhances the effects of gluon saturation: the relavant  saturation momentum $\tilde{Q}_s(x, A,Y_{\mathbb P})$
 increases with $Y_{\mathbb P}$. The inclusion of the high-energy evolution is quite
straightforward for the problem at hand: this refers to the evolution of gluon dipole amplitude 
$\mcal{T}_g(R,Y_{\mathbb P})$, which is in turn determined (at least in the limit of a large number of colours) by
the BK equation\footnote{More precisely, the BK equation is usually written for the $S$-matrix $\mcal{S}(\bm{R})$
of a colour dipole in the fundamental representation; but the respective quantity $\mcal{S}_g(\bm{R})$ for a gluon-gluon
dipole can be exactly related to  $\mcal{S}(\bm{R})$. In the large-$N_c$ limit of interest, one simply has
$\mcal{S}_g(\bm{R})\simeq [\mcal{S}(\bm{R})]^2$.} \cite{Balitsky:1995ub,Kovchegov:1999yj}.
In what follows, we shall numerically solve this equation for $Y_{\mathbb P} > Y_{0}$, with the 
initial condition at $Y_0$ provided by the MV model. (In practice, one generally chooses a value
 $Y_0\sim 4$, corresponding to $x_0\sim 10^{-2}$.) The effects of the evolution depend only upon the rapidity
 difference $\Delta Y_{\mathbb P}\equiv Y_{\mathbb P} - Y_{0}$ and in what follows we shall present our 
 results for the special value  $\Delta Y_{\mathbb P}=3$ (which should be in the ballpark of the diffractive kinematics
expected at the EIC).
 
 Before we comment on these numerical results, let us consider two special limits which are analytically under control.
Clearly, for very low momenta $K_{\perp} \ll \tilde{Q}_s(x, Y_{\mathbb P})$, Eq.~\eqref{pomugdlow} is not changed by the evolution.
 since this result is fully determined by the black disk limit $ \mcal{T}_g(R,Y_{\mathbb P})=1$. When $K_{\perp} \gg \tilde{Q}_s(x,Y_{\mathbb P})$, we must resort to the amplitude for values of $R$ such that $R \ll 1/Q_s(Y_{\mathbb P})$. For asymptotically high $Y_{\mathbb P}\gg Y_0$, the BK equation leads to the well-known scaling form
\begin{align}
	\label{TBFKL}
	\mcal{T}_g(R,Y_{\mathbb P}) = 
	\big[R^2 Q_s^2(Y_{\mathbb P})\big]^{\gamma} 
	\ln \frac{1}{R^2 Q_s^2(Y_{\mathbb P})}
	\quad  
	\mathrm{for} 
	\quad R \ll 1/Q_s^2(Y_{\mathbb P}),
\end{align}
where $\gamma \simeq 0.63$ for leading logarithmic evolution, while it is slightly modified when including higher order corrections. Inserting the above to Eq.~\eqref{Gscalarnew} we obtain
\begin{align}
	\label{GBFKL}
	\mcal{G}(K_{\perp}, x,Y_{\mathbb P}) =
	c_{\gamma}(x)
	(1-x) 
	\left[\frac{\tilde{Q}_s^2(x,Y_{\mathbb P})}
	{K_{\perp}^2}\right]^\gamma 
	\ln \frac{K_{\perp}^2}
	{\tilde{Q}_s^2(x,Y_{\mathbb P})}
	\quad  
	\mathrm{for} 
	\quad K_{\perp} \gg \tilde{Q}_s(x,Y_{\mathbb P}),
\end{align}
that is, a slightly softer tail in comparison to the MV model one. The function $c_{\gamma}(x)$ varies smoothly between the values $c_{\gamma}(0) = 4^\gamma \Gamma(1+\gamma)/\Gamma(2-\gamma)$ and $c_{\gamma}(1) = 4^\gamma \Gamma(1+\gamma) \Gamma(3+\gamma)/2$ as $x$ changes from zero to one (cf.~the corresponding factor $1+2x$ in Eq.~\eqref{pomugdhigh} in the case of the MV model). In the right panel of Fig.~\ref{fig:dxGP} we see that the high energy evolution does not alter the basic features of the Pomeron UGD which we derived for the MV model. That said, one clearly sees the onset of the anomalous dimension
in the tail at $K_{\perp}$: the decrease with increasing $K_{\perp}$ is indeed slower than in the left figure.

\subsection{The Pomeron gluon distribution} 
\label{sect:pomgd}

As explained towards the end of Sect.~\ref{sec:fact}, it is also interesting to study the 
integrated version of the Pomeron gluon distribution, that is, the function $xG_{\mathbb{P}}(x, x_{\mathbb{P}}, P_{\perp}^2)$
defined in Eq.~\eqref{xGP}.  As before, the most interesting kinematics is such that
 $P_{\perp}^2\sim Q^2 \gg \tilde{Q}_s^2(x, Y_ {\mathbb P})$ and $x_{\mathbb{P}}\ll 1$. Given that the integrand in Eq.~\eqref{xGP} is rapidly falling when $K_{\perp}\gg \tilde{Q}_s$
(like $1/K_{\perp}^4$ within the MV model, cf.~Eq.~\eqref{pomugdhigh}),
the respective integral has only a weak dependence upon its upper cutoff. To get a parametric estimate, we shall therefore let
$P_{\perp}\to\infty$ and use the piecewise approximation for the Pomeron UGD shown in \eqn{Philimits}; we thus find
\begin{align}
	\label{xGPsplit}
	 \int_0^{\infty}\!
	\dif K_{\perp}^2\, 
	\Phi(x,x_{\mathbb P},K_{\perp})
	\sim
	(1-x)
	\left(  
	\int_{0}^{\tilde{Q}_s^2}
	\dif K_{\perp}^2 +
	\int_{\tilde{Q}_s^2}^{\infty}
	\dif K_{\perp}^2\,
	\frac{\tilde{Q}_s^4}{K_{\perp}^4}
	\right)
	\sim\, (1-x)\tilde{Q}_s^2(x,Y_ {\mathbb P}),
\end{align}
where the final result could have been anticipated on dimensional arguments.
The first integral in Eq.~\eqref{xGPsplit} is dominated by its upper limit, the second integral by its lower limit,
and they both yield contributions of the same order of magnitude. We thus conclude that the integral is truly 
controlled by momenta $K_{\perp} \sim \tilde{Q}_s$. 
This conclusion is not restricted to the MV model, for example it is easy to check that it also holds when using the result of the BK equation, as shown  in Eq.~\eqref{GBFKL}.
More generally, it does not depend on the fine details of the scattering amplitude $\mcal{T}_g(R,Y_{\mathbb P})$,
so long as it satisfies general properties like colour transparency with ``anomalous dimension'' $1-\gamma < 0.5$
(cf. Eq.~\eqref{GBFKL}) and unitarity.
This is worth emphasising, since it justifies our use of the $gg$ dipole picture whenever the jets are hard enough compared to the saturation momentum, even after the dijet imbalance has been integrated over. 

\eqn{xGPsplit} already encodes the dominant $x$--dependence of the Pomeron gluon distribution. On
top of that, we expect some
additional dependence which is comparatively weak, as demonstrated e.g. by the factor $1+2x$ in 
Eq.~\eqref{pomugdhigh}, and also ``model--dependent'', in the sense that it depends upon our model, or
approximations, for the gluon dipole amplitude $\mcal{T}_g(R,Y_{\mathbb P})$.
To allow for that, we shall write
\begin{align}
	\label{xGPhigh}
	xG_{\mathbb{P}}(x, x_{\mathbb{P}}, P_{\perp}^2\to \infty)
	=
	\frac{S_\perp (N_c^2-1)}{4\pi^3}\,
	\kappa(x)
	(1-x)\, \tilde{Q}_s^2(x,Y_ {\mathbb P}),
\end{align} 
with $\kappa(x)$ a slowly varying function, to be later specified in special cases\footnote{Strictly speaking,
$\kappa(x)$ is a function of $x$ alone only in the GBW model, which involves a single intrinsic momentum scale, namely
$Q_s$. In more general cases, like the MV model, it will also depend upon dimensionless ratios of
the available momentum scales, e.g. $Q_s^2/\Lambda^2$. Besides, adding the BK evolution
will make this function (slowly) dependent upon the rapidity $Y_ {\mathbb P}$. That said, all such 
dependences are indeed weak: when $P_{\perp}^2 \gg \tilde{Q}_s^2(x,Y_ {\mathbb P})$, the dominant 
functional dependences of the
Pomeron gluon distribution are those encoded in the product $(1-x)^2{Q}_s^2(Y_ {\mathbb P})$.}.
Recalling \eqn{Qstilde}, one sees that the r.h.s. vanishes like $(1-x)^2$ when $x\to 1$, as first observed in
\cite{Iancu:2021rup} and independently confirmed in \cite{Hatta:2022lzj}. This power-like behaviour is reminiscent
of the constituent quark model prediction for the gluon distribution of a pion \cite{Mueller:1981sg}: in that context, 
the power 2 is simply the number
of constituent quarks (it would be 3 for a proton). This is known as the ``quark counting rule'',
but what truly matters is the {\it number} of partonic constituents, and not their spins or other quantum numbers. 
In the present case, one can think of the Pomeron as being built with 2 constituent gluons.

Eq.~\eqref{xGPhigh} shows that, at least under the current assumptions, the Pomeron gluon distribution 
 is controlled by the physics of saturation, including in the hard DIS regime at $Q^2\gg Q_s^2(Y_{\mathbb P})$.
 This should be contrasted to the inclusive dijet production, where the sensitivity to saturation is essentially lost after 
 integrating over the dijet imbalance $K_{\perp}$. Indeed, the corresponding TMD (the WW UGD of the hadronic target) 
 has a slowly-decreasing tail $\propto 1/K_{\perp}^2$ at large momenta, cf.  \eqn{PhiWW}.
  Accordingly, the result of the respective integral over $K_{\perp}$ ---
   the standard gluon distribution  $xG(x, P_{\perp}^2)$ --- is
   sensitive to all the intermediate momenta $Q_s^2\ll K_{\perp}^2 \ll P_\perp^2$ and logarithmically increasing
 with $P_{\perp}^2$.

Returning to the case of the diffractive distribution $xG_{\mathbb{P}}(x, x_{\mathbb{P}}, P_{\perp}^2)$,
 it is also possible to estimate its mild dependence upon the resolution scale $P_{\perp}^2$.
 By using \eqn{xGPhigh}, one can write
\begin{align}
	\label{xGPfinite}
	\pi \int_0^{P_{\perp}^2}\!
	\dif K_{\perp}^2\,
	[\mcal{G}(x,x_{\mathbb P},K_{\perp})]^2 = 
	2 \pi \kappa(x) (1-x)^2 \tilde{Q}_s^2(x,Y_{\mathbb P}) - 
	\pi \int_{P_{\perp}^2}^{\infty}\!
	\dif K_{\perp}^2\, 
	[\mcal{G}(x,x_{\mathbb P},K_{\perp})]^2.
\end{align}
Given the asymptotic behavior of $\mcal{G}(x,x_{\mathbb P},K_{\perp})$ (see e.g. Eqs.~\eqref{pomugdhigh} and \eqref{GBFKL}),
 we expect the second term to be of the order of $(1-x)^2\tilde{Q}_s^2 (\tilde{Q}_s^2/P_{\perp}^2)^{2\gamma-1}$, 
 with $\gamma$ equal to one for the GBW and MV models, but smaller than 1 for the solution \eqref{GBFKL} to the 
 BK equation.
 
\begin{figure}[t] 
\centerline{\includegraphics[width=0.8\columnwidth]{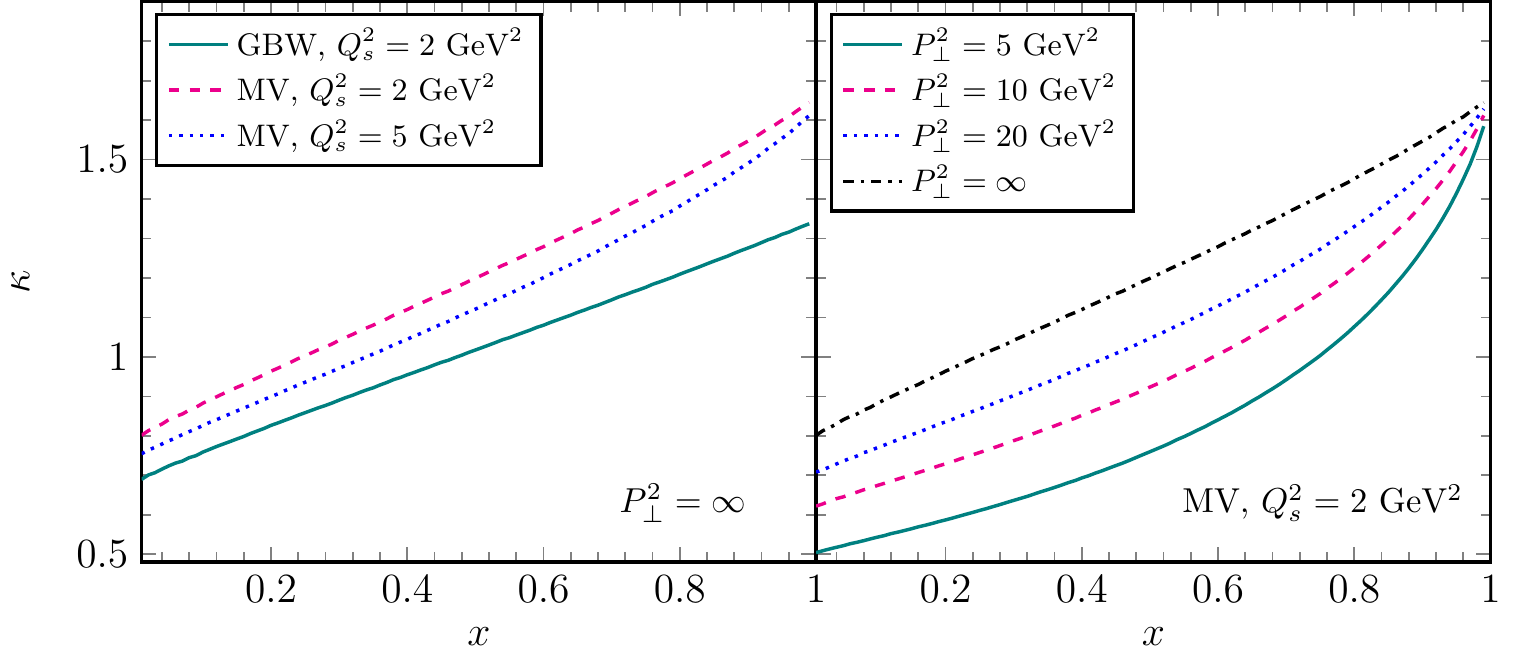}}
\caption{\small Left panel: The smooth function $\kappa(x)$ appearing in the integrated Pomeron gluon distribution in Eq.~\eqref{xGPhigh} for the GBW model (for which it is independent of $Q_s$) and for the MV model for two different saturation scales. Right panel: The reduced Pomeron gluon distribution of Eq.~\eqref{xGkappa} for various values of $P_{\perp}^2$ (with $P_{\perp}^2>Q_s^2$) as a function of $x$ in the MV model with $Q_s^2=2$ GeV$^2$. }
\label{fig:kappx}
\end{figure}

 In order to gain more intuition, let us study the limits $x \ll 1$ and $1-x \ll 1$ by using 
 the GBW model, $\mcal{T}_g(R) = 1 - e^{-R^2 Q_s^2/4}$, which allows for analytic results.
When $x\ll 1$, the respective estimate for the Pomeron UGD has been given in Eq.~\eqref{pomugdsat0}. The integration over the transverse momentum can be done in terms of the exponential integral function,  and we obtain 
\begin{align}
	\label{xGPGBW0}
	\hspace*{-0.6cm}
	\pi \int_0^{P_{\perp}^2}\!
	\dif K_{\perp}^2\,
	[\mcal{G}(x,x_{\mathbb P},K_{\perp})]^2 &\,=
	\pi Q_s^2
	\left[
	2 \ln 2
	- \frac{Q_s^2}{P_{\perp}^2}
	\left( 
	1 - e^{-P_{\perp}^2/Q_s^2} 
	\right)^2 +
	2 E_1\!\left(\frac{2 P_{\perp}^2}{Q_s^2}\right) -
	2 E_1\!\left(\frac{P_{\perp}^2}{Q_s^2}\right)
	\right]\nonumber\\
	&\,\simeq \pi Q_s^2
	\left[
	2 \ln 2
	- \frac{Q_s^2}{P_{\perp}^2}\right],
\end{align}
where the expression in the first line is valid for any $P_{\perp}$, while the second line holds at high $P_{\perp}\gg \tilde{Q}_s(x).$
 Similarly, when $1-x\ll 1$, one can make use of Eq.~\eqref{pomugdsat1}
to deduce\footnote{The (indefinite) integrals involving a single $E_1$ are straightforward to be done by integration by parts. The integral involving the square of $E_1$ can be reduced to the definite integral $\int_0^{\infty} \dif k\, e^{2k} E_1^2(k) = \pi^2/4$ given in 8.219.2 in \cite{GR7}.} 
\begin{align}
	\label{xGPGBW1}
	\hspace*{-0.6cm}
	\pi \int_0^{P_{\perp}^2}\!
	\dif K_{\perp}^2\,
	[\mcal{G}(x,x_{\mathbb P},K_{\perp})]^2 \simeq
	\pi (1-x)^2\tilde{Q}_s^2(x)
	\left[
	\frac{45 \pi^2 - 272}{64}
	- \frac{9 \tilde{Q}_s^2(x)}{P_{\perp}^2}
	\right]
	\quad
	\mathrm{for}
	\quad
	P_{\perp}\gg \tilde{Q}_s(x).
\end{align}
In particular, by taking the limit $P_{\perp}^2 \to \infty$ of these results, we can infer the behaviour of the unknown function
$\kappa(x)$ for limiting values of its argument (and for the GBW model, of course). One thus finds that, with increasing $x$,
$\kappa(x)$ increases from $\kappa(0) \simeq 0.693$ to $\kappa(1) \simeq 1.345$, i.e.~roughly by a factor of 2. 
These numbers were first obtained in \cite{Buchmuller:1998jv} and
 are consistent with the recent analysis in \cite{Hatta:2022lzj}.
They confirm that $\kappa(x)$ is a slowly increasing function, in qualitative agreement with Fig.~\ref{fig:dxGP} 
(consider  the area under the various curves in that figure). This is further demonstrated by the curves in the
 left panel in Fig.~\ref{fig:kappx}, which show the numerical results for $\kappa(x)$ obtained with the GBW model
 and with 2 versions of the MV model, which differ in the value of the saturation momentum $Q_s$.
 
For the plotting purposes, at least, it is convenient to introduce the  dimensionless quantity
$\kappa\big(x, Y_{\mathbb P}, P_{\perp}^2/ \tilde{Q}_s^2(x,Y_{\mathbb P})\big)$ via the straightforward generalisation 
of \eqn{xGPhigh}, that is
\begin{align}
	\label{xGkappa}
	xG_{\mathbb{P}}(x, x_{\mathbb{P}}, P_{\perp}^2)
	=
	\frac{S_\perp (N_c^2-1)}{4\pi^3}\,
	\kappa\left(x, Y_{\mathbb P}, P_{\perp}^2/ \tilde{Q}_s^2(x,Y_{\mathbb P})\right)
	(1-x)\, \tilde{Q}_s^2(x,Y_ {\mathbb P}).
	\end{align} 
This can be viewed as reduced gluon distribution, which is local in impact parameter
and encodes the subdominant functional dependences upon $x$, $Y_{\mathbb P}$, and $P_\perp^2$
(in the interesting regime at large  $P_\perp^2\gg \tilde{Q}_s^2(x,Y_ {\mathbb P})$).
In the right panel in Fig.~\ref{fig:kappx}, we illustrate (for the case of the MV model) the approach of 
this function to its asymptotic value at $P_{\perp}^2\to\infty$
with increasing $P_{\perp}^2$.

\begin{figure}[t] 
\centerline{\includegraphics[width=0.8\columnwidth]{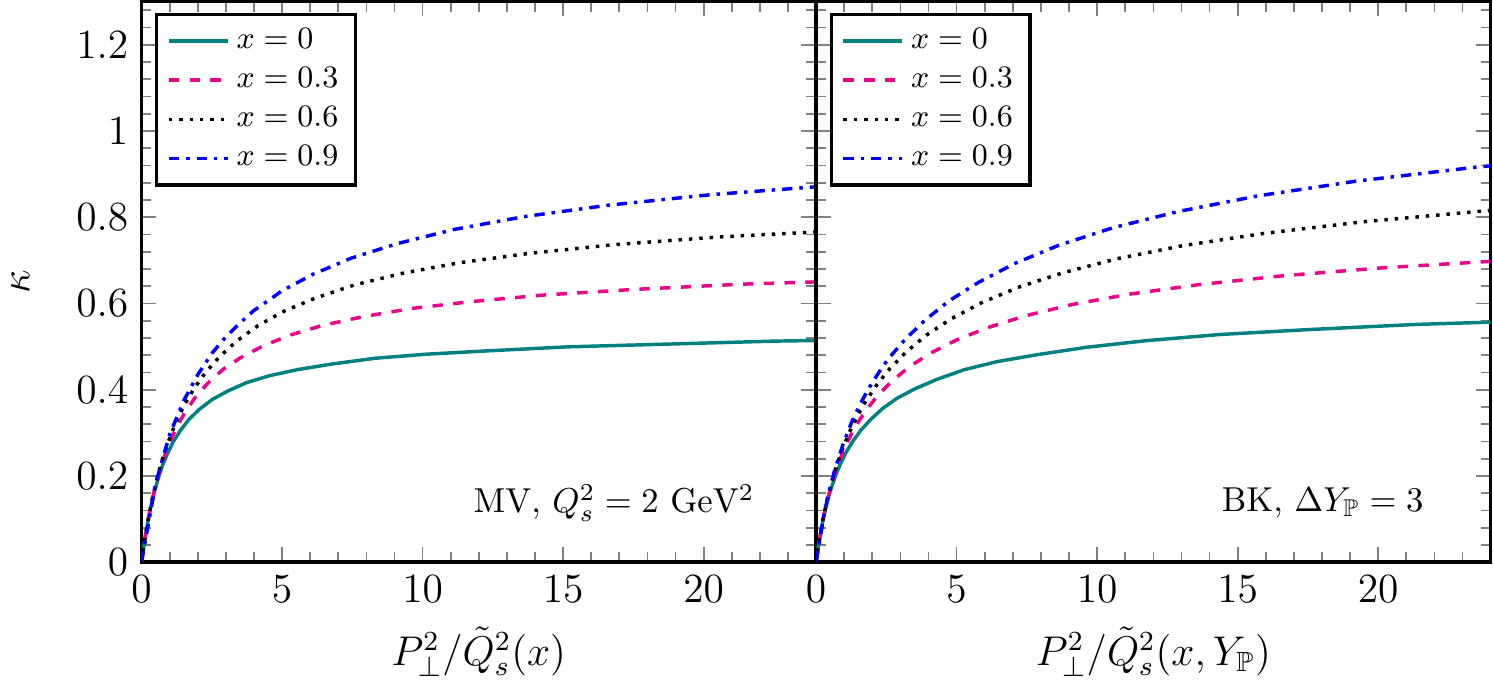}}
\caption{\small The reduced Pomeron gluon distribution $\kappa\big(x, Y_{\mathbb P}, P_{\perp}^2/ \tilde{Q}_s^2(x,Y_{\mathbb P})\big)$,
as introduced in \eqn{xGkappa}, is plotted as a function of the rescaled momentum $P_{\perp}^2/ \tilde{Q}_s^2(x,Y_{\mathbb P}) $
for different values of $x$ and 2 values of $\Delta Y_{\mathbb P}\equiv Y_{\mathbb P}-Y_0$. 
Left panel: The tree-level result at $\Delta Y_{\mathbb P}=0$, where  $\mathcal{T}_{g}({R})$ is given by the MV model. 
 	Right panel:  The result of the BK evolution up to $\Delta Y_{\mathbb P}=3$.
	$\mathcal{T}_{g}({R,Y_{\mathbb P}})$ is obtained from the solution to the BK equation with the initial condition at 
	$\Delta Y_{\mathbb P}=0$ given by the MV model.}
\label{fig:kappa}
\end{figure}

\begin{figure}[t] 
\centerline{\includegraphics[width=0.8\columnwidth]{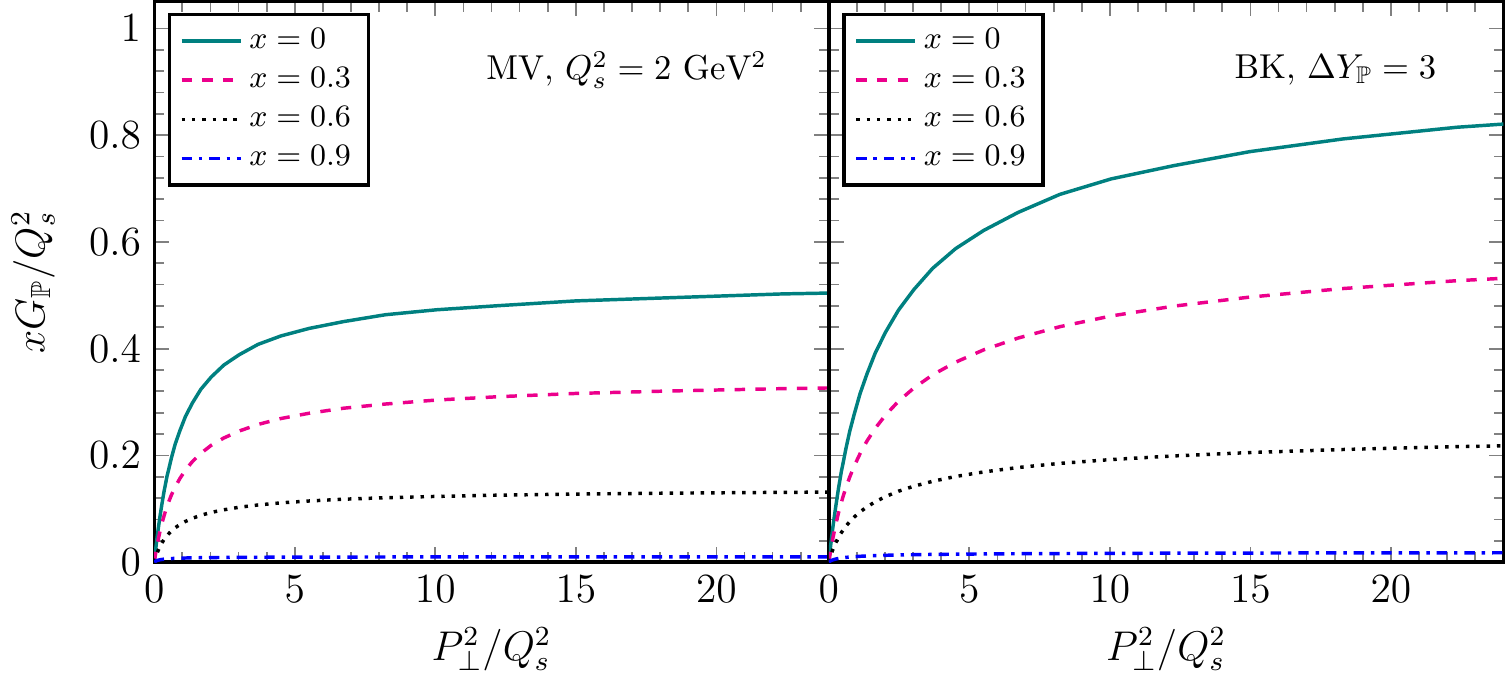}}
\caption{\small The  Pomeron gluon distribution $xG_{\mathbb{P}}(x, x_{\mathbb{P}}, P_{\perp}^2)$
plotted as a function of  $P_{\perp}^2/ {Q}_s^2$, where $Q_s^2=2$~GeV$^2$ is the
saturation scale in the MV model. We more precisely plot the dimensionless quantity obtained by dividing 
 $xG_{\mathbb{P}}(x, x_{\mathbb{P}}, P_{\perp}^2)$ with $Q_s^2$ and by omitting a factor $S_{\perp}(N_c^2 -1)/4\pi^3$.
Left panel:  the tree-level result at $\Delta Y_{\mathbb P}=0$, 
as computed with the MV model.  Right panel:  The result of the BK evolution up to $\Delta Y_{\mathbb P}=3$.}
\label{fig:xGP}
\end{figure}

To study the effects of the high-energy evolution with increasing the rapidity gap $Y_{\mathbb P}$, we compare
in Fig.~\ref{fig:kappa} the results for the reduced gluon distribution as obtained in the MV model 
(the left panel) and after including the effects of the BK evolution up to $\Delta Y_{\mathbb P}=3$ (the right panel).
This comparison shows the expected features: \texttt{(i)} the flattening of the distribution with increasing  $P_{\perp}$  
occurs at a scale of the order of $ \tilde{Q}_s(x,Y_{\mathbb P}) $, for any $x$ and $Y_{\mathbb P}$; \texttt{(ii)} the
 effects of the high-energy evolution on this reduced distribution are quite small.
 
To  visualise 
 the main effect of the BK evolution --- namely, the increase of the saturation momentum with $Y_{\mathbb P}$ ---,
 it is convenient to also plot the Pomeron gluon distribution $xG_{\mathbb{P}}(x, x_{\mathbb{P}}, P_{\perp}^2)$ {\it per se},
 instead of the reduced distribution. Since proportional to $(1-x)^2{Q}_s^2(Y_{\mathbb P})$, cf. \eqn{xGkappa},
 this quantity should rapidly increase with $Y_{\mathbb P}$ and decrease with $1-x$.
 This is indeed confirmed by the numerical results in Fig.~\ref{fig:xGP}, that should be compared to those for the
 reduced distribution in  Fig.~\ref{fig:kappa}.

\subsection{Adding DGLAP evolution}
\label{sect:dglap}

 So far,  the Pomeron gluon distribution $xG_{\mathbb{P}}(x, x_{\mathbb{P}}, P_{\perp}^2)$ has been computed under the
 assumption that the gluon emitted by the Pomeron with a typical transverse momentum $ K_\perp\sim Q_s(Y_{\mathbb{P}})$
  is directly absorbed by the $q\bar q$ pair, without further evolution.
 However, when $P_{\perp}^2\gg Q_s^2(Y_{\mathbb{P}})$, the $q\bar q$ pair can resolve the partonic substructure
 of the exchanged gluon, as generated via the DGLAP evolution --- that is, via subsequent partonic emissions which are
 strongly ordered in transverse momenta, within the range $Q_s^2(Y_{\mathbb{P}}) < k_\perp^2 < P_{\perp}^2$. 
We consider only gluon emissions, for simplicity (dynamical quarks can be trivially added, if needed) 
and we omit the factor $S_{\perp} (N_c^2-1)/ 4\pi^3$ appearing in front of the integral in Eq.~\eqref{xGP} ---
this should be reinserted in the final results as a multiplicative constant. 
The tree-level contribution\footnote{Note that in this context, the ``tree-level approximation'' also includes the
effects of the high-energy evolution internal to the Pomeron --- that is, the
BK evolution of the gluon dipole amplitude.}
 acts as a source term, so the relevant version of the DGLAP equation reads
\begin{align}
	\label{DGLAP}
	\frac{\rmd 
	xG_{\mathbb{P}}(x, x_{\mathbb{P}}, K_{\perp}^2)}
	{\rmd \ln K_{\perp}^2} = 
	\pi K_{\perp}^2 
	\Phi_{\mathbb{P}}(x,x_{\mathbb{P}}, K_{\perp}^2)\,+
	\frac{\alpha_s (K_{\perp}^2)}{2\pi} 
	\int_x^1 \rmd z\,
	P_{gg} (z) \,
	\frac{x}{z}\,
	G_{\mathbb{P}}
	\left(\frac{x}{z}, x_{\mathbb{P}}, K_{\perp}^2
	\right)\,,
\end{align}
with the splitting function $P_{gg} (z)$
\begin{align}
	\label{Pgg}
	P_{gg} (z) = 2N_c 
	\left[
	\frac{z}{(1-z)_+} + 
	\frac{1-z}{z} + z (1-z) \right] + 
	\frac{11N_c-2N_f}{6}\,
	\delta (1-z),
\end{align}
which includes contributions from both real and virtual emissions.
We recall the definition of the plus prescription
\begin{align}
	\label{plus}
	\int_x^1 \dif z\, 
	\frac{f(z)}{(1-z)_+} = 
	\int_x^1 \dif z\,
	\frac{f(z)-f(1)}{1-z}
	+ f(1) \ln(1-x).
\end{align}
Eq.~\eqref{DGLAP} includes the QCD running coupling, assumed to be evaluated at one loop order:
\begin{align}
	\label{1LRC}
  	\alpha_s(K_{\perp}^2)=
  	\frac{1}{b
  	\ln({K_{\perp}^2}/{\Lambda^2})}
	\quad \mathrm{with} \quad
	b = \frac{11 N_c - 2  N_f}{12 \pi}.
\end{align}
We expect the solution to the DGLAP equation \eqref{DGLAP} to exhibit a very different behavior in the two limiting cases $x \to 0$ and $x \to 1$, so we shall separately discuss these two cases.

When $x\ll 1$, the dominant contribution to the evolution can be obtained by
 keeping only the singular part of the splitting function, proportional to $1/z$. With $P_{\perp}^2 \gg Q_s^2$, we iterate the tree level Pomeron gluon distribution --- which with the present normalisation reads $\kappa(0) Q_s^2(Y_{\mathbb P})$ ---, to obtain 
\begin{align}
	\label{DGLAPx0}
	xG_{\mathbb{P}}(x, x_{\mathbb{P}}, P_{\perp}^2)
	=\kappa(0)
	Q_s^2(Y_{\mathbb P})\,
	I_0
	\left(\sqrt{\frac{4 N_c}{\pi b}
	\ln \frac{\alpha_s(Q_0^2)}
	{\alpha_s(P^2_{\perp})}
	\ln\frac{1}{x}}
	\right)
	\quad 
	\mathrm{for} 
	\quad x \ll 1,
\end{align}
where $Q_0^2$ is the scale at which we start the evolution. Eq.~\eqref{DGLAPx0} has the typical form of a double logarithmic evolution:
$xG_{\mathbb{P}}$ increases with both $P_{\perp}^2$ and $1/x$ since it is dominated by the real gluons emissions. 

When $1-x$ is very small, the second term in Eq.~\eqref{plus}, which originates from virtual graphs, 
becomes the dominant one due to the presence of the \emph{negative} logarithmic factor $\ln(1-x)$. Since this term
is independent of $z$, the longitudinal dependence in the DGLAP equation becomes trivial. 
Again it is straightforward to iterate the corresponding tree level distribution
$\kappa(1) Q_s^2(Y_{\mathbb P}) (1-x)^2$, to get
\begin{align}
	\label{DGLAPx1}
	xG_{\mathbb{P}}(x, x_{\mathbb{P}}, P_{\perp}^2)
	=\kappa(1)
	Q_s^2(Y_{\mathbb P})
	(1-x)^{\textstyle 2 + 
	\frac{N_c}{\pi b} 
	\ln \frac{\alpha_s(Q_0^2)}
	{\alpha_s(P^2_{\perp})}}
	\quad 
	\mathrm{for} 
	\quad  1-x \ll 1.
\end{align}
This result shows an increased suppression when $x\to 1$ as compared to the tree-level result. It further shows  that,
when $x$ is close to 1, $xG_{\mathbb{P}}$ decreases with increasing $P_{\perp}^2$ (unlike for $x \ll 1$), a property which was of course expected since the evolution for $1-x \ll 1$ is determined by the virtual terms only.  Altogether, this is physically reasonable: 
via successive parton branchings, the gluons disappear at large $x$, but they accumulate at small $x$.

This argument also suggests that there should be some intermediate regime in $x$, 
 in which real and virtual contributions approximately cancel. In such a regime the Pomeron gluon 
 distribution is not altered by the DGLAP evolution, and thus it is equal to the tree level result to a good accuracy.
 The numerical results confirm that such a ``neutral'' regime exists indeed, around $x \simeq 0.1$.

Before going on to present the numerical solution, which will exhibit the features discussed above, let us mention
 that in the limit $x \to 1$ there are additional double logarithmic terms. 
 They have not appeared in our calculation so far, because they are not generated by the solution to the DGLAP equation
--- rather,  they are part of the coefficient function in the operator product expansion.  
These terms are well known in the literature (see e.g. \cite{Mueller:1981sg}). For completeness,
they will be computed in App.~\ref{sect:x1}, for the simpler case of a fixed coupling. This leads to the result in Eq.~\eqref{xgpall},
whose generalisation to a running coupling is pretty straightforward and reads (compare to Eq.~\eqref{DGLAPx1})
\begin{align}
	\label{DGLAPpDL}
	xG_{\mathbb{P}}(x, x_{\mathbb{P}}, P_{\perp}^2)
	=& \,\kappa(1)
	Q_s^2(Y_{\mathbb P})\,
	(1-x)^2
	\nn 
	& \times e^{\textstyle -\frac{N_c}{\pi b}
	\left(\ln\frac{P_{\perp}^2}{\Lambda^2} 
	\ln\ln\frac{P_{\perp}^2}{\Lambda^2}
	- \ln\frac{P_{\perp}^2(1-x)}{\Lambda^2}
	\ln\ln\frac{P_{\perp}^2(1-x)}{\Lambda^2}
	-\ln\frac{1}{1-x} \ln \ln \frac{1}{1-x}
	\right)}
\end{align}
where for simplicity we have neglected terms involving $\ln Q_0^2/\Lambda^2$.

\begin{figure}[t] 
\centerline{\includegraphics[width=0.8\columnwidth]{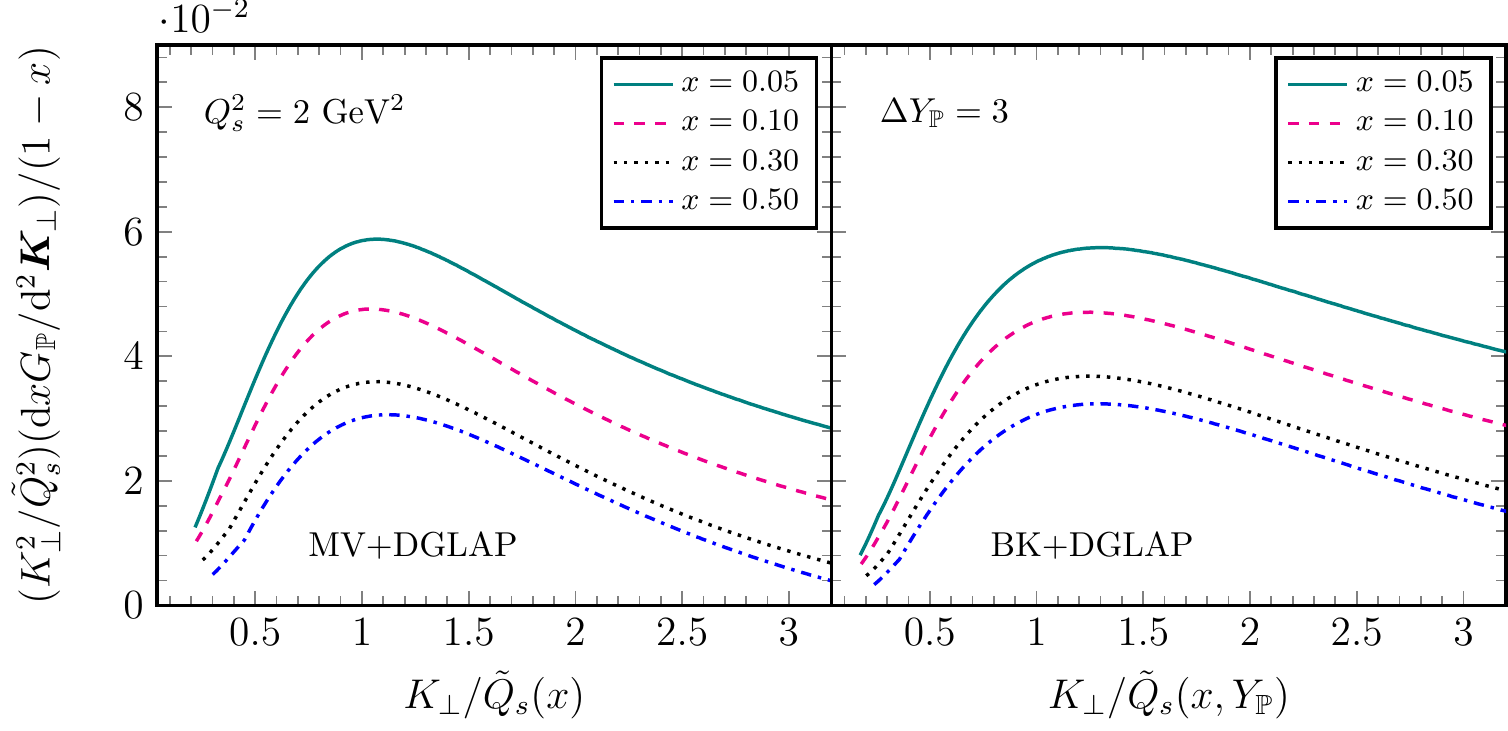}}
\caption{\small The derivative of the Pomeron gluon distribution $\dif xG_{\mathbb{P}}/\dif^2\bK$, as determined from the solution to the DGLAP equation and scaled with $[K_{\perp}/\tilde{Q}_s(x,Y_{\mathbb P})]^2/(1-x)$, is plotted as a function of the rescaled momentum $K_{\perp}/ \tilde{Q}_s(x,Y_{\mathbb P})$ for various values of $x$ and $Y_\mathbb{P}$. A factor $S_{\perp}(N_c^2 -1)/4\pi^3$  has been omitted. Left panel: The source term in the DGLAP equation is evaluated with the MV model. Right panel: The source also includes the BK evolution of the gluon dipole amplitude.}
\label{fig:K2dxGP-DGLAP}
\end{figure}

To suggestively visualise the effects of the DGLAP evolution, it is preferable to plot the derivative 
${\rmd 	xG_{\mathbb{P}}(x, x_{\mathbb{P}}, K_{\perp}^2)}/	{\rmd K_{\perp}^2}$ of the Pomeron gluon distribution,
which gives us an indication about the gluon distribution in the transverse momentum $ K_{\perp}$. At tree-level at least, 
this derivative yields the Pomeron UGD \eqref{pomugddef}, or the gluon occupation number \eqref{occup}. 
Similar to the case in Fig.~\ref{fig:dxGP}, we multiply the result by a factor $(K_{\perp}/\tilde{Q}_s(x))^2/(1-x)$.
We show our results in Fig.~\ref{fig:K2dxGP-DGLAP}, both at tree level (left panel) and after adding the DGLAP evolution (right panel).
(The left plot is of course very similar to those in Fig.~\ref{fig:dxGP}.) By comparing these 2 figures, one can notice that
the approximate scaling that is visible at tree-level is in fact washed out by the DGLAP evolution.
Perhaps this should not come as surprise given the different ways how the DGLAP evolution acts at large and, respectively, 
small values of $x$, as previous discussed. In particular, the effects of this evolution appear to be negligible around
$x \sim 0.1$ --- in agreement with the previous discussion.

Remarkably, the plots in Fig.~\ref{fig:K2dxGP-DGLAP} show that the peak in the UGD
at  $K_{\perp} \sim \tilde{Q}_s(x)$ is preserved by the DGLAP evolution. That is, the solution to the DGLAP 
equation shows a  high-momentum tail which decreases very fast at $K_{\perp} > \tilde{Q}_s(x)$ --- 
faster than the standard tail $\sim 1/K_\perp^2$
that would be generated by solving this equation with an initial condition at $Q^2=Q^2_0$.
Clearly, this behaviour is introduced by the source term in the r.h.s. of \eqn{DGLAP}. 
This is important since it means that the prominence of the saturation physics for this diffractive
problem is  in fact preserved by the DGLAP evolution: the bulk of the gluon distribution remains at transverse momenta
of order $\tilde{Q}_s(x)$.

\begin{figure}[t] 
\centerline{\includegraphics[width=0.8\columnwidth]{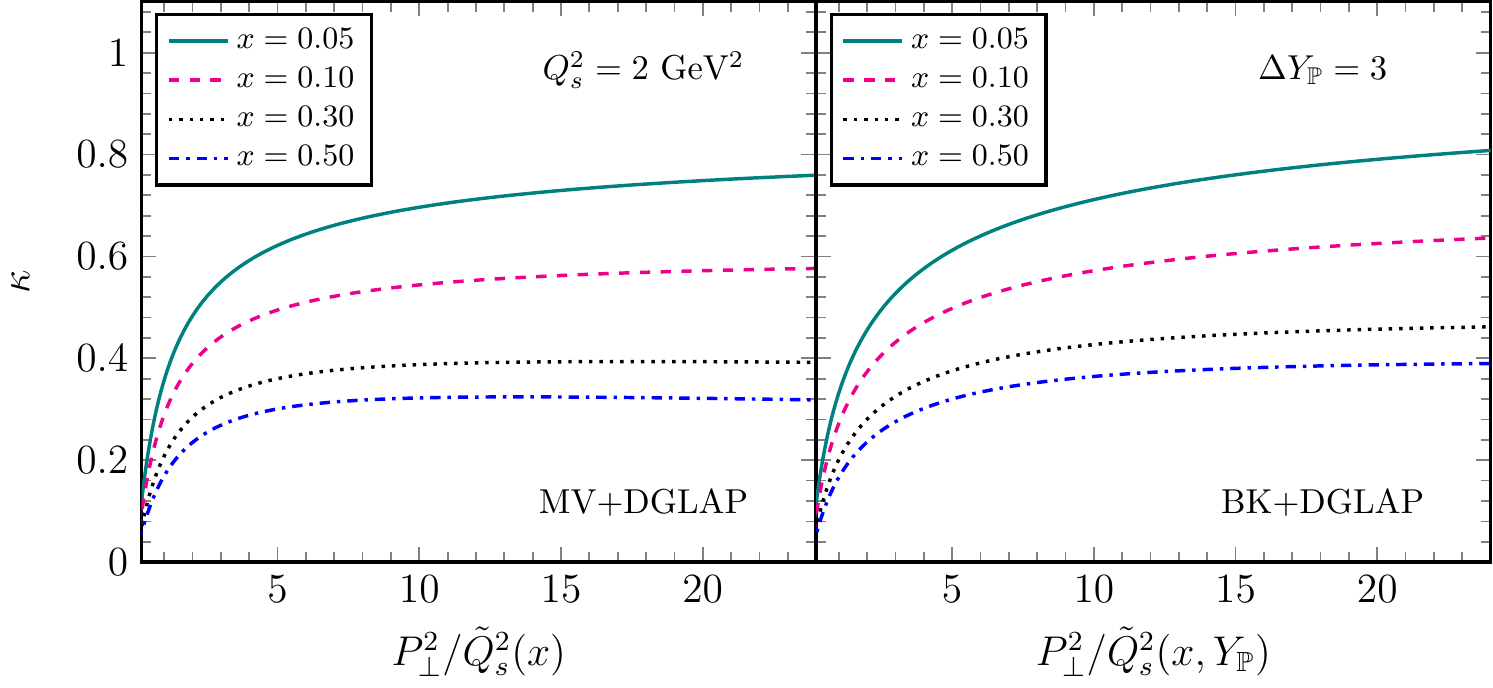}}
\caption{\small The reduced Pomeron gluon distribution $\kappa\big(x, Y_{\mathbb P}, P_{\perp}^2/ \tilde{Q}_s^2(x,Y_{\mathbb P})\big)$, defined in Eq.~\eqref{xGkappa} and obtained from the solution to the DGLAP equation, is plotted as a function of the rescaled momentum $P_{\perp}^2/ \tilde{Q}_s^2(x,Y_{\mathbb P})$ for various values of $x$ and $Y_\mathbb{P}$. Left panel: The source term in the DGLAP equation is evaluated with the MV model. Right panel: The source also includes the BK evolution of the gluon dipole amplitude.}
\label{fig:xGP-DGLAP}
\end{figure}

In Fig.~\ref{fig:xGP-DGLAP} we  show the corresponding results for the (integrated) Pomeron gluon distribution 
$xG_{\mathbb{P}}(x, x_{\mathbb{P}}, P_{\perp}^2)$. This is seen to flatten out above $P_{\perp}^2\sim 4\tilde{Q}_s^2$,
in the same as in the absence of the DGLAP evolution (compare to Fig.~\ref{fig:kappa}) --- in agreement with the previous
discussion of Fig.~\ref{fig:K2dxGP-DGLAP}.

\begin{figure}[t] 
\centerline{\includegraphics[width=0.8\columnwidth]{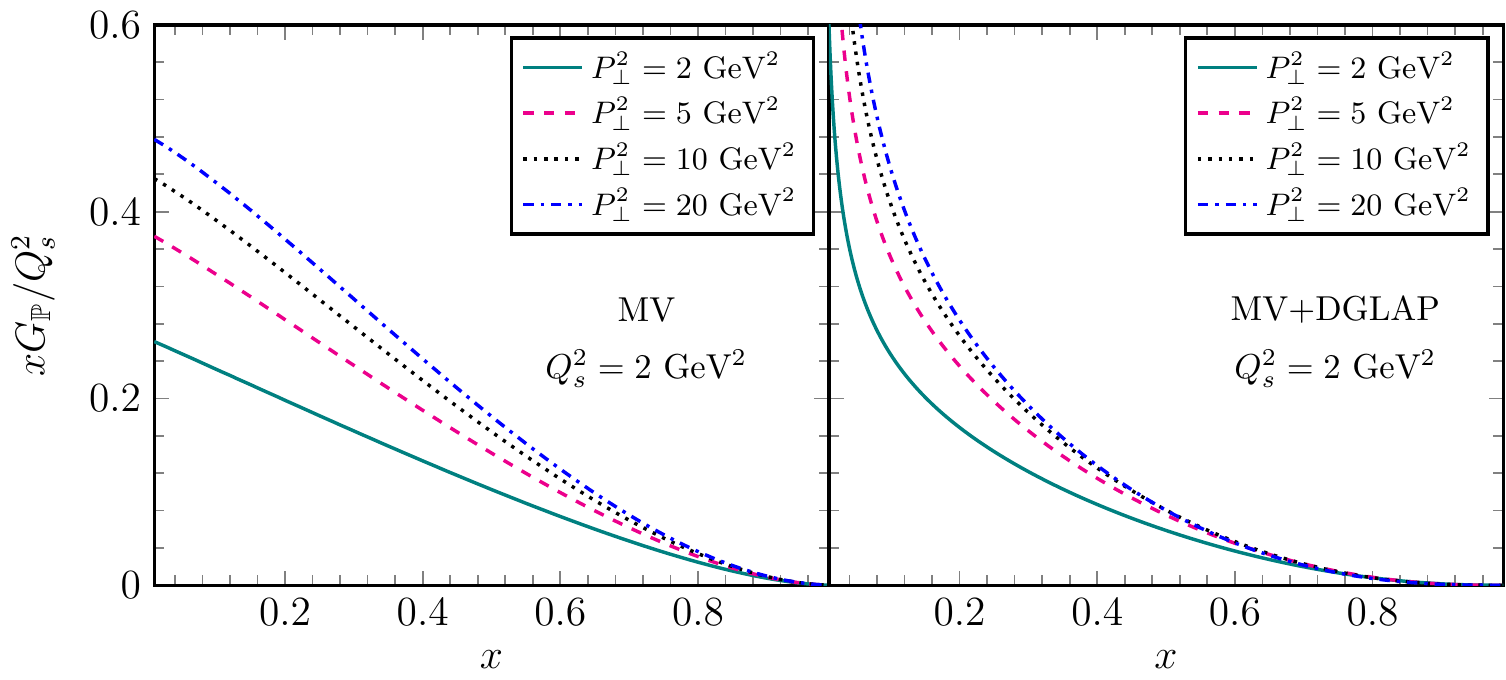}}
\caption{\small A dimensionless Pomeron gluon distribution obtained by dividing with $Q_s^2$ and by omitting a factor $S_{\perp}(N_c^2 -1)/4\pi^3$. Left panel: ``Tree-level'' result (with $\mcal{T}_g(R)$ given by the MV model) as a function of $x$, for various values of $P_{\perp}^2$. Right panel: The same as obtained from the solution to the DGLAP equation \eqref{DGLAP} with the source term evaluated with the MV model.}
\label{fig:xGP-DGLAP-x}
\end{figure}

\begin{figure}[t] 
\centerline{\includegraphics[width=0.8\columnwidth]{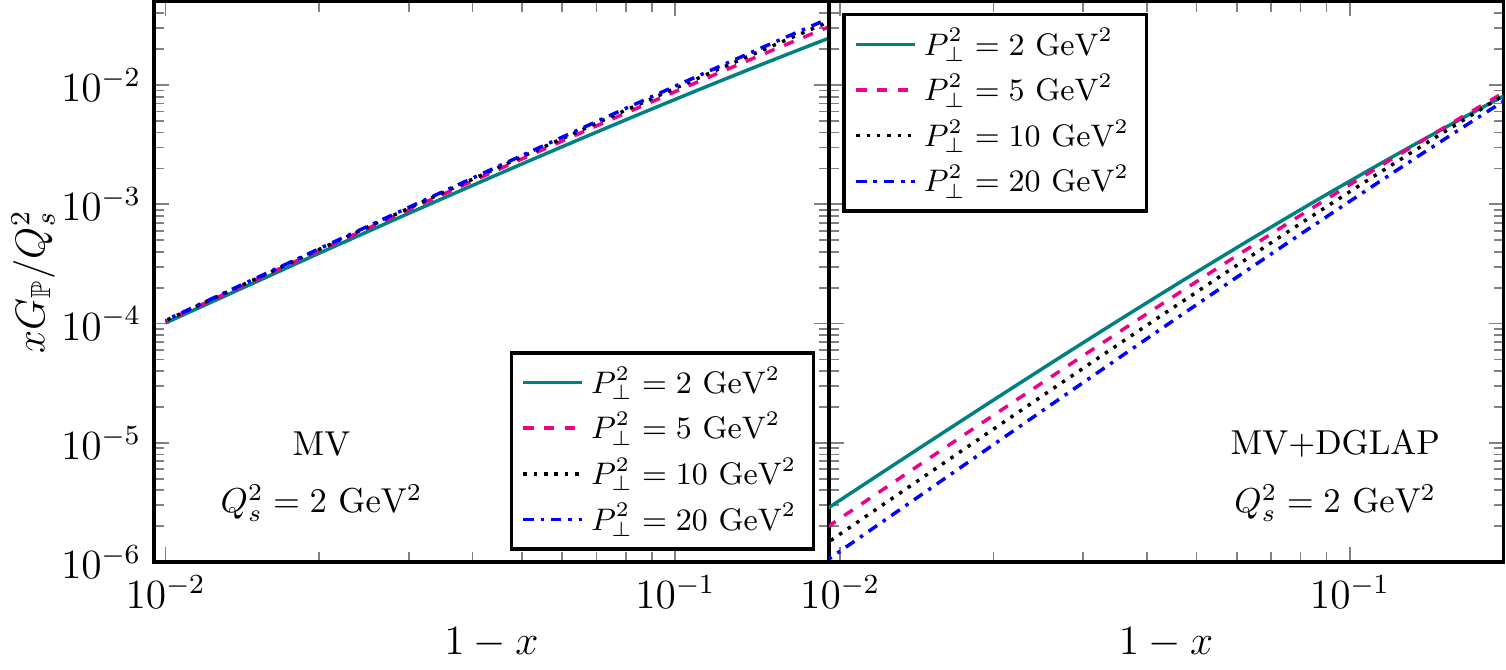}}
\caption{\small Logarithmic plot of a dimensionless Pomeron gluon distribution obtained by dividing with $Q_s^2$ and by omitting a factor $S_{\perp}(N_c^2 -1)/4\pi^3$ when $x \to 1$. Left panel: Tree-level result (with $\mcal{T}_g(R)$ given by the MV model) as a function of $x$, for various values of $P_{\perp}^2$. Right panel: The same as obtained from the solution to the DGLAP equation \eqref{DGLAP} with the source term evaluated with the MV model.}
\label{fig:xGP-DGLAP-x-loglog}
\end{figure}

In Figs.~\ref{fig:xGP-DGLAP-x} and \ref{fig:xGP-DGLAP-x-loglog} we study the $x$--dependence of the 
function $xG_{\mathbb{P}}(x, x_{\mathbb{P}}, P_{\perp}^2)$, for various values of $P_{\perp}^2$.
Fig.~\ref{fig:xGP-DGLAP-x} shows the whole range in $x$, while Fig.~\ref{fig:xGP-DGLAP-x-loglog} focuses
the end-point behaviour at $x \to 1$. Here, we compare tree-level results (left plots) to those including the DGLAP evolution
(right plots). In particular, the right plot in Fig.~\ref{fig:xGP-DGLAP-x} exhibits the rapid increase at small $x$ that is naturally
generated by the perturbative QCD evolution. Furthermore, the results  in Fig.~\ref{fig:xGP-DGLAP-x-loglog} 
confirm the fact that $xG_{\mathbb{P}}(x, x_{\mathbb{P}}, P_{\perp}^2)$ vanishes like $(1-x)^{a(P_{\perp}^2)}$. 
While at tree level we have $a=2$ (as it can be checked in the left plot), the DGLAP evolution produces a further suppression, 
since gluons at large-$x$ get depleted by radiating softer ones. This additional suppression is indeed visible
in the curves from the right plot.

\section{The gluon contribution to the diffractive structure function}
\label{sec:qqgFD}

Given the differential cross-section \eqref{3jetsD1} for diffractive 2+1 jet production, one can deduce the
$q\bar qg$ contribution to the diffractive structure function by integrating out the kinematics of the 3 final jets.
A priori, this is a higher-order correction: it is suppressed by a power of $\alpha_s$ compared to the respective
leading-order contribution, as given by the (strong) elastic scattering of the $q\bar q$ pair alone.
Yet, the $q\bar q$  piece is concentrated at large\footnote{Notice that, for the exclusive $q\bar q$ production,
the variable $\beta$ corresponds to $\beta=\xbj/x_{q\bar q}=\bar Q^2/(\bar Q^2+P_\perp^2)$. One typically
has $P_\perp^2\sim \bar Q^2$, hence  $\beta\sim 1/2$. See also the discussion in App.~\ref{sec:el}.}
 $\beta\sim 1$, so one must allow for (at least) one gluon emission
in order to obtain a non-trivial diffractive cross-section at smaller values $\beta < 1$.
Moreover, as originally observed in \cite{Wusthoff:1997fz,GolecBiernat:1999qd, Hebecker:1997gp,Buchmuller:1998jv}, 
the $q\bar qg$ contribution is
enhanced by the logarithm  $\ln Q^2/ {Q}_s^2$,  which can be numerically large.
(As before, we assume a hard DIS process, with $Q^2\gg {Q}_s^2(Y_ {\mathbb P})$.)
In the original analysis
\cite{Wusthoff:1997fz,GolecBiernat:1999qd, Hebecker:1997gp,Buchmuller:1998jv}, this logarithm is generated 
by $q\bar qg$ configurations which are similar to our 2+1 jets in the sense of the  transverse momenta
(but not also for the longitudinal momenta; see below). So, it is interesting to see if and how
such a contribution emerges when integrating out our (2+1)-jet final state over the independent 
kinematical variables $P_\perp$, $K_\perp$ and $\vartheta_1$.

Clearly, when performing these integrations, one must respect the kinematical constraints underlying the previous
derivation of \eqn{3jetsD1}, notably the fact that the $q\bar q$ dijets are much harder than both the gluon jet and
the target saturation momentum: $P_\perp^2\gg {Q}_s^2(Y_ {\mathbb P})$ and $P_\perp^2\gg K_\perp^2/(1-x)$
(recall \eqn{PvsKx}). As shown in \cite{Wusthoff:1997fz,GolecBiernat:1999qd}, the logarithmically-enhanced contribution
only occurs for the case of a virtual photon with transverse polarisation, so we shall concentrate on this case.
Starting with \eqn{3jetsD1}, the $q\bar qg$ contribution to the total diffractive cross-section is obtained as
\begin{align}
	\label{qqgDIFFR}
	\frac{\rmd \sigma_{{\rm D}}^{q\bar q g}}{\rmd Y_{\mathbb P}} 
	&\,=\int \rmd \vartheta_1
  	\rmd \vartheta_2
  	\rmd^{2}\!\bm{P}
  	\rmd^{2}\!\bm{K}\,
	\frac{\rmd \sigma_{\rm D}^{\gamma_{T}^* A
	\rightarrow q\bar q g A}}
  	{\rmd \vartheta_1
  	\rmd \vartheta_2
  	\rmd^{2}\!\bm{P}
  	\rmd^{2}\!\bm{K}
  	\rmd Y_{\mathbb P}} \nonumber\\*[0.2cm]
       &\,	=  \int \rmd \vartheta_1
  	\rmd \vartheta_2
  	\rmd^{2}\!\bm{P}
  	\rmd^{2}\!\bm{K}\,\int_\beta^1\rmd x\,\delta\left(x-\beta\, \frac{\bar Q^2+P_\perp^2}{\bar Q^2}\right)
  	 H_{T}(\vartheta_1,\vartheta_2, {Q}^2, P_{\perp}^2)\,
  	 \frac{\dif xG_{\mathbb{P}}(x, x_{\mathbb{P}}, K_\perp^2)}
  	 {\dif^2\bm{K}}\,,
 \end{align}
where in the second line we found it convenient to introduce unity in the form of an integral over  
a $\delta$--function. This is useful in view of changing one of the integration variables, from $P_\perp^2$ to $x$.
Using this $\delta$--function to perform the integration over $P_\perp^2$ and observing that
 \beq\label{HFx}
\frac{P_{\perp}^4 + \bar{Q}^4}
	{(P_{\perp}^2 + \bar{Q}^2)^4}\,=\,\frac{1}{\bar Q^4}\left(\frac{\beta}{x}\right)^2\left[\left(1-\frac{\beta}{x}\right)^2+
	\left(\frac{\beta}{x}\right)^2\right],
\eeq
one deduces (we renote $\vartheta_1\to\vartheta$ for simplicity)
\begin{align}
	\label{qqgDIFFR1}
	\frac{\rmd \sigma_{{\rm D}}^{q\bar q g}}{\rmd Y_{\mathbb P}} 
	&\,={\alpha_{em}\alpha_s}
	\Big(\sum e_{f}^{2}\Big) \,\frac{\pi}{\beta Q^2}\int_0^1 \rmd \vartheta\left(\frac{1-\vartheta}{\vartheta}
	+ \frac{\vartheta}{1-\vartheta}\right)\int_\beta^1\rmd x
	\left(\frac{\beta}{x}\right)^2\left[\left(1-\frac{\beta}{x}\right)^2+
	\left(\frac{\beta}{x}\right)^2\right]
	 \nonumber\\*[0.2cm]
       &\quad\times \int\rmd^{2}\!\bm{K}\,\frac{\dif xG_{\mathbb{P}}(x, x_{\mathbb{P}}, K_\perp^2)}
  	 {\dif^2\bm{K}}
	 \,\Theta\left(\vartheta(1-\vartheta)Q^2\frac{(1-x)(x-\beta)}{\beta}- {K_\perp^2}\right)\,,
 \end{align}
where the $\Theta$--function in the second line implements the condition $P_\perp^2> K_\perp^2/(1-x)$.
Without this $\Theta$--function, the integral over $\vartheta$ would develop logarithmic singularities from
the endpoints at $\vartheta=0$ and $\vartheta=1$. Let us now assume that
\beq\label{thetamin}
\vartheta_{\rm min}\,\equiv\,\frac{K_\perp^2}{Q^2}\,\frac{\beta}{(1-x)(x-\beta)}\,\ll\,1.
\eeq
(We shall later check that this condition is indeed satisfied by the dominant part of the phase-space.)
Then the integral over $\vartheta$ has logarithmic domains at $\vartheta_{\rm min}\ll \vartheta\ll 1$ and also
at $\vartheta_{\rm min}\ll 1-\vartheta\ll 1$, which yield identical contributions (to leading logarithmic
accuracy). This implies\footnote{Note that, in the case of a longitudinal photon, the respective integral over 
$\vartheta$ is not logarithmic anymore, due to the prefactor $\vartheta(1-\vartheta) $ in the structure
of the hard factor $H_L$, cf. \eqn{HardL}.}
\begin{align}
	\label{qqgDIFFR2}
	\frac{\rmd \sigma_{{\rm D}}^{q\bar q g}}{\rmd Y_{\mathbb P}} 
	&\,={\alpha_{em}\alpha_s}
	\Big(\sum e_{f}^{2}\Big) \,\frac{2\pi}{\beta Q^2}\int_\beta^1\rmd x
	\left(\frac{\beta}{x}\right)^2\left[\left(1-\frac{\beta}{x}\right)^2+
	\left(\frac{\beta}{x}\right)^2\right]
	 \nonumber\\*[0.2cm]
       &\quad\times \int\rmd^{2}\!\bm{K}\,\frac{\dif xG_{\mathbb{P}}(x, x_{\mathbb{P}}, K_\perp^2)}
  	 {\dif^2\bm{K}}\,\ln\left(\frac{Q^2}{K_\perp^2}\,\frac{(1-x)(x-\beta)}{\beta}\right),
	 \end{align}
	where in the second line a $\Theta$--function is implicitly understood, which enforces the condition that
	the argument of the logarithm be larger than unity. In turn, this $\Theta$--function implies an upper cutoff
	on the integral over $K_\perp^2$: $K_\perp^2 < Q^2(1-x)(x-\beta)/\beta$.
	
By also recalling the expression  of the Pomeron UGD, that is, \eqn{pomugddef} with the scalar function 	
$\mcal{G}(K_{\perp},x,Y_{\mathbb P})$ from  \eqn{Gscalarnew}, one can check that our above result for
the $q\bar qg$ contribution to the diffractive structure function is indeed equivalent (up to an overall normalisation
factor, due to the different conventions) to that originally obtained in  \cite{Wusthoff:1997fz,GolecBiernat:1999qd}
 (see Eq.~(39) in \cite{GolecBiernat:1999qd}). In particular, the integrand of \eqn{qqgDIFFR2} exhibits the
 expected logarithmic enhancement in the limit $Q^2(1-x)\gg K_\perp^2$. As shown by the above derivation,
this enhancement is generated by integrating over  ``aligned-jet'' configurations,
i.e. $q\bar q$ pairs which are very asymmetric in longitudinal momentum: $\vartheta\ll 1$ or $1-\vartheta\ll 1$.
We shall shortly return to a physical discussion of these configurations.

Before that, we would like to make one more step and rewrite
 \eqn{qqgDIFFR2} in such a way that  {\it collinear factorisation} becomes manifest.
Following the discussion in Sect.~\ref{sect:pomeron}, we expect 
the integral over $K_\perp$ in \eqn{qqgDIFFR2} to be dominated by values
$K_\perp\sim \tilde{Q}_s(x,Y_ {\mathbb P})$ (see also below). To the logarithmic accuracy of interest, we can therefore replace
$K_\perp^2\to \tilde{Q}_s^2(x,Y_ {\mathbb P})=(1-x)Q_s^2(Y_ {\mathbb P})$ 
within the argument of the logarithm in \eqn{qqgDIFFR2} and thus recover collinear factorisation, as anticipated:
\begin{align}
	\label{qqgDIFFRcoll}
	\frac{\rmd \sigma_{{\rm D}}^{q\bar q g}}{\rmd Y_{\mathbb P}} 
	\,=\frac{2\pi{\alpha_{em}\alpha_s}}{Q^2}\,\Big(\sum e_{f}^{2}\Big)
	\int_\beta^1\frac{\rmd x}{x}\,\frac{\beta}{x}&\left[\left(1-\frac{\beta}{x}\right)^2+
	\left(\frac{\beta}{x}\right)^2\right]\,\ln\left(\frac{Q^2}{Q_s^2(Y_ {\mathbb P})}\,\frac{x-\beta}{\beta}\right)\nn
	 &\quad xG_{\mathbb{P}}\left(x, x_{\mathbb{P}}, Q^2\frac{(1-x)(x-\beta)}{\beta}\right).
	 \end{align}
This result can be understood as follows (see also \cite{Buchmuller:1998jv}). 
In the target infinite momentum frame, where the collinear factorisation
admits a partonic interpretation, the cross-section ${\rmd \sigma_{{\rm D}}^{q\bar q g}}/{\rmd Y_{\mathbb P}}$ is proportional
to the quark DPDF. Then \eqn{qqgDIFFRcoll} represents the particular contribution to
this distribution where the quark has been generated from the gluon DPDF (our Pomeron gluon distribution 
$xG_{\mathbb{P}}\left(x, x_{\mathbb{P}}, P_\perp^2\right)$) via one-step in the DGLAP evolution. The expression 
inside the square brackets is recognised as the DGLAP splitting function
for the branching of a gluon into a quark-antiquark pair, with splitting
fractions ${\beta}/{x}$ and  $1-{\beta}/{x}$, respectively. And the logarithm $\ln(Q^2/Q_s^2(Y_ {\mathbb P}))$ is the
transverse phase-space available to this branching. The measured quark carries a longitudinal momentum fraction $\beta$ w.r.t. the Pomeron, that is, $\beta x_{\mathbb{P}}=\xbj$ w.r.t. the nucleon target.

As a final consistency check, let us verify that the integral in \eqn{qqgDIFFRcoll} is indeed
controlled by values of $x$ such that the argument of the logarithm within the integrand is much larger than one.
Since we already assumed that $Q^2\gg {Q}_s^2(Y_ {\mathbb P})$, it is enough to check that this integral
is not specially sensitive to its endpoints at $x\to \beta$ and $x\to 1$. So, let us study the integrand in these 2 limits.

The Pomeron gluon distribution in \eqn{qqgDIFFRcoll} is evaluated at a resolution scale 
$P^2\equiv Q^2(1-x)(x-\beta)/\beta$, which vanishes when either $x\to \beta$, or $x\to 1$. Hence,
to study its behaviour near the endpoints, one can estimate $ xG_{\mathbb{P}}(x, x_{\mathbb{P}}, P^2)$
by using the black disk limit \eqref{pomugdlow}, which is valid when  $P^2\ll \tilde{Q}_s^2(x,Y_ {\mathbb P})$. This yields
\begin{align}
	\label{xGPsmallP}	
	xG_{\mathbb{P}}(x, x_{\mathbb{P}}, P^2)
	\simeq 
	\frac{S_\perp (N_c^2-1)}{4\pi^3} \,\frac{1-x}{2}\,P^2\quad
	\mathrm{for} 
	\quad
	P^2 \ll \tilde{Q}_s^2(x,Y_ {\mathbb P}).
 \end{align}
For the relevant value $P^2= Q^2(1-x)(x-\beta)/\beta$, the distribution \eqref{xGPsmallP} vanishes fast enough
when either  $x\to \beta$, or $x\to 1$, to ensure that the endpoints contributions to the integral \eqref{qqgDIFFRcoll}
are indeed negligible. Besides justifying the leading-logarithmic approximation leading to
Eq.~\eqref{qqgDIFFRcoll}, this conclusion also allows us to use the large-$P^2$ limit of 
$xG_{\mathbb{P}}(x, x_{\mathbb{P}}, P^2)$ in \eqn{xGPhigh} to get a parametric estimate:
\begin{align}
	\label{qqgDIFFRparam}
	\frac{\rmd \sigma_{{\rm D}}^{q\bar q g}}{\rmd Y_{\mathbb P}} \,\sim\,
       S_\perp\alpha_{em}\alpha_s\Big(\sum e_{f}^{2}\Big)(N_c^2-1)\,\mathcal{F}(\beta)\,\frac{{Q}_s^2(Y_ {\mathbb P})}{Q^2}\,
        \ln\frac{Q^2}{Q_s^2(Y_ {\mathbb P})},
 \end{align}
with $\mathcal{F}(\beta)$ a slowly varying function, as obtained after performing the integral over $x$ in \eqn{qqgDIFFRcoll}.

We conclude this section with a couple of comments on the aligned-jet configurations. 
Albeit relatively abundant,
in the sense that they provide the dominant $q\bar q g$ contribution \eqref{qqgDIFFRcoll}  to the diffractive structure function,
these configurations are not so interesting for a measurement of {\it jets} in the final state. 
Indeed, the dijets initiated by asymmetric  $q\bar q$ pairs with $\vartheta(1-\vartheta)  \ll 1$
are relatively soft, with transverse momenta  $P_\perp^2\sim \vartheta(1-\vartheta)Q^2\ll Q^2$. One of them
also has a small longitudinal momentum.
This is not a very favorable situation for jet observation --- especially in the small-$x$ kinematics at the EIC,
where $Q^2$ itself is not very large: for $\xbj\sim 10^{-3}$, one should have $Q^2\lesssim 10$~GeV$^2$.
This is the reason why, in most part of this paper, we rather focused on quasi-symmetric 
dijets, with $\vartheta\sim \order{1}$.

Our last point refers to a comparison between the $q\bar q g$ contribution to  the diffractive structure function,
cf.  \eqn{qqgDIFFRcoll}, and the respective contribution of the  $q\bar q$ pair alone
\cite{GolecBiernat:1999qd}, which counts at zeroth order in $\alpha_s$.
 (The calculation of the latter is briefly reviewed in App.~\ref{sec:el} for
convenience.) In both cases, the
dominant contributions come from aligned-jet $q\bar q$ pairs with $\vartheta(1-\vartheta)  \ll 1$. Yet,
whereas in the first case ($q\bar q g$ fluctuation), this  $q\bar q$ pair is still quite hard, 
 $\vartheta(1-\vartheta)Q^2\gg Q_s^2(Y_ {\mathbb P})$, and thus scatters only weakly, in the second
 case ($q\bar q$ fluctuation),  the $q\bar q$ dipole is {\it semi-hard}, 
 $\vartheta(1-\vartheta)Q^2\lesssim Q_s^2(Y_ {\mathbb P})$, so its scattering is strong (see App.~\ref{sec:el}
for details).
  In both cases, the respective contribution to the diffractive cross-section is controlled by the black disk limit, 
  but the partonic (sub)system which participates in the scattering is not the same: for the LO contribution,
   this is the  $q\bar q$ pair itself, whereas for  the $q\bar q g$ contribution, it is rather the semi-hard gluon 
   (or, equivalently, the effective gluon-gluon dipole).

 \section{Conclusions and perspectives}
 \label{sec:conc}

In this paper, we have studied the opportunities offered by diffractive jet production in photon-nucleus collisions 
as a probe of gluon saturation in QCD at high energies. Our main observation is that an interesting process to that
purpose is the production of three jets via coherent diffraction, in a special kinematics that we refer to as
``2+1 jets in the correlation limit''.
This is the regime where two of the jets (initiated by a quark-antiquark pair) are {\it hard}, 
with transverse momenta much larger than the target saturation scale $Q_s(A,Y_ {\mathbb P})$,
and with large and comparable longitudinal momenta, whereas the third, gluon,  jet is {\it semi-hard}, 
with a transverse momentum  of the order of $Q_s(A,Y_ {\mathbb P})$ and a small longitudinal momentum fraction
$\xi\ll 1$ w.r.t. the virtual photon. 
The third, gluon,  jet may be too soft to be directly observed in the experiments, yet it should
leave important imprints on the final event, that could be observed.  Below are some examples.

First, it is the presence of this semi-hard
jet which ensures that the overall process exists {\it at leading-twist}:  the cross-section for the {\it exclusive}
production of two hard jets ($P_\perp \gg Q_s$, with $P_\perp$ the dijet relative momentum) is
suppressed, roughly, by a factor $Q_s^2/P_\perp^2$, compared to that for diffractive 2+1 jets (see the discussion
in App.~\ref{sec:el}). This suppression 
is due to the colour transparency of small colour dipoles combined with the
elastic nature of the diffractive process, which enhances its sensitivity to the strength
of the scattering. Accordingly, most of
the diffractive  ``hard dijets'' events to be observed at high $P_\perp$ should truly be 2+1 jet events, 
even when the softer jet is not visible in the detector.

Second, this (2+1)--jets diffractive process is controlled by the physics of saturation, hence its
cross-section can be computed in perturbation theory so long as the target saturation momentum
is sufficiently large, $Q_s^2(A,Y_ {\mathbb P})\gg \Lambda^2$ --- as is expected to be the case
for a very large nucleus (Pb, Au...) even if the rapidity gap is only moderately large (say, 
$Y_ {\mathbb P}\gtrsim 3$). In particular, the dependence of the total cross-section \eqref{gluondipColl} upon 
 $Y_ {\mathbb P}$ is controlled by the respective dependence of the saturation momentum $Q_s(A,Y_ {\mathbb P})$,
 hence by the high-energy, BK/JIMWLK, evolution in pQCD.
 
Third, the semi-hard jet controls the transverse momentum imbalance $K_\perp$ between the two hard jets
at the time of scattering: $K_\perp =k_{3\perp}\sim Q_s(A,Y_ {\mathbb P})$. So, one may be tempted to directly measure
$Q_s$ as the dijet imbalance in the final state.  In practice though, this is complicated by the ``Sudakov effect''
(radiation by the hard jets occurring after the scattering), whose strength should be proportional to 
$\alpha_s\ln^2(P_\perp^2/Q_s^2)$. This effect is presumably less important at the Electron-Ion Collider, where the 
 ``hard dijets'' are not {\it much} harder than the saturation scale: in e+Pb events with $\xbj\!=\!10^{-3}$,
 one should be able to access virtualities  $Q^2\le 10$~GeV$^2$ and the measured dijets should not be
 much harder    \cite{Accardi:2012qut,Aschenauer:2017jsk}.  By comparison, 
  the nuclear saturation momentum is estimated as $Q_s^2\simeq 1.5\div 2$~GeV$^2$. On the other hand, the final-state radiation might become the dominant effect in the context of
ultraperipheral heavy ion collisions (UPC) at the LHC, where the transverse momenta of the measured jets are 
much higher, in the ballpark of dozens or even hundreds of GeV.
In order to verify such expectations and also to have a better theoretical control over the dijet imbalance, it would be
important to compute the Sudakov effect for this particular process. We let this to further work.

\begin{figure}[t] 
\centerline{\includegraphics[width=0.7\columnwidth]{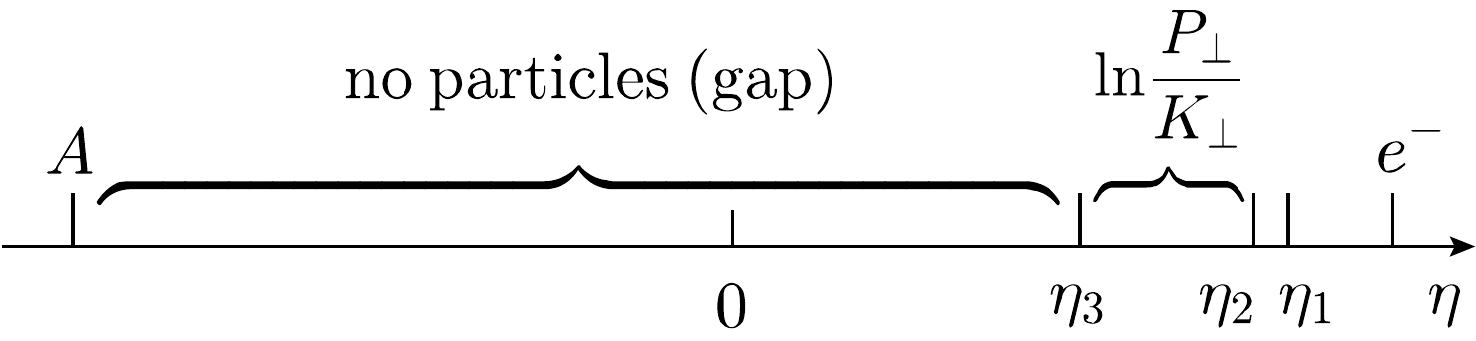}}
\caption{\small The pseudo-rapidity distribution expected for the final state of diffractive 2+1 jet production in DIS.
A similar distribution should be observed for the corresponding events in UPCs, except for replacing the final electron
($e^-$) by the nucleus which emits the photon.}
\label{fig:eta}
\end{figure}

Fourth,  the semi-hard jet will also have an imprint on the pattern of the rapidity gap.  Since propagation angles
are easier to measure, let us formulate our argument in terms of the pseudo-rapidity $\eta=-\ln\tan(\theta/2)$, with $\theta$ 
the angle made by a particle (hadron or jet) w.r.t. the collision axis. For massless particles, this can also be computed
as $\eta=\ln(\sqrt{2}p^+/p_\perp)$. In a Lorentz frame where the virtual photon is ultrarelativistic, 
 like the laboratory frame at the EIC, the two hard jets should emerge at large and comparable pseudo-rapidities:
 $\eta_1\sim\eta_2 \sim \ln(q^+/P_\perp) \gg 1$. The third jet,  which is considerably softer
--- both in longitudinal momentum  ($k^+_3=\xi q^+$ with $\xi\ll 1$) and in transverse momentum 
($k_{3\perp}\sim Q_s\ll P_\perp$) ---,  will also have a significantly
  lower pseudo-rapidity, as shown by the following argument: as emphasised at
  several places, the gluon formation time should not exceed the coherence time of the virtual photon,
  $\tau_3 \lesssim \tau_\gamma$.    This condition can be rewritten as
  \beq
  \frac{k^+_3}{k_{3\perp}^2}\,\lesssim\,\frac{q^+}{Q^2}\,\sim \,\frac{\vartheta(1-\vartheta)q^+}{P_\perp^2}
  \,\lesssim\,\frac{k^+_{\rm min}}{P_\perp^2}\ \ \Longrightarrow\ \ \frac{k^+_3}{k_{3\perp}}\,\lesssim\,
  \frac{k^+_{\rm min}}{P_\perp}\,\frac{k_{3\perp}}{P_\perp}
  \,,\eeq
  where   $k^+_{\rm min}\equiv{\rm min}\{k^+_1,\,k^+_2\}$ and we have used $P_\perp^2\sim 
  \vartheta(1-\vartheta)Q^2$ for typical hard dijets. After also taking the logarithm and using 
  $k_{3\perp}\sim Q_s$, one finds
  \beq\label{deltaeta}
  \eta_3=\ln\frac{\sqrt{2}k^+_3}{k_{3\perp}}\ \lesssim \ \eta_{\rm min} - \ln\frac{P_\perp}{Q_s}\,,\eeq
   where $\eta_{\rm min}\equiv{\rm min}\{\eta_1,\,\eta_2\}$. Since smaller than $\eta_{\rm min}$,
   this $\eta_3$ is also the upper limit of the {\it pseudo}--rapidity gap between 
  the emerging target and the diffractive system (see Fig.~\ref{fig:eta}).
   Hence, even if the third jet is not explicitly measured, 
  one should observe a sizeable pseudo-rapidity separation  $\Delta\eta \gtrsim \ln(P_\perp/Q_s)> 1$ 
  between the upper edge  of the rapidity gap and the two hard jets.
  
  Although our analysis has focused on DIS and was primarily motivated by the forthcoming experimental
 program at the EIC, it should be clear that most of our conclusions should also apply to photon-nucleus
 interactions in the context of ultraperipheral nucleus-nucleus collisions (notably at the LHC). 
 This could be very interesting for studies of gluon saturation, since the energies available in UPCs at the LHC 
 are much higher than the projections for the EIC. The photons exchanged in UPCs are quasi-real and 
 thus transversely polarised. So, the respective  cross-section can be obtained by taking the  $Q^2\to 0$ limit of
  \eqn{3jetsD1} and convoluting the result  with the photon flux describing the energy ($q^+$) distribution 
  of the photons radiated by the ``projectile'' nucleus.
    Recent analyses \cite{ATLAS:2017kwa,CMS:2020ekd,ATLAS:2022cbd} demonstrate that the experimental facilities 
     at the LHC have the capability to distinguish coherent diffraction from the inelastic one and to accurately measure
    rapidity gaps and distributions, together with transverse momentum distributions in wide range, from tens to hundreds of GeV.
    In fact, preliminary results for dijet and multi-jet production via coherent
    diffraction in Pb+Pb UPCs have already been reported in \cite{ATLAS:2022cbd}.

\section*{Acknowledgements} 
We would like to thank Yoshitaka Hatta, Bowen Xiao, and Feng Yuan for insightful discussions and for
informing  us about their recent work in  \cite{Hatta:2022lzj} prior to its publication. Also, we are grateful
to Tuomas Lappi for constructive comments on our manuscript.
The work of E.I. is supported in part by the Agence Nationale de la Recherche project 
 ANR-16-CE31-0019-01.   The work of A.H.M.
is supported in part by the U.S. Department of Energy Grant \# DE-FG02-92ER40699.

\appendix

\section{Exclusive dijet production}
\label{sec:el}

In this Appendix, we shall briefly review the calculation of the cross-section for exclusive dijet production in DIS at small
Bjorken $x$ ($\xbj\ll 1$) and relatively high virtuality $Q^2\gg Q_s^2(Y_ {\mathbb P})$, with two main purposes: \texttt{(i)} to demonstrate
that the production of {\it hard} dijets with $P_\perp^2\gg Q_s^2(Y_ {\mathbb P})$ is strongly suppressed (by an additional
inverse power of $P_\perp^2$ and/or $Q^2$ as compared to the diffractive production of 2+1 jets), and  \texttt{(ii)} that the
$q\bar q$ contribution to the diffractive structure function is controlled by semi-hard, aligned-jet, configurations, with
$P_\perp^2\sim\vartheta(1-\vartheta) Q^2\sim Q_s^2(Y_ {\mathbb P})$, which are sensitive to gluon saturation in the target,
but lead to a pair of relatively soft and very asymmetric jets in the final state.

 By {\it exclusive} dijet production we mean a coherent 
process where the scattering is elastic on both the projectile (the $q\bar q$ pair) and the target side. For this process,
the relevant value of $x_{\mathbb P}$ (the fraction of the target longitudinal momentum $P_N^-$ which is transmitted
to the diffractive system) is $x_{q\bar q}$  as defined in \eqn{xqqdef}; hence, the rapidity gap reads
$Y_ {\mathbb P}=Y_{q\bar{q}}\equiv \ln(1/x_{q\bar q})$, while the complimentary interval  is $\Ybj-Y_{q\bar{q}}=\ln(1/\beta)$ with
$\beta=\xbj/x_{q\bar q}\simeq Q^2/(Q^2+ M_{q\bar q}^2)$ and $M_{q\bar q}^2\simeq P_\perp^2/[\vartheta(1-\vartheta)]$.

For more clarity, we start by displaying the cross section for {\it inclusive} dijet production at leading order;
for the case of a transverse photon, this reads \cite{Dominguez:2011wm}
\begin{align}
  \label{2jetinc}
  \frac{\rmd \sigma^{\gamma_T^* A
  \rightarrow q\bar qX}}
  {\rmd k_1^+
  \rmd k_2^+
  \rmd^{2}\bm{k}_{1}
  \rmd^{2}\bm{k}_{2}} 
  =\frac{\alpha_{em}}{2 \pi^2 q^+}
  \left(\sum e_{f}^{2}\right)
  \left(\vartheta_1^2+\vartheta_2^2\right)
  \delta(q^{+}-k_1^+-k_2^+)
  \big \langle \tilde{A}^l_{\alpha\beta} \tilde{A}^{l*}_{\beta\alpha} \big \rangle_{Y_{q\bar{q}}}, 
\end{align}
with the following amplitude ($\alpha$ and $\beta$ are color indices in the fundamental representation):
\begin{align}
        \label{app:tildeA}
        \tilde{A}_{\alpha\beta}^{l} = 
        &\int\frac{\rmd^2\bx}{2\pi}
        \int\frac{\rmd^2\by}{2\pi}\,
        e^{-i \bk_1\cdot \bx - i \bk_2\cdot \by}\,
        \frac{r^l}{r}\,
        \bar{Q} K_1(\bar{Q}r)\,
        \big[V(\bx)V^{\dagger}(\by)-1\big]_{\alpha\beta}.
\end{align}

In order to pass from the inclusive cross section to the elastic one, we require that the final state be a color singlet, that is, the quark and the antiquark must recombine into a colour dipole after the scattering. This projection is achieved with the replacement 
(cf. \eqn{projD})
\begin{align}
        \label{vvel}
        \big[V(\bx)V^{\dagger}(\by)-1\big]_{\alpha\beta}
        \to 
        \frac{1}{N_c}
        {\rm tr} 
        \big[
        V(\bx)V^{\dagger}(\by) - 1 
        \big]
        \delta_{\alpha\beta},
\end{align}
meaning that the CGC average of the scattering operator reduces to a product of dipole amplitudes
for colour dipoles in the fundamental representation
(we denote $T(\bx,\by)\equiv {S}_{q\bar{q}}(\bm{x}, \bm{y})$, cf. \eqn{proj1}):
\begin{align}
        \label{ttave}
        \big\langle
        T(\bx,\by)\,
        T^*(\bar{\bx},\bar{\by})
        \big\rangle = 
        \mcal{T}(\bx,\by)\,
        \mcal{T}^*(\bar{\bx},\bar{\by})
        + \mcal{O}(1/N_c^2).
\end{align}
The first, leading in $N_c$, contribution corresponds to a coherent collision for the nucleus. The $1/N_c^2$ piece, which can be also explicitly expressed in terms of the average colour dipole in the Gaussian approximation \cite{Dumitru:2011vk,Iancu:2011ns,Iancu:2011nj,Dominguez:2011wm}, corresponds to the case when the nucleus breaks up \cite{Mantysaari:2019hkq}. In what follows we shall focus on the coherent part of the process, for which it becomes clear that the cross section factorizes between the DA and the CCA. That is, it can be written as the modulus squared of an average amplitude 
describing both the decay of the virtual photon and the elastic scattering of the ensuing $q\bar q$ pair 
off the hadronic target with the latter remaining intact.

Now we do the usual change of variables given in Eqs.~\eqref{PandK}-\eqref{bandrinv} (where we let $\vartheta_1+ \vartheta_2=1$) so that the exponent in the double Fourier transform of the DA becomes
\begin{align}
        \label{phase2jet}
        \bm{k}_{1}\cdot\bm{x}+ 
        \bm{k}_{2}\cdot\bm{y}=
        \bm{P}\cdot\bm{r}+ 
        \bm{K}\cdot\bm{b}
\end{align}
and similarly for the CCA. We further assume that the target is homogeneous, so that the scattering amplitude depends only on the separation $\br$. Then the integrations over the impact parameters $\bm{b}$ and $\bar{\bm{b}}$ can be done (like in Eq.~\eqref{intb}) to give 
\begin{align}
        \label{intb2jet}
        \int \frac{\rmd^2 \bar{\bb}}{2\pi}\,
        e^{i \bm{K}\cdot\bar{\bm{b}}} 
        \int \frac{\rmd^2 \bb}{2\pi}\,
        e^{-i \bm{K}\cdot\bm{b}} 
        =\int \rmd^2 \bar{\bb}\,
        \delta^{(2)}(\bm{K})=
        S_{\perp} \delta^{(2)}(\bm{K}),
\end{align}
so that there is no net transfer of transverse momentum  from the target to the $q\bar q$ pair and the final jets propagate back-to-back in the transverse plane\footnote{This back-to-back peak can be broadened by  the Sudakov effect associated with final-state radiation 
 \cite{Mueller:2013wwa}, which is not included in this calculation.}.

Putting everything together and integrating the trivial $\delta^{(2)}(\bm{K})$ dependence, we arrive at the following expression
for the elastic dijet differential cross section,
\begin{align}
  \label{2jetel}
  \frac{\rmd \sigma_{\rm el}^{\gamma_T^* A
  \rightarrow q\bar q A}}
  {\rmd \vartheta_1
  \rmd \vartheta_2
  \rmd^{2}\bm{P}} 
  =\frac{S_{\perp} \alpha_{em} N_c}{2 \pi^2}
  \left(\sum e_{f}^{2}\right)
  \left(\vartheta_1^2+\vartheta_2^2\right)
  \delta(1 - \vartheta_1 - \vartheta_2)
  \big| \mcal{A}_{\rm el}^l(\bP,\bar{Q})\big|^2,
\end{align}
with the reduced amplitude
\begin{align}
        \label{ael}
        \mcal{A}_{\rm el}^l(\bP,\bar{Q}) =
        \int 
        \frac{\rmd^2 \bm{r}}{2\pi}\,
        e^{- i \bm{P} \cdot \bm{r}}
        \frac{r^l}{r}\,
        \bar{Q}
        K_{1}\big(\bar{Q}r\big)
        \mcal{T}(r)= 
        -\frac{i \bar{Q} P^l}{P_{\perp}}
        \int_0^{\infty} \rmd r\, r J_1(P_{\perp} r)
        K_1(\bar{Q} r) \mcal{T}(r).
\end{align} 
In writing the last equality, we also assumed that the scattering amplitude $\mcal{T}$ depends on the dipole separation only through its magnitude $r$.

For definiteness, we consider that $\mcal{T}(r)$ is given by the MV model, that is, by  \eqn{MVgg} adapted to a colour dipole
in the fundamental representation (meaning that we replace $Q_{gA}^2\to Q_{A}^2$, with $Q_{A}^2 = (C_F/N_c) Q_{gA}^2$).
 For this model, the integration over $r$ in \eqn{ael} can be 
easily done in the two limiting regimes $P_{\perp} \gg Q_s$ and  $P_{\perp} \ll Q_s$. We find
\begin{align}
        \label{aellimits}
        \mcal{A}_{\rm el}^l(P_{\perp},\bar{Q})
        \simeq
    \begin{cases}
    \displaystyle{- 2 i\,
    \frac{\bar{Q}^2 P^l}
    {(P_\perp^2 + \bar{Q}^2)^3}}\,
    Q_A^2 
    \ln \frac{P_\perp^2}{\Lambda^2}\, &
    \mathrm{for} \quad P_\perp\gg Q_{s},
    \\*[0.4cm]
    \displaystyle{ - i \,
    \frac{P^l}{P_\perp^2 + \bar{Q}^2}}\, &
        \mathrm{for} \quad P_\perp\ll Q_{s}
    \end{cases}        
\end{align}
from which the amplitude squared $\big| \mcal{A}_{\rm el}^l(\bP,\bar{Q}) \big|^2$ is trivially obtained. Focusing on the limit $P_\perp\gg Q_{s}$ and recalling that the MV model gluon distribution in this regime reads
\begin{align}
        \label{xgmv}
        xG(x,Q^2) =
        \frac{S_{\perp} N_c}{2\pi^2 \alpha_s}\,
        Q_A^2
        \ln\frac{Q^2}{\Lambda^2},
\end{align}
we have
\begin{align}
        \label{ael2}
        \big| \mcal{A}_{\rm el}^l(P_{\perp},\bar{Q}) \big|^2 = 
        (2\pi)^4
        \frac{\bar{Q}^4 P_{\perp}^2}
        {(P_\perp^2 + \bar{Q}^2)^6}
    \left[\frac{\alpha_s}{N_c} 
    \frac{xG(x,P_{\perp}^2)}
    {S_{\perp}} 
    \right]^2
        \quad 
        \mathrm{for} 
        \quad P_\perp \gg Q_{s}.
\end{align}
This function is peaked at  $P_{\perp} \sim \bar{Q}$, so the typical values of $P_{\perp}^2$ are of the order of
 $ \bar{Q}^2=\vartheta_1\vartheta_2Q^2$.
When $P_{\perp} \sim \bar{Q} \gg Q_s$, the elastic cross section falls like $1/P_{\perp}^6$ and hence it is of higher twist when compared either to the inclusive dijet cross section \cite{Dominguez:2011wm}
or to the diffractive dijet one in which the hard $q\bar{q}$ pair is accompanied by a semi-hard gluon that has been integrated over (cf.~Eqs.~\eqref{HardT} and \eqref{gluondipColl}). 

Now we would like to comment on how the above results get modified in the limit $Q^2 \to 0$ (a kinematic regime which is of interest, for example, in ultra peripheral heavy ion collisions). The second equality in Eq.~\eqref{ael} reduces to
\begin{align}
        \label{aelq0}
        \mcal{A}_{\rm el}^l(P_{\perp},\bar{Q}=0)
        = -\frac{i P^l}{P_{\perp}}
        \int_0^{\infty} \rmd r J_1(P_{\perp} r)
        \mcal{T}(r),
\end{align} 
so that when $P_{\perp} \ll Q_s$, the result of the integration coincides with the $\bar{Q}^2 \to 0$ limit of the lower case in Eq.~\eqref{aellimits}. However, this is not true when $P_{\perp} \gg Q_s$, since in the upper case in Eq.~\eqref{aellimits} we have shown only the logarithmically enhanced contribution, but which now vanishes in the limit of interest since proportional to $\bar{Q}^2$. From \eqref{aelq0} we find
\begin{align}
        \label{aelq0high}
        \mcal{A}_{\rm el}^l(P_{\perp},\bar{Q}=0) 
        \simeq  - \frac{i Q_A^2 P^l}{P_{\perp}^4}
\quad \text{for} \quad  P_\perp \gg Q_{s},
\end{align}
from which it is straightforward to get $\big| \mcal{A}_{\rm el}^l(P_{\perp},\bar{Q}=0) \big|^2  \simeq Q_A^4/P_{\perp}^6$.

For completeness, we add that the longitudinal cross section exhibits the same qualitative behavior when $P_{\perp} \sim \bar{Q} \gg Q_s$, i.e.~it falls like $1/P_{\perp}^6$. On the contrary, it vanishes (quadratically) when $Q^2 \to 0$, as expected for a longitudinal photon.

We now move to the second point that we would like to discuss in the Appendix, namely the contribution of the exclusive
dijet production to the diffractive structure function at $\xbj\ll 1$ and $Q^2\gg Q_s^2(Y_ {\mathbb P})$. For a transverse photon, 
this is obtained by integrating the differential cross-section \eqref{2jetel} over $\vartheta_1$, $\vartheta_2$ and $\bP$ under the
 constraint that $Y_ {\mathbb P}=Y_{q\bar{q}}$ or, equivalently, $\beta= \bar Q^2/(\bar Q^2+ P_\perp^2)$.
\begin{align}
  \label{F2el}
  \frac{\rmd \sigma_{{\rm el}}^{q\bar q}}{\rmd Y_{\mathbb P}} 
	&\,=\int \rmd \vartheta_1
  	\rmd \vartheta_2
  	\rmd^{2}\!\bm{P}\,  \beta\delta\left(\beta-\frac{\bar Q^2}{\bar Q^2+ P_\perp^2}\right)
  \frac{\rmd \sigma_{\rm el}^{\gamma_T^* A
  \rightarrow q\bar q A}}
  {\rmd \vartheta_1
  \rmd \vartheta_2
  \rmd^{2}\bm{P}}\,.
\end{align}
Up to an overall factor which can be easily reinserted in the final result, this yields 
\begin{align}
  \label{F2el2}
  \frac{\rmd \sigma_{{\rm el}}^{q\bar q}}{\rmd Y_{\mathbb P}} 
	&\,=\int \rmd \vartheta  	
	\rmd^{2}\!\bm{P}\, \beta\delta\left(\beta-\frac{\bar Q^2}{\bar Q^2+ P_\perp^2}\right)
\left(\vartheta^2+(1-\vartheta)^2\right)
   \big| \mcal{A}_{\rm el}^l(\bP,\bar{Q})\big|^2,\end{align}
   where we have trivially integrated over $\vartheta_2$ and denoted $\vartheta_1\equiv \vartheta$; so, in particular,
    $ \bar{Q}^2=\vartheta(1-\vartheta)Q^2$. At this level, it is convenient to use the above $\delta$-function to perform
    the integral over $\vartheta$, by writing
  \begin{align}  \label{delta}
  \beta\delta\left(\beta-\frac{\bar Q^2}{\bar Q^2+ P_\perp^2}\right)&\,=\frac{\vartheta(1-\vartheta)}{1-\beta}\,
  \delta\left(\vartheta(1-\vartheta)-\frac{\beta}{1-\beta}\frac{P_\perp^2}{Q^2}\right)\nn
  &\,\simeq \frac{2\vartheta}{1-\beta}\,
  \delta\left(\vartheta-\frac{\beta}{1-\beta}\frac{P_\perp^2}{Q^2}\right),
  \end{align}
where the approximation in the second line can be understood as follows: as we shall shortly see, the dominant contribution
comes from the aligned $q\bar q$ configurations with $\vartheta(1-\vartheta)\ll 1$, meaning from the corners of the longitudinal
phase-space at $\vartheta\ll 1$, or $1-\vartheta\ll1$. Both corners give identical contributions, hence it suffices to choose e.g.
$\vartheta\ll 1$ and multiply the result by 2. This is what we have done in the second line of \eqn{delta}. Then \eqn{F2el2}
implies
\begin{align}
  \label{F2el3}
  \frac{\rmd \sigma_{{\rm el}}^{q\bar q}}{\rmd Y_{\mathbb P}} 
	&\,\simeq \frac{2\pi}{1-\beta}\int\rmd P_\perp^2\,\vartheta^*
	  \big| \mcal{A}_{\rm el}^l(\bP,\bar{Q^*})\big|^2,\end{align}
    where $\bar{Q^*}^2\equiv \vartheta^*Q^2$ and
    \beq
    \vartheta^*\equiv \frac{\beta}{1-\beta}\frac{P_\perp^2}{Q^2}
    \eeq
is indeed small, $\vartheta^*\ll 1$, since, typically,  $P_\perp^2\sim Q_s^2 \ll Q^2$, as we shall shortly argue. Since we only aim
at parametric estimates, one can use the limiting behaviours of the elastic amplitude exhibited in \eqn{aellimits}.
Consider e.g. the low momentum regime at $P_\perp\ll Q_s$, where the second line in  \eqn{aellimits} applies. By also
cutting the integral at $P_\perp\sim Q_s$, one easily finds
\begin{align}
  \label{F2el4}
  \frac{\rmd \sigma_{{\rm el}}^{q\bar q}}{\rmd Y_{\mathbb P}} \,\simeq\,2\pi\beta\,\frac{Q_s^2}{Q^2}\,,
  \end{align}
where the numerical factor is strictly speaking not under control. The part of this result that one can trust, is the fact
that the dominant contribution comes from the upper limit of the integral over $P_\perp$, i.e. from $P_\perp\sim Q_s$, 
and also the proportionality with $\beta$ --- i.e. the fact the contribution of exclusive $q\bar q$ production to the diffractive
structure function becomes very small when $\beta\ll 1$. Similar conclusions can be obtained if one starts with the 
large momentum regime at $P_\perp\gg Q_s$, where the first  line in  \eqn{aellimits} applies. In that case, the integral
in \eqn{F2el3} is dominated by its lower limit, i.e. by $P_\perp\sim Q_s$ again, and the result has the same
parametric behaviour as shown in \eqn{F2el4}. Restoring the overall factor from \eqn{2jetel}, we finally arrive at the fiollowing
parametric estimate for the $q\bar q$ contribution to the diffractive cross-section at $\xbj\ll 1$ and $Q^2\gg Q_s^2(Y_ {\mathbb P})$:
\begin{align}
  \label{F2qq}
  \frac{\rmd \sigma_{{\rm el}}^{q\bar q}}{\rmd Y_{\mathbb P}} \,\sim\, {S_{\perp} \alpha_{em} N_c}
    \left(\sum e_{f}^{2}\right)\beta\,\frac{Q_s^2(Y_ {\mathbb P})}{Q^2}\,.
  \end{align}
 As demonstrated by the previous arguments, this dominant contribution comes from relatively large $q\bar q$ dipoles, with
 transverse sizes $r\sim 1/P_\perp \sim 1/Q_s$, which undergo strong scattering: $\mathcal{T}(r)\sim 1$.

\section{The instantaneous pieces of the virtual photon LCWF}
\label{sect:lcwf}

For convenience, let us summarise here the contributions to the $q\bar q g$ Fock state of a transverse virtual photon
which involve the instantaneous piece of the (anti)quark propagator. We take these results from Ref.~\cite{Beuf:2017bpd}
and adapt them to the present notations and to limit where the gluon is soft ($\xi\ll 1$).

The instantaneous gluon emission by the antiquark yields a contribution
\begin{align}
	\label{psiinstq}
	\Psi^{ij,\,\alpha\beta}_{\bar q-{\rm inst}, \,
	\lambda_1\lambda_2} = 
	\delta_{\lambda_{1}\lambda_{2}}
	\frac{e e_f g q^+}{2(2\pi)^6}\,
	\frac{t^a_{\alpha\beta}}{\sqrt{\xi}}\,
	\frac{\delta^{ij} + 
	2i\varepsilon^{ij} 
	\lambda_{1}} {\vartheta_2}\,
	\frac{\xi}{k_{3\perp}^2 + \mcal{M}^2},
\end{align}
whereas the corresponding contribution by the quark reads
\begin{align}
	\label{psiinstqbar}
	\Psi^{ij,\,\alpha\beta}_{q-{\rm inst}, 
	\,\lambda_1\lambda_2} = -
	\delta_{\lambda_{1}\lambda_{2}}
	\frac{e e_f g q^+}{2(2\pi)^6}\,
	\frac{t^a_{\alpha\beta}}{\sqrt{\xi}}\,
	\frac{\delta^{ij} - 
	2i\varepsilon^{ij}
	\lambda_{1}} {\vartheta_1}\,
	\frac{\xi}{k_{3\perp}^2 + \mcal{M}^2}.
\end{align}
Putting these two parts together and using $\vartheta_1+\vartheta_2\simeq 1$ together with the definition
\eqref{phidef} of the helicity function $\varphi_{\lambda_{1}\lambda_2}^{ij}$, one finally
 obtains
\begin{align}
	\label{psiinstqq}
	\Psi^{ij,\,\alpha\beta}_{{\rm inst}, \,
	\lambda_1\lambda_2} = 
	\frac{e e_f g q^+}{(2\pi)^6}\,
	\frac{t^a_{\alpha\beta}}{\sqrt{\xi}}\,
	\frac{\xi}{2\vartheta_1\vartheta_2}\,
	\frac{\varphi_{\lambda_{1}\lambda_2}^{ij}
	(\vartheta_1)}{k_{3\perp}^2 + \mcal{M}^2},
\end{align}
which, remarkably, has the same helicity structure as the regular terms in e.g. Eq.~\eqref{Psieik}.  Up to an overall
factor (that was systematically excluded in the manipulations in Sect.~\ref{sect:noneik}),
this is the same as the result previously shown in \eqn{psiinst}.

\section{Double logarithmic terms in the limit $x \to 1$}
\label{sect:x1}

As evident in Eq.~\eqref{plus}, one step in DGLAP evolution is accompanied by an extra logarithmic factor $\ln(1-x)$ which can be large in the limit $x \to 1$. Here, we would like to take into account double logarithmic terms which are missed by DGLAP. 

\begin{figure}[t] 
\centerline{
\includegraphics[width=0.47\textwidth]{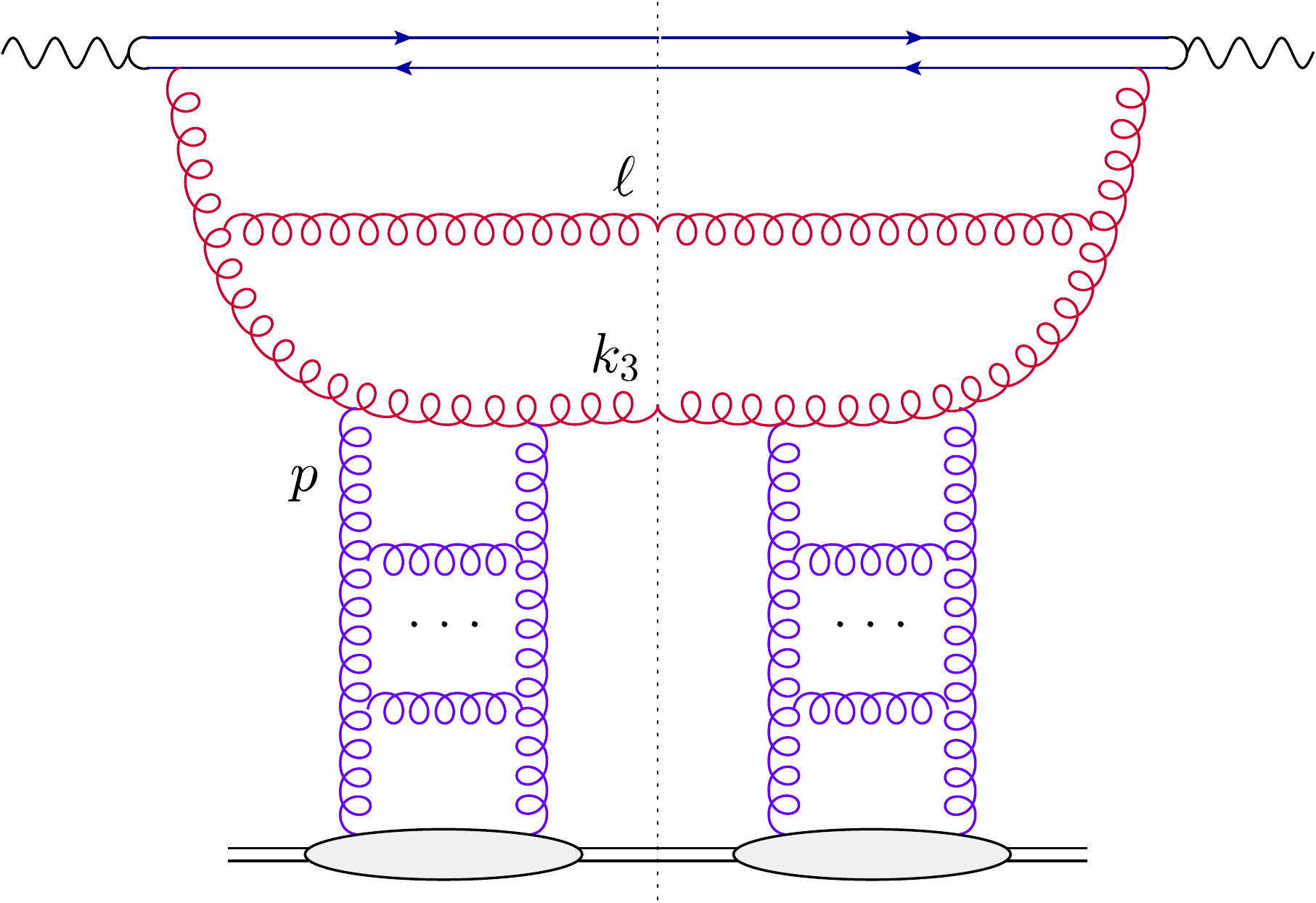}
\hspace*{0.05\textwidth}
\includegraphics[width=0.47\textwidth]{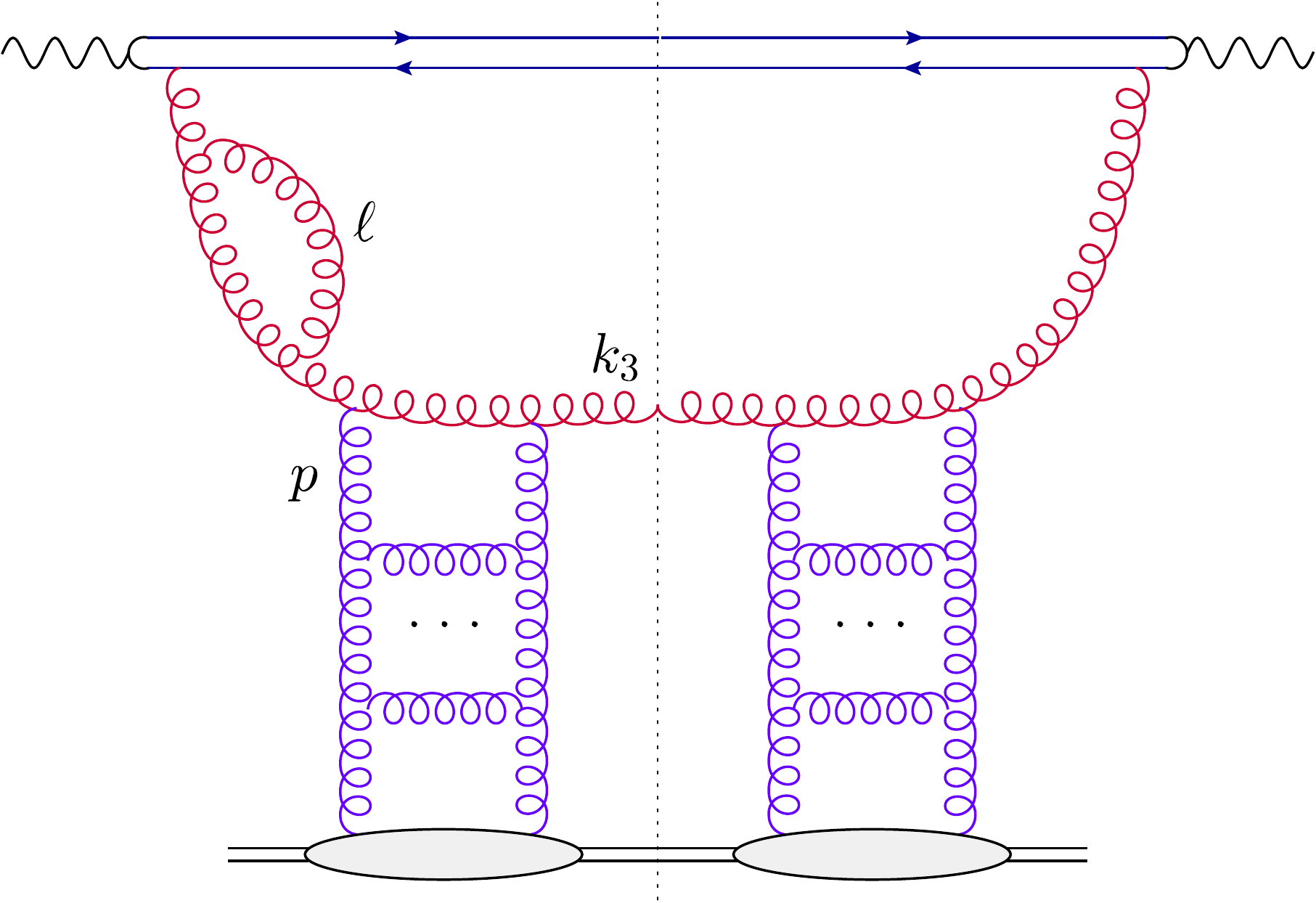}}
\caption{\small Diagrams with an additional real (left) or virtual (right) gluon which contribute to the diffractive process at NLO.}
\label{fig:xto1}
\end{figure}

For the problem under consideration it is natural to work in the light cone gauge of the target nucleus which we recall is a left mover, that is we take $A_a^{-}=0$. The relevant graphs for the 3-jet diffractive process under consideration at next-to-leading order (NLO) are shown in Fig.~\ref{fig:xto1}. For both the real and virtual processes at NLO there is a logarithmic regime of integration in the transverse $\bm{\ell}$ and longitudinal $\ell^-$ momenta of the additional gluon emitted, i.e.~we obtain 
\begin{align}
        \label{xgp1g}
        xG_{\mathbb P}^{(1)} = 
        xG_{\mathbb P}^{\rm tree} \times
        \frac{\alpha_s N_c}{\pi}
        \int \frac{\dif \ell_{\perp}^2}{\ell_{\perp}^2}
        \int \frac{\dif \ell^-}{\ell^-},
\end{align}
with 
\begin{align}
        \label{xgptree}        
        xG_{\mathbb P}^{\rm tree} = 
        \kappa(1) Q_s^2(Y_{\mathbb P}) (1-x)^2
\end{align}
and where, for the sake of simplicity, we have assumed the coupling to be fixed. The non trivial task is to determine the limits of integration. 

Regarding the real term, a lower limit for the $\ell^-$ integration arises from the requirement that the soft gluon $\ell$ be emitted before the $q\bar{q}$ dijet, more precisely (recall that for a left-moving system the light cone time is $x^-$ which is dual to the light cone plus momentum)
\begin{align}
        \label{llow}
        \frac{2 \ell^-}{\ell_{\perp}^2} \gg \frac{2 p^-}{Q^2},
\end{align}
where $p^-$ refers to the longitudinal momentum of the Pomeron. Un upper limit in the $\ell^-$ integration is imposed from the fact that $Y_{\mathbb P}$ and the diffractive mass are fixed by the minus longitudinal momentum of the gluon with momentum $\bm{k}_3$, that is we need
\begin{align}
        \label{lhigh}
        \ell^- \ll k_3^- = (1-x) p^-.
\end{align}
The lower limit in the transverse momentum $\ell_{\perp}^2$ integration is $\mu^2$, which is either a hadronic non-pertubative scale or $Q_s^2(Y_{\mathbb P})$ in the presence of saturation. The upper limit is $Q^2(1-x)$ (and not $Q^2$) since this is the maximal allowed value for which the lower limit in the $\ell^-$-integration is smaller than the upper one. Thus, putting everything together we obtain the real term
\begin{align}
        \label{xgp1real}
        xG_{\mathbb P}^{(1)} \big |_{\rm real} = 
        xG_{\mathbb P}^{\rm tree} \times
        \frac{\alpha_s N_c}{\pi}
        \int_{\mu^2}^{Q^2(1-x)} 
        \frac{\dif \ell_{\perp}^2}{\ell_{\perp}^2}
        \int_{(\ell_{\perp}^2/Q^2)p^-}^{(1-x)p^-}
        \frac{\dif \ell^-}{\ell^-} 
        =
        xG_{\mathbb P}^{\rm tree} \times
        \frac{\alpha_s N_c}{2 \pi}
        \ln^2\frac{Q^2(1-x)}{\mu^2}.
\end{align}

For the virtual term, the lower limit in the integration over $\ell^-$ remains as in Eq.~\eqref{llow}. However, the gluon is virtual and it cannot take away any minus longitudinal momentum, so the upper limit constraint in Eq.~\eqref{lhigh} does not apply in this case. Still, in order that the logarithmic integration in \eqref{xgp1g} exist, the $\ell$-denominator ($\ell^+$) must dominate the $k_3$-denominator ($k_3^+$), that is 
\begin{align}
        \label{lhighv}
        \frac{2 \ell^-}{\ell_{\perp}^2} \ll
        \frac{2 (1-x) p^-}{\mu^2},
\end{align}
where we have assumed $k_{3\perp}^2 \sim Q_s^2(Y_{\mathbb P}) \sim \mu^2$. Since $\ell_{\perp}^2 > \mu^2$ this is a less stringent constraint compared to the one for the real term, thus, as expected, we shall obtain a negative correction when combining both terms. One must be careful when $\ell^2_{\perp} > \mu^2/(1-x)$ to impose the necessary condition $\ell^- < p^-$ since it is not guaranteed by Eq.~\eqref{lhighv}. We split the integration in two regimes to get
\begin{align}
        \label{xgp1virt}
        \hspace*{-0.5cm}
        xG_{\mathbb P}^{(1)} \big |_{\rm virt} = &-  
        xG_{\mathbb P}^{\rm tree} \times
        \frac{\alpha_s N_c}{\pi}
        \left[
        \int_{\mu^2}^{\mu^2/(1-x)} 
        \frac{\dif \ell_{\perp}^2}{\ell_{\perp}^2}
        \int_{(\ell_{\perp}^2/Q^2)p^-}^{(\ell_{\perp}^2/\mu^2)(1-x)p^-}
        \frac{\dif \ell^-}{\ell^-}
        +
        \int_{\mu^2/(1-x)}^{Q^2} 
        \frac{\dif \ell_{\perp}^2}{\ell_{\perp}^2}
        \int_{(\ell_{\perp}^2/Q^2)p^-}^{p^-}
        \frac{\dif \ell^-}{\ell^-}
        \right]
        \nn
        = & 
        - xG_{\mathbb P}^{\rm tree} \times
        \frac{\alpha_s N_c}{\pi}
        \left[
        \ln\frac{Q^2(1-x)}{\mu^2}\ln \frac{1}{1-x}
        +
        \frac{1}{2}
        \ln^2\frac{Q^2(1-x)}{\mu^2}
        \right].
\end{align}   

The real and virtual pieces in Eqs.~\eqref{xgp1real} and \eqref{xgp1virt} combine to give
\begin{align}
        \label{xgp1tot}
        xG_{\mathbb P}^{(1)} = -  
        xG_{\mathbb P}^{\rm tree} \times
        \frac{\alpha_s N_c}{\pi}
        \ln\frac{Q^2(1-x)}{\mu^2}\ln \frac{1}{1-x},
\end{align}
and it is not hard to convince oneself that taking into account higher orders the NLO correction exponentiates, that is 
\begin{align}
        \label{xgpall}
        xG_{\mathbb P} = & \, 
        xG_{\mathbb P}^{\rm tree} \times
        \exp 
        \left[ -\frac{\alpha_s N_c}{\pi}
        \ln\frac{Q^2(1-x)}{\mu^2}\ln \frac{1}{1-x}
        \right]
        \nn
        = & \, xG_{\mathbb P}^{\rm tree} \times
        \exp 
        \left[ -\frac{\alpha_s N_c}{\pi}
        \ln\frac{Q^2}{\mu^2}\ln \frac{1}{1-x}
        + 
        \frac{\alpha_s N_c}{\pi}
        \ln^2(1-x)
        \right].
\end{align}
In the second line in the above we have decomposed the double logarithm in the exponent into two pieces. The first one is the contribution generated by DGLAP evolution (under the assumption that the coupling is fixed), while the second one is part of the coefficient function in the operator product expansion. Since such a term is independent of $Q^2$, it is clear that $xG_{\mathbb P}$ as given in Eq.~\eqref{xgpall} still satisfies the DGLAP equation. Repeating the calculation presented in this Appendix with the running of the coupling included, one finds that Eq.~\eqref{xgpall} gets replaced with Eq.~\eqref{DGLAPpDL}.

\bibliographystyle{utcaps}

\providecommand{\href}[2]{#2}\begingroup\raggedright\endgroup

\end{document}